\input harvmac.tex
\input epsf
\input graphicx

\noblackbox
\def\figin{\epsfcheck\figin}\def\figins{\epsfcheck\figins}
\def\epsfcheck{\ifx\epsfbox\UnDeFiNeD
\message{(NO epsf.tex, FIGURES WILL BE IGNORED)}
\gdef\figin##1{\vskip2in}\gdef\figins##1{\hskip.5in}
\else\message{(FIGURES WILL BE INCLUDED)}%
\gdef\figin##1{##1}\gdef\figins##1{##1}\fi}
\def\DefWarn#1{}
\def\figinsert{\goodbreak\midinsert}
\def\ifig#1#2#3{\DefWarn#1\xdef#1{fig.~\the\figno}
\writedef{#1\leftbracket fig.\noexpand~\the\figno}%
\figinsert\figin{\centerline{#3}}\medskip\centerline{\vbox{\baselineskip12pt
\advance\hsize by -1truein\noindent\footnotefont{\bf Fig.~\the\figno } \it#2}}
\bigskip\endinsert\global\advance\figno by1}


\def\encadremath#1{\vbox{\hrule\hbox{\vrule\kern8pt\vbox{\kern8pt
 \hbox{$\displaystyle #1$}\kern8pt}
 \kern8pt\vrule}\hrule}}
 %
 %
 

 \font\cmss=cmss10
 \font\cmsss=cmss10 at 7pt
 \def\rlx{\relax\leavevmode}
 \def\inbar{\vrule height1.5ex width.4pt depth0pt}
 \def\IC{\relax\,\hbox{$\inbar\kern-.3em{\rm C}$}}
 \def\IN{\relax{\rm I\kern-.18em N}}
 \def\IP{\relax{\rm I\kern-.18em P}}

\def\ZZ{\rlx\leavevmode\ifmmode\mathchoice{\hbox{\cmss Z\kern-.4em Z}}
  {\hbox{\cmss Z\kern-.4em Z}}{\lower.9pt\hbox{\cmsss Z\kern-.36em Z}}
  {\lower1.2pt\hbox{\cmsss Z\kern-.36em Z}}\else{\cmss Z\kern-.4em Z}\fi}
 \def\IZ{\relax\ifmmode\mathchoice
 {\hbox{\cmss Z\kern-.4em Z}}{\hbox{\cmss Z\kern-.4em Z}}
 {\lower.9pt\hbox{\cmsss Z\kern-.4em Z}}
 {\lower1.2pt\hbox{\cmsss Z\kern-.4em Z}}\else{\cmss Z\kern-.4em Z}\fi}
 \def\IZ{\relax\ifmmode\mathchoice
 {\hbox{\cmss Z\kern-.4em Z}}{\hbox{\cmss Z\kern-.4em Z}}
 {\lower.9pt\hbox{\cmsss Z\kern-.4em Z}}
 {\lower1.2pt\hbox{\cmsss Z\kern-.4em Z}}\else{\cmss Z\kern-.4em Z}\fi}

 \def\narrowplus{\kern -.04truein + \kern -.03truein}
 \def\narrowminus{- \kern -.04truein}
 \def\narrowminussub{\kern -.02truein - \kern -.01truein}

 \def\ep{{\epsilon}}

 \def\frac#1#2{{#1\over #2}}

 \def\IZ{\relax\ifmmode\mathchoice
 {\hbox{\cmss Z\kern-.4em Z}}{\hbox{\cmss Z\kern-.4em Z}}
 {\lower.9pt\hbox{\cmsss Z\kern-.4em Z}}
 {\lower1.2pt\hbox{\cmsss Z\kern-.4em Z}}\else{\cmss Z\kern-.4em Z}\fi}
 \def\IB{\relax{\rm I\kern-.18em B}}
 \def\IC{{\relax\hbox{$\inbar\kern-.3em{\rm C}$}}}
 \def\Ic{{\relax\hbox{$\inbar\kern-.22em{\rm c}$}}}
 \def\ID{\relax{\rm I\kern-.18em D}}
 \def\IE{\relax{\rm I\kern-.18em E}}
 \def\IF{\relax{\rm I\kern-.18em F}}
 \def\IG{\relax\hbox{$\inbar\kern-.3em{\rm G}$}}
 \def\IGa{\relax\hbox{${\rm I}\kern-.18em\Gamma$}}
 \def\IH{\relax{\rm I\kern-.18em H}}
 \def\II{\relax{\rm I\kern-.18em I}}
 \def\IK{\relax{\rm I\kern-.18em K}}
 \def\IP{\relax{\rm I\kern-.18em P}}

 \font\cmss=cmss10 \font\cmsss=cmss10 at 7pt
 \def\IR{\relax{\rm I\kern-.18em R}}

 %

 %
 %
 \def\eqnn#1{\xdef
#1{(\secsym\the\meqno)}\writedef{#1\leftbracket#1}%
 \global\advance\meqno by1\wrlabeL#1}
 \def\eqna#1{\xdef
#1##1{\hbox{$(\secsym\the\meqno##1)$}}

\writedef{#1\numbersign1\leftbracket#1{\numbersign1}}%
 \global\advance\meqno by1\wrlabeL{#1$\{\}$}}
 \def\eqn#1#2{\xdef
#1{(\secsym\the\meqno)}\writedef{#1\leftbracket#1}%
 \global\advance\meqno by1$$#2\eqno#1\eqlabeL#1$$}

\newdimen\tableauside\tableauside=1.0ex
\newdimen\tableaurule\tableaurule=0.4pt
\newdimen\tableaustep
\def\phantomhrule#1{\hbox{\vbox to0pt{\hrule height\tableaurule width#1\vss}}}
\def\phantomvrule#1{\vbox{\hbox to0pt{\vrule width\tableaurule height#1\hss}}}
\def\sqr{\vbox{%
  \phantomhrule\tableaustep
  \hbox{\phantomvrule\tableaustep\kern\tableaustep\phantomvrule\tableaustep}%
  \hbox{\vbox{\phantomhrule\tableauside}\kern-\tableaurule}}}
\def\squares#1{\hbox{\count0=#1\noindent\loop\sqr
  \advance\count0 by-1 \ifnum\count0>0\repeat}}
\def\tableau#1{\vcenter{\offinterlineskip
  \tableaustep=\tableauside\advance\tableaustep by-\tableaurule
  \kern\normallineskip\hbox
    {\kern\normallineskip\vbox
      {\gettableau#1 0 }%
     \kern\normallineskip\kern\tableaurule}%
  \kern\normallineskip\kern\tableaurule}}
\def\gettableau#1 {\ifnum#1=0\let\next=\null\else
  \squares{#1}\let\next=\gettableau\fi\next}

\tableauside=1.0ex
\tableaurule=0.4pt

\def\IE{\relax{\rm I\kern-.18em E}}
\def\IP{\relax{\rm I\kern-.18em P}}

\lref\tv{
  M.~Aganagic, A.~Klemm, M.~Marino, C.~Vafa,
  ``The Topological vertex,''
Commun.\ Math.\ Phys.\  {\bf 254}, 425-478 (2005).
[hep-th/0305132].
}

\lref\dvI{
R.~Dijkgraaf and C.~Vafa,
``Matrix models, topological strings, and supersymmetric gauge theories,''
Nucl.\ Phys.\ B {\bf 644}, 3 (2002)
[arXiv:hep-th/0206255].
}

\lref\dvII{
R.~Dijkgraaf and C.~Vafa,
``On geometry and matrix models,''
Nucl.\ Phys.\ B {\bf 644}, 21 (2002)
[arXiv:hep-th/0207106].
}

\lref\akmv{
M.~Aganagic, A.~Klemm, M.~Mari\~no and C.~Vafa,
``The topological vertex,''
arXiv:hep-th/0305132.
}

\lref\GGV{
  L.~F.~Alday, D.~Gaiotto, S.~Gukov, Y.~Tachikawa, H.~Verlinde,
  ``Loop and surface operators in N=2 gauge theory and Liouville modular geometry,''
JHEP {\bf 1001}, 113 (2010).
[arXiv:0909.0945 [hep-th]].
}

\lref\adkmv{ M.~Aganagic, R.~Dijkgraaf, A.~Klemm, M.~Marino, C.~Vafa,
  ``Topological strings and integrable hierarchies,''
Commun.\ Math.\ Phys.\  {\bf 261}, 451-516 (2006).
[hep-th/0312085].
}

\lref\no{
N.~Nekrasov and A.~Okounkov,
``Seiberg-Witten theory and random partitions,''
arXiv:hep-th/0306238.
}

\lref\toda{
  R.~Dijkgraaf, C.~Vafa,
  ``Toda Theories, Matrix Models, Topological Strings, and N=2 Gauge Systems,''
[arXiv:0909.2453 [hep-th]].
}

\lref\mmb{M. Marino, "Chern-Simons Theory, Matrix Models, And Topological Strings" (Oxford University Press, 2005).}

\lref\lmv{J. M. F. Labastida, M. Marino, and C. Vafa, �Knots, Links, and Branes At Large N,� JHEP 0011(2000) 007, hep-th/0010102.}

 \lref\AY{M. Aganagic and M. Yamazaki, �Open BPS Wall Crossing and M-theory,� Nucl. Phys. B834 (2010) 258�272, 0911.5342.}

\lref\ems{
  S.~Elitzur, G.~W.~Moore, A.~Schwimmer, N.~Seiberg,
  ``Remarks on the Canonical Quantization of the Chern-Simons-Witten Theory,''
Nucl.\ Phys.\  {\bf B326}, 108 (1989).
}
\lref\mm{
  M.~Marino,
  ``Chern-Simons theory, matrix integrals, and perturbative three-manifold invariants,''
Commun.\ Math.\ Phys.\  {\bf 253}, 25-49 (2004).
[hep-th/0207096].
}

\lref\coloredtrefoil{
H. Fuji, S. Gukov and P. Sulkowski,
  ``Super-A-polynomial for knots and BPS states,''
[arXiv:1205.1515].
}

\lref\ovknot{H.~Ooguri and C.~Vafa,
``Knot invariants and topological strings,''
Nucl.\ Phys.\ B {\bf 577}, 419 (2000)
[arXiv:hep-th/9912123].
}

\lref\gr{N. M. Dunfield, S. Gukov, and J. Rasmussen, �The Superpotential For Knot Homologies,� Experiment. Math. 15 (2006) 129, math/0505662.}

\lref\rasb{{{Rasmussen}, J.}, "{Some differentials on Khovanov-Rozansky homology}", {arXiv:math/0607544}.}

\lref\rasa{J. Rasmussen. Khovanov-Rozansky homology of two-bridge knots and links. arXiv:math.GT/0508510,
2005.}

\lref\gvI{R.~Gopakumar and C.~Vafa,
``On the gauge theory/geometry correspondence,''
Adv.\ Theor.\ Math.\ Phys.\  {\bf 3}, 1415 (1999)
[arXiv:hep-th/9811131].
}
\lref\dimofte{
  T.~Dimofte, S.~Gukov, L.~Hollands,
  ``Vortex Counting and Lagrangian 3-manifolds,''
[arXiv:1006.0977 [hep-th]].
}
\lref\taki{
  M.~Taki,
  ``Flop Invariance of Refined Topological Vertex and Link Homologies,''
[arXiv:0805.0336 [hep-th]].
}

\lref\gvII{
  R.~Gopakumar, C.~Vafa,
  ``M theory and topological strings. 1.,''
[hep-th/9809187];
  R.~Gopakumar, C.~Vafa,
  ``M theory and topological strings. 2.,''
[hep-th/9812127].
}

\lref\wrr{
D. Gaiotto and E. Witten, �Supersymmetric Boundary Conditions In N = 4 Super Yang-Mills Theory,� J. Stat. Phys. (2009) 135 789-855. arXiv:0804.2902.}

\lref\wrrr{ D. Gaiotto and E. Witten, �Janus Configurations, Chern-Simons Couplings, And The Theta-Angle in N=4 Super Yang-Mills Theory,� JHEP 1006 2010 097, arXiv:0804.2907.}

\lref\wp{E. Witten, �A New Look At The Path Integral Of Quantum Mechanics,� arXiv:1009.6032.}

\lref\nikitaa{
N.~A.~Nekrasov,
``Seiberg-Witten prepotential from instanton counting,''
arXiv:hep-th/0206161.
}
%


\lref\kh{
M. Khovanov, �A Categorification Of The Jones Polynomial,� Duke. Math. J. 101 (2000) 359-426.}

\lref\hiv{
T.~J.~Hollowood, A.~Iqbal, C.~Vafa,
``Matrix models, geometric engineering and elliptic genera,''
JHEP {\bf 0803}, 069 (2008).
[hep-th/0310272].
}

\lref\macda{Macdonald, I.G., A new class of symmetric functions, Publ. I.R.M.A. Strasbourg, 372/S- 20, Actes 20 Seminaire Lotharingien (1988), 131-171.}

\lref\macdb{Macdonald, I.G., "Orthogonal polynomials associated with root systems", preprint (1988).}


\lref\ek{Etingof, P. and Kirillov, Jr, A., "On Cherednik-Macdonald-Mehta identities", q-alg 9712051, Electr.Res.An., v.4(1998),p.43-47.
}

\lref\eka{Etingof, P.I. and Kirillov, A.A., Macdonald�s polynomials and representations of quantum groups, Math. Res. Let. 1 (1994), 279�296}

\lref\ekb{Etingof, P.I. and Kirillov, A.A., Representation-theoretic proof of inner product and symmetry identities for Mac- donald�s polynomials, hep-th/9410169, to appear in Comp. Math. (1995).}

\lref\kirila{Kirillov, Jr, A, "Lectures on affine Hecke algebras and Macdonald�s conjectures", Bull. Amer. Math. Soc. 34 (1997), 251�292.}

\lref\kirillov{ Kirillov, Jr., A.,
"{On inner product in modular tensor categories. I}",
 arXiv:q-alg/9508017}

\lref\cheredniko{Cherednik, I., "{Macdonald's Evaluation Conjectures and Difference Fourier Transform}",
arXiv:q-alg/9412016}

\lref\cherednikt{ Cherednik, I.,"Double Affine Hecke Algebras and Macdonald's Conjectures",
The Annals of Mathematics,
Second Series, Vol. 141, No. 1 (Jan., 1995), pp. 191-216}

\lref\co{I. Cherednik, V. Ostrik, "From Double Affine Hecke Algebra to Fourier Transform", Selecta Math. (N.S.) 9, no. 2, 161249, (2003).}

\lref\taubes{C. Taubes, �Lagrangians for the Gopakumar-Vafa conjecture,� math.DG/0201219.}

\lref\refmm{
  A.~Brini, M.~Marino, S.~Stevan,
  ``The Uses of the refined matrix model recursion,''
[arXiv:1010.1210 [hep-th]].
}

\lref\bcov{
M.~Bershadsky, S.~Cecotti, H.~Ooguri and C.~Vafa,
``Kodaira-Spencer theory of gravity and exact results for quantum string amplitudes,''
hep-th/9309140,
Commun.\ Math.\ Phys.\  {\bf 165} (1994) 311.}


\lref\amv{M.~Aganagic, M.~Mari\~no and C.~Vafa,
``All loop topological string amplitudes from Chern-Simons theory,''
hep-th/0206164.}

\lref\akmv{
  M.~Aganagic, A.~Klemm, M.~Marino, C.~Vafa,
  ``Matrix model as a mirror of Chern-Simons theory,''
JHEP {\bf 0402}, 010 (2004).
[hep-th/0211098].
}

\lref\gsv{
S. Gukov, A. S. Schwarz, and C. Vafa, �Khovanov-Rozansky Homology And Topological Strings,� Lett. Math. Phys. 74 (2005) 53-74, hep-th/0412243.
}

\lref\DVV{
  R.~Dijkgraaf, C.~Vafa, E.~Verlinde,
  ``M-theory and a topological string duality,''
[hep-th/0602087].
}

\lref\macdonald{I.G. Macdonald, {\it Symmetric functions and Hall polynomials},
Oxford University Press, 1995.}

  \lref\dbn{D. Bar-Natan, �On Khovanov�s Categorification Of The Jones Polynomial,� arXiv:math/0201043, Alg. Geom. Topology 2 (2002) 337-370.}

  \lref\tur{{Turner}, P., "{Five Lectures on Khovanov Homology}"
 {arXiv:math/0606464}.}

 \lref\ak{{{Asaeda}, M. and {Khovanov}, M.}  "{Notes on link homology}", {arXiv/0804.1279}.}

\lref\hk{{{Huerfano}, R.~S. and {Khovanov}, M.},
 "{Categorification of some level two representations of sl(n)}",
{arXiv:math/0204333}.} \lref\kc{{{Khovanov}, M.},
 "{Categorifications of the colored Jones polynomial}",
{arXiv:math/0302060}.}\lref\web{{{Webster}, B.},
 "{Knot invariants and higher representation theory I: diagrammatic and geometric categorification of tensor products}",
arXiv/1001.2020.}

 \lref\ps{P. Seidel and I. Smith, �A Link Invariant From The Symplectic Geometry Of Nilpotent Slices,� arXiv:math/0405089.}

 \lref\km{P. B. Kronheimer and T. S. Mrowka, �Knot Homology Groups From Instantons,� arXiv:0806.1053}

 \lref\kmt{ P. B. Kronheimer and T. S. Mrowka, �Khovanov Homology Is An Unknot-Detector,� arXiv:1005.4346}

\lref\ik{
A.~Iqbal and A.~K.~Kashani-Poor,
``$SU(N)$ geometries and topological string amplitudes,''
arXiv:hep-th/0306032.
}

\lref\jones{
Jones V.F.R. (1983) Index for subfactors, Invent. Math. 72, 1-25;  Jones V.F.R. (1985) A polynomial invariant for knots via von Neumann
algebras, Bull. Amer. Math. Soc. 12 103-112; Jones, V. F. R. Hecke algebra representations of braid groups and link
polynomials. Ann. of Math. (2) 126 (1987), no. 2, 335�388.}

\lref\wcs{
  E.~Witten,
  ``Quantum Field Theory and the Jones Polynomial,''
Commun.\ Math.\ Phys.\  {\bf 121}, 351 (1989).
}

\lref\giv{S. Gukov, A. Iqbal, C. Kozcaz, and C. Vafa, "Link Homologies and the Refined Topological Vertex," arXiv:0705.1368.}

\lref\civ{A. Iqbal, C. Kozcaz, C. Vafa, "The Refined Topological Vertex", hep-th/0701156.}

\lref\ikhopf{A. Iqbal, C. Kozcaz, "Refined Hopf Link Revisited", arXiv:1111.0525.}

\lref\wst{
  E.~Witten,
 ``Chern-Simons gauge theory as a string theory,''
Prog.\ Math.\  {\bf 133}, 637-678 (1995).
[hep-th/9207094].
}

\lref\wr{
  E.~Witten,
  ``Fivebranes and Knots,''
[arXiv:1101.3216 [hep-th]].
}

\lref\awata{  H.~Awata, H.~Kanno,
 ``Refined BPS state counting from Nekrasov's formula and Macdonald functions,''
Int.\ J.\ Mod.\ Phys.\  {\bf A24}, 2253-2306 (2009).
[arXiv:0805.0191 [hep-th]].
}

\lref\acdkv{
  M.~Aganagic, M.~C.~N.~Cheng, R.~Dijkgraaf, D.~Krefl, C.~Vafa,
  ``Quantum Geometry of Refined Topological Strings,''
[arXiv:1105.0630 [hep-th]].
}
\def\Title#1#2{\nopagenumbers\abstractfont\hsize=\hstitle\rightline{#1}
\vskip .5in\centerline{\titlefont #2}\abstractfont\vskip .5in\pageno=0}

{
\Title
{\vbox{
 \baselineskip12pt
}}
{\vbox{
\centerline{Knot Homology }
\vskip 0.4 cm
\centerline{from }
\vskip 0.4cm
\centerline{Refined Chern-Simons Theory}
}}
 \centerline{{\bf Mina Aganagic}$^{a,b}$ and {\bf Shamil Shakirov}$^{a,c}$}
\vskip 0.5cm

\centerline{$^a$ \it Department of Mathematics, University of California, Berkeley, USA}
\centerline{$^b$ \it Center for Theoretical Physics, University of California, Berkeley, USA}
\centerline{$^c$ \it Institute for Theoretical and Experimental Physics, Moscow, Russia}

\smallskip
\vskip 0.1cm
\centerline{\bf Abstract}
\vskip 0.2cm

We formulate a refinement of $SU(N)$ Chern-Simons theory on a three-manifold via the refined topological string and the $(2,0)$ theory on $N$ M5 branes. The refined Chern-Simons theory is defined on any three-manifold with a semi-free circle action. We give an explicit solution of the theory, in terms of a one-parameter refinement of the $S$ and $T$ matrices of Chern-Simons theory, related to the theory of Macdonald polynomials. The ordinary and refined Chern-Simons theory are similar in many ways; for example, the Verlinde formula holds in both. Refined Chern-Simons theory gives rise to new topological invariants of Seifert three-manifolds and torus knots inside them. We conjecture that the invariants are certain indices on knot homology groups. For knots in $S^3$ colored by fundamental representation, the theory ends up computing the Poincar\'e polynomials of the knot homology theory categorifying the HOMFLY polynomial.  As a byproduct, we show that our theory on $S^3$ has a large-$N$ dual which is the refined topological string on $X={\cal O}(-1)\oplus {\cal O}(-1)\rightarrow \IP^1$; this supports the conjecture by Gukov, Schwarz and Vafa relating the spectrum of BPS states on $X$ to $sl(n)$ knot homology. We also provide a matrix model description of some amplitudes of the refined Chern-Simons theory on $S^3$.

\Date{May 2011}

\goodbreak
\vfill
\eject
}
\newsec{Introduction}

One of the beautiful stories in the marriage of mathematics and physics developed from Witten's realization \wcs\ that
three dimensional Chern-Simons theory on $S^3$ computes the polynomial invariant of knots constructed by Jones in \jones. While Jones constructed an invariant $J(K, {\bf q})$ of knots in three dimensions, his construction relied on projections of knots to two dimensions. This obscured the three dimensional origin of the Jones polynomial.
The fact that Chern-Simons theory is a topological quantum field theory in three dimensions made it  manifest that the Jones polynomial is an invariant of the knot, and independent of the two dimensional projection. Moreover, it also gave rise to new topological invariants of three-manifolds and knots in them. For any three-manifold $M$ and a knot in it, Chern Simons path integral, with Wilson loop observable inserted along the knot, gives a topological invariant that depends only on $M$, $K$ and the representation of the gauge group. Moreover, Chern-Simons theory gives a whole family of invariants associated to $M$ and $K$, by changing the gauge group $G$ and the representation $R$ on the Wilson line. Jones polynomial $J(K, {\bf q})$ corresponds to $G=SU(2)$, and $R$ the fundamental, two dimensional representation of $G$.  Taking $G=SU(n)$ instead, one computes the HOMFLY polynomial $H(K, {\bf q}, {\bf a})$ \ref\HOMFLY{Freyd, P.; Yetter, D., Hoste, J., Lickorish, W.B.R., Millett, K., and Ocneanu, A. (1985). "A New Polynomial Invariant of Knots and Links". Bulletin of the American Mathematical Society 12 (2): 239�246. doi:10.1090/S0273-0979-1985-15361-3.} evaluated at ${\bf a}={\bf q}^n.$
The work in \wcs\ was made even more remarkable by the fact that it explained how to solve Chern-Simons theory for any $M$ and collection of knots in it.

A mystery left open by \wcs\ is the integrality of the coefficients of the Jones and HOMFLY polynomials. They are both Lauren't polynomials in ${\bf q}$, and in the latter case ${\bf a}$, with integer coefficients. While Chern-Simons theory gives means of computing knot invariants, it gives no insight into question why the coefficients are integers. An answer to this question was provided by \kh .  Khovanov associates a bi-graded (co)homology theory to a knot $H^{i,j}(K)$, in such a way that its Euler characteristic is the Jones polynomial,
$$J(K,{\bf q}) = \sum_{i,j} (-1)^i{\bf q}^j\, {\rm dim}\,H^{i,j}(K).$$
Interpreted in this way, the integrality of the coefficients is manifest, since they are counting dimensions of knot homology groups. This gives rise to a refinement of the Jones  polynomial, where one computes the Poincar\'e polynomial instead,
$$
Kh(K, {\bf q}, {\bf t}) =    \sum_{i,j} {\bf t}^i{\bf q}^j\, {\rm dim}\,H^{i,j}(K).
$$
This depends on one extra parameter ${\bf t},$ and reduces to the Jones polynomial at ${\bf t}=-1$. The Poincar\'e polynomial has more information about the knot than the Euler characteristic. In particular, it is better at distinguishing knots.
Subsequent generalizations of \kh\ include categorification of HOMFLY polynomial at ${\bf a}=q^n$, by Khovanov and Rozansky \ref\kro{Khovanov, M. and Rozansky, L.,"Matrix factorizations and link homology", arXiv:math/0401268.}\ref\krt{{{Khovanov}, M. and {Rozansky}, L.}, "{Matrix factorizations and link homology II}", {arXiv:math/0505056}.}. 
Making contact with a theory in three or more dimensions is needed, since just like the Jones' original construction, Khovanov's approach has the drawback that it also relies on the two dimensional projections of knots.

Around the time of Khovanov's work,  Ooguri and Vafa \ovknot , using the results of \refs{\gvI,\gvII} provided an explanation for the integrality of the Jones polynomial using topological string and M-theory. Namely, the invariants of knot $K$ in Chern-Simons theory on $S^3$ get related, via the large $N$ duality, to computing topological string partition function on $X={\cal O}(-1)\oplus {\cal O}(-1)\rightarrow {\IP^1}$ with topological branes on a Lagrangian $L_K$ related to the knot. The latter is the same as the partition function of M-theory on $(X \times\, {\IC}^2 \times S^1)_q$, with M5 branes on $(L_K\times {\,\IC}\times S^1)_q$ where one is computing an index, the number BPS states of M2 branes ending on the M5 branes, counted with $(-1)^F$.
Thus topological string relates the Jones polynomial of a knot $K$ on the $S^3$, to an index counting BPS states of M2 branes. Since BPS degeneracies are manifestly integral, this provides an explanation for integrality of the Jones polynomial. In addition, M-theory gives a precise prediction for the integrality structure of the Jones polynomial. Many checks of this were performed in \ref\LabastidaZP{
  J.~M.~F.~Labastida, M.~Marino,
  ``Polynomial invariants for torus knots and topological strings,''
Commun.\ Math.\ Phys.\  {\bf 217}, 423-449 (2001).
[hep-th/0004196].
}\lmv.

In \gsv\ a relation between the work of \ovknot\ and \kh\ was proposed. Gukov, Vafa and Schwarz conjectured that to obtain the knot homology itself, instead of its Euler characteristic, one simply needs to consider the spaces of BPS states of M2 branes, instead of computing the index ${\rm Tr}(-1)^F$.
It follows that  one can obtain the Poincar\'e polynomial of the knot homology, by considering a refined counting of M2 brane BPS states. Moreover, \gsv\ showed that the conjecture implies certain regularities in the Khovanov-Rozansky knot homology that were indeed observed \refs{\gr,\rasa,\rasb}.

In \wr\ another way to obtain knot homologies from string theory was proposed, by studying the way finite $N$ Chern-Simons theory arises in string theory, following \refs{\wst, \ovknot}. One studies  theory $N$ M5 branes on three-manifolds $M \times {\IC}\times S^1$, in $T^*M\times {\IC}^2\times S^1$. Computing the index $Tr(-1)^F$ of the theory, Chern-Simons partition function can be recovered \wr. Considering the BPS states themselves, one conjecturally obtains knot homology groups.

In general, there is no path integral way to obtain the Poincar\'e polynomial on the space of BPS states, as the Poincar\'e polynomial is not an index. This implies that one cannot relate it to the partition function of M-theory, as in any attempt to do so, non-BPS states would contribute.
In the context of closed topological sting, there is a well known way to circumvent this, and obtain a refined index, counting of BPS states as a partition function of the theory. Namely, when the M-theory on a Calabi-Yau $Y$ has an extra $U(1)_R$ symmetry, one can obtains a new index, the M-theory partition function on $(Y\times {\;\IC}^2 \times S^1)_{q,t}$ where the subscript denotes that, as one goes around the $S^1$, the two ${\,\IC}$ planes get rotated by $(z_1, z_2) \rightarrow ( q z_1,  t^{-1}z_2)$ accompanied by the extra $U(1)_R$ symmetry rotation needed to preserve supersymmetry. This is just Nekrasov's Omega-background \refs{\nikitaa, \no}. 
This defines a refinement of the partition function of the closed topological string \refs{\hiv,\civ,\acdkv}, depending on the extra parameter $t$ and agreeing with the partition function of the ordinary topological string at $q=t$.

This can be extended to refine the open topological string.\foot{For other work on refined open topological string partition functions see \refs{\taki, \awata,\dimofte}.}, corresponding in M-theory to adding M5 branes on the Lagrangian $M$ in $Y$.
The case of most interest to us corresponds to taking $Y=T^*M$, with $N$ M5 branes on $(M\times \;\IC \times S^1)_{q,t}.$ This is just the setup of \wr .
The theory has the extra $U(1)_R$ symmetry provided $M$ admits a semi-free $U(1)$ action (this is a free action, where some elements of finite order are allowed to have fixed points). The $U(1)$ in question is not the $R$ symmetry itself, but it implies its existence\foot{We are grateful to C. Vafa and E. Witten for discussions regarding the relevant indices and the construction of the required $U(1)_R$ symmetry.}.  In this case the partition function of $N$ M5 branes on
$(M\times \,{\IC}\times S^1)_{q,t}$ is an index. This index defines the partition function of the refined Chern-Simons theory on $M$ with $SU(N)$ gauge group.  This generalizes the relation of the open topological string with D-branes on $M$ in $T^*M$ to the ordinary Chern-Simons theory on $M$.
The $U(1)$ symmetry is present and the refined Chern-Simons theory can be formulated when $M$ is a Seifert 3-manifold. Seifert 3-manifolds are $S^1$ bundles over Riemann surfaces $\Sigma$; a simple example of this is the $S^3$.

We will show that we can use M-theory to not only formulate, but also solve the refined Chern-Simons theory. We cut up the three-manifold $M$, and with it $T^*M$, into simple pieces on which the refined Chern-Simons amplitudes are computable by elementary means involving counting the few BPS states that end up contributing to the refined index. From this, by gluing we get everything else. As explained in \wcs, to solve a topological field theory on any three-manifold with arbitrary knots inside one needs the $S$ and $T$ matrices, representing the action of $SL(2, \IZ)$ on the Hilbert space of the theory on $T^2$, and the braiding matrix acting on the Hilbert space of the four-punctured sphere. In the present case, braiding appears necessary only in cases where the $U(1)$ symmetry is absent, so to solve the refined Chern-Simons theory on any relevant 3 manifold with knots in them, the $S$ and $T$ matrices are all that is needed.
We will be able to deduce them from $M$ theory. We derive a matrix integral expression for the $S$ matrix of the theory, refining the Chern-Simons matrix model of \refs{\mm,\amv }. This generalizes the beta-deformed matrix models of \toda\ to the topological A-model.

The theory depends on an extra parameter ${t}$, but shares many of the same properties as the ordinary Chern-Simons theory.  In particular, in the $SU(N)$ theory the Hilbert space on ${T^2}$ is finite dimensional, and labeled by level $k$ integrable highest weight representations of $SU(N)_k$. When $q$ and $t$ are given by $q = e^{2 \pi i \over k+\beta N}$ and $t = e^{2 \pi i \beta \over k+\beta N}$, the $S$ and $T$ matrices we derived satisfy
$$
S^4=1, \qquad (ST)^3 = S^2,
$$
and provide a unitary representation of the $SL(2,\IZ)$ action on ${\cal H}_{T^2}$  (with this parameterization, the unrefined case is $\beta=1$).  The unitarity of $S$ and $T$ matrices is necessary for topological invariance. The $S$ matrix in addition satisfies the Verlinde formula \ref\VerlindeSN{
  E.~P.~Verlinde,
  ``Fusion Rules and Modular Transformations in 2D Conformal Field Theory,''
Nucl.\ Phys.\  {\bf B300}, 360 (1988).
}\ref\ms{
  G.~W.~Moore, N.~Seiberg,
  ``Lectures On Rational Conformal Field Theory,'' in {\it Strings '89}, Proceedings
of the Trieste Spring School on Superstrings,
3-14 April 1989, M. Green, et. al. Eds. World
Scientific, 1990.
},
$$
S_{{\bar k }i}\, S_{{\bar k} j}/S_{{\bar k} 0 } = \sum_{\ell}{ N^{\ell}}_{i j} S_{{\bar k }\ell}
$$
with fusion coefficients $N^{i}_{jk}$ computed by considering the theory on $S^2\times S^1$ with Wilson loops in representations ${\bar R}_i$, $R_j$ and $R_k$ inserted. The fusion coefficients share all the properties of the fusion coefficients of the WZW model, except integrality, as they depend on $q, { t}$ explicitly.

Our theory gives rise to three-manifold and knot invariants which are explicitly computable, analogously to the ordinary Chern-Simons theory.\foot{The deformation of $S$ and $T$ matrices that arises in the refined Chern-Simons theory corresponds to
replacing everywhere, the $SU(N)$ characters by Macdonald polynomials. The fact that they provide a representation of $SL(2,Z),$ and generalize the Verlinde formula has been discovered earlier in \refs{\cheredniko,\cherednikt,\kirillov,\co,\ek,\eka,\ekb, \kirila} in the context of proving the Macdonald conjectures. However, it has not been noticed that this gives rise to Seifert and three manifold invariants. Moreover,  the connection to knot homology has not been made previously.}  Using the operator formulation of the theory in terms of the $S$ and $T$ matrices, we compute invariants of many non-trivial torus knots, starting with the trefoil knot. The expectation value of a Wilson loop in representation $R$, on a torus knot $K$ in the $S^3$, is given by
$$
Z(S^3, K, R_i) = \sum_{j,k,\ell} K_{0k}{N^{k}}_{ij}{(K^{-1})^{j}}_{\ell} {S^{\ell}}_p
$$
where the matrix $K$ is a product of $S$ and $T$ matrices depending on the knot. The formula is standard in Chern-Simons theory, here we need to interpret it using the refined $S$, $T$ and $N$ matrices depending on $q$,$t$. We define the normalized knot invariant, where we set the expectation value of the unknot to $1$,
$$
{\cal P}_{R_i}(K)= Z(S^3, K, R_i)/Z(S^3, \bigcirc , R_i)
$$

The refined Chern-Simons theory, per definition, computes indices. This leads us to a remarkable conjecture:
From the mathematical perspective, we predict that, when the three-manifold $M$ and knots in it admit a semi-free $U(1)$ action, the $SL_N$ knot homology groups (corresponding to knots colored by arbitrary representation) admit an additional grading,
$$
{\cal H}_{ij} = \oplus_{k}{\cal H}_{ijk},
$$
which allows one to define a refined index written in terms of knot theory variables abstractly as,
\eqn\rcci{
{\cal P}_{R_i}(K)=\sum_{i,j,k} (-1)^k {\bf q}^i {\bf t}^{j+k} {\rm {dim}} {\cal H}_{ijk}.
}
Morever, we conjecture that the refined index is computed by the refined $SU(N)$ Chern-Simons theory.
The refined index has more information about knot homology than the Euler characteristic computed by the ordinary Chern-Simons theory. The index \rcci\ reduces to the Euler characteristic only upon setting ${\bf t}=-1.$
While the Poincar\'e polynomial of the knot homology theory
$$
\sum_{i,j, k} {\bf q}^i {\bf t}^{j} {\rm {dim}} {\cal H}_{ijk} =
\sum_{i,j} {\bf q}^i {\bf t}^{j} {\rm {dim}}{ \cal H}_{ij}  .
$$
has yet more information than the index \rcci , computing it is hard. The index can, by contrast, be obtained simply, by cutting and gluing, from refined Chern-Simons theory. More generally, as we will discuss, we expect that this can be extended to the general ADE case.

In some special cases, it can happen that the index and the Poincar\'e polynomials agree (for a famous example, see \ref\VafaGR{
  C.~Vafa,
  ``Black holes and Calabi-Yau threefolds,''
Adv.\ Theor.\ Math.\ Phys.\  {\bf 2}, 207 (1998).
[hep-th/9711067].
}, and for more recent examples \ref\GaiottoBE{
  D.~Gaiotto, G.~W.~Moore and A.~Neitzke,
  ``Framed BPS States,''
[arXiv:1006.0146 [hep-th]].
}). We will provide ample evidence that for knots colored by fundamental representation, the Chern-Simons knot invariant computes the Poincar\'e polynomial of the knot homology categorifying the colored HOMFLY polynomial. Based on many examples, we conjecture that  ${\cal P}_{\tableau{1}}(K) ({\bf q}, {\bf t}, {\bf a})$, with
${\bf t } = \sqrt{t}$ , ${\bf q} = -\sqrt{q/t}$, and ${\bf a} = {\sqrt{\lambda}}$, where $\lambda=t^{N} t^{1/2} q^{-1/2}$,
computes the superpolynomial of \gr. We show that the conjecture holds for the $(2,2m+1)$ torus knots for any $m$, and for the $(3,3m+1), (3,3m+2)$ torus knots for $m = 1,2$, computed previously in \refs{ \gr, \rasa,\rasb}. The cases of $(2,2m+1)$ and $(3,4)$ knots are described in detail in section 7 below; the detailed description of $(3,5), (3,7)$ and $(3,8)$ cases is omitted, since the explicit polynomials in these cases become increasingly lengthy.

Our work also provides strong evidence for the conjecture of \gsv\ relating Khovanov homology to spaces of BPS states of M2 branes ending on M5 branes wrapping $L_K$ in $X = {\cal O}(-1)\oplus {\cal O}(-1)\rightarrow \IP^1. $
The Gopakumar-Vafa duality, relating Chern-Simons theory on $S^3$ to topological strings on $X$ was
expected to extend to the refined topological string \refs{\gsv,\civ}. Having solved the refined Chern-Simons theory, we are able to show that this is indeed the case, at the level of the partition functions. The partition function of the refined Chern-Simons theory on $S^3$ is the vacuum matrix element of the $S$ matrix, $Z(S^3) = S_{00}.$  We will show that, at large $N$, this equals the partition function of the refined topological string on $X$ \civ , with the identification $e^{-Area({\IP^1})} = \lambda =t^{N} t^{1/2} q^{-1/2}$.
The results of our paper thus give very strong evidence for the conjecture of \gsv .  

\lref\AwataSZ{
  H.~Awata, H.~Kanno,
  ``Macdonald operators and homological invariants of the colored Hopf link,''
[arXiv:0910.0083 [math.QA]].
}\lref\yy{Y. Yonezawa, Quantum (sln, mVn) link invariant and matrix factorizations,� arXiv:0906.0220[math.GT].}

The organization of the paper is as follows\foot{Note added: In the second version of the paper, we put more emphasis on the homological meaning of the refined index, and discuss in more detail relation of our work to \wr . Version one incorrectly stated the relationship between Khovanov-Rozansky $SL_N$ knot homology and HOMFLY knot homology; the two equal only at sufficiently large $N$.  It is possible that the refined index computes Poincare polynomials more generally: for the Hopf link and $(2,2n+1)$ knots \coloredtrefoil\ they equal for all totally symmetric or anti-symmetric representations. }. In section 2 we review the relation of Chern-Simons theory to open topological A-model string.
In section 3 we review the relation of topological string to counting of BPS states in M-theory.
In section 4, we use M-theory to compute the refined Chern-Simons partition function on $S^3$, as a variant of a beta-deformed matrix integrals of \toda . We also derive the $S$ and $T$ matrices of the refined Chern-Simons theory from M-theory.
In section 5 we show that the $S$ and $T$ matrices that we derived via M-theory correspond indeed to the $S$ and $T$ matrices of a topological field theory in three dimensions.
In section 6 we show that the large $N$ dual of the refined Chern-Simons theory on the $S^3$ is the refined topological string on $X= {\cal O}(-1)\oplus {\cal O}(-1) \rightarrow {\IP}^1$.
In section 7 we explain the connection to Khovanov homology, from the mathematical side, and counting of BPS states of M2 branes on $X$ from the physics side.
We present numerous examples. In all the cases that we have checked the knot polynomial we obtain agrees with the superpolynomial of \refs{\gr,\rasa,\rasb}.
In section 8 we end with a discussion of some directions for future research, including a generalization of our work to the refined Chern-Simons theory based on an arbitrary gauge group $G$.

\newsec{ Chern Simons Theory and Topological String}

In \wcs\ Witten explained how to obtain knot and three-manifold invariants from Chern-Simons theory in three dimensions. The Chern-Simons path integral on a three-manifold $M$ is
 $$
 Z_{CS} (M)= \int {\cal D} {A} \;e^{{i k \over 4 \pi}S_{CS}({\cal A})}
 $$
where
\eqn\scs{S_{CS} = \int_M {\rm Tr} \Bigl(A\wedge dA + {2\over 3} A\wedge A \wedge A\Bigr),
}
$k$ is an integer, and $A$ is a connection of gauge group $G$. In this paper, we will mainly work with $G=SU(N)$. The path integral is independent of the metric on $M,$ and so it associates to $M$ a topological invariant, its value $Z_{CS}(M)$. The theory also has topologically invariant observables corresponding to knots in $M$. The Wilson loop observable in representation $R$ associated to a knot $K$
$$
{\cal O}_R(K) = {\rm Tr}_R P \exp i\oint_{K} A
$$
is also independent of the metric on $M$.
The path integral,
$$
Z_{CS} (M; K_1, \ldots, K_n)= \int {\cal D} {A} \;e^{{i k \over 4 \pi}S_{CS}({\cal A})}{\cal O}_{R_1}(K_1)\ldots {\cal O}_{R_n}(K_n)
$$
with insertions of observables ${\cal O}_R(K)$ leads to invariants of knots in the three-manifold $M$.

Witten also explained how to solve the theory exactly. The data needed are provided by the relation, discovered in \wcs , of any three dimensional topological theory with a two dimensional rational conformal field theory. The Hilbert space of the three dimensional theory and operators acting on it can be constructed from conformal blocks of the CFT, and from representations of the corresponding modular group. In the Chern-Simons case, the relevant conformal field theory is the $SU(N)_k$ WZW model. More precisely, the $S$, $T$ and braiding matrices are all that one needs to solve the theory on any manifold. In this way, the knot invariants that arise from Chern-Simons theory can be explicitly computed. The polynomial knot invariants considered earlier by Jones \jones\ correspond to $G=SU(2)$ and Wilson lines in fundamental representation, on $M=S^3$. More generally, one finds that the expectation values
$$
\langle {\cal O}_{R_1}(K_1)\ldots {\cal O}_{R_n}(K_n)\rangle_{S^3} = P_{R_1, \ldots, R_n}(S^3, K_1, \ldots K_n)
$$
are polynomials in terms of the variable
$$q = e^{2\pi i \over k+N},
$$
with integer coefficients. These are known as HOMFLY polynomials, constructed in \HOMFLY\ from the mathematical point of view.

Chern-Simons theory is closely related to topological strings and also to M-theory computations on certain Calabi-Yau threefolds. While most topological theories are solvable in principle, the fact that Chern-Simons theory is solvable explicitly, and at the same time dual to string theory and M-theory, is the key insight that lead to solution of topological strings and counting of BPS states of M-theory in many different contexts, see for example \gvI\ovknot\ref\AganagicGS{
  M.~Aganagic, C.~Vafa,
  ``Mirror symmetry, D-branes and counting holomorphic discs,''
[hep-th/0012041].
}. We will review some aspects of this below. In the following sections we will explain that one can turn this around and use M-theory and the refined topological string to deduce the three-dimensional topological field theory that refines Chern-Simons theory and computes homological knot invariants.

\subsec{Topological string and Chern Simons theory}

In \wst , Witten explained that, for any three-manifold $M$, the open topological $A$-model on
$$Y= T^*M,
$$
with $N$ topological D-branes on ${M}$ is the same as $SU(N)$ Chern-Simons theory on $M$. The string coupling and the level of Chern-Simons get related to the topological string coupling $g_s$ by
$$
g_s = {2\pi i \over k+N}.
$$
The A-model topological string on a Calabi-Yau manifold $Y$ counts holomorphic maps from Riemann surfaces into $Y$.
When $Y=T^*M$, there are no holomorphic curves of any kind, so only the degenerate maps can contribute. In the case of the open A-model, the degenerate maps precisely reproduce the Feynman graphs of the Chern-Simons theory.
In particular, this implies that the Chern-Simons partition function on $M$, and the open topological string partition function on $Y$ with $N$ D-branes on $M$, $Z^{top}_{open}(T^*M)$ are the same:
$$
Z_{CS}(M) = Z^{top}_{open}(T^*M).
$$

Adding a knot $K$ to $M$, in some representation ${R}$ of the gauge group also has a topological string interpretation \ovknot, of adding D-branes wrapping a non-compact Lagrangian $L_K$. $L_K$ is a rank 2 bundle over
the knot $K$, constructed as follows.
Take a point on $K$ and the vector $V$ tangent to $K$ at that point in $M$. One obtains a two-plane in the fiber of $T^*M$ consisting of the cotangent vectors, orthogonal to $V$ in the pairing between the cotangent and tangent vectors provided by the symplectic form on $Y$ (see \refs{\ovknot,\taubes}). Such $L_K$ is topologically $\IR^2\times S^1$, and
$$L_K \cap M = K.
$$
Since the $b_1(L_K)=1$, the Lagrangian $L_K$ has a modulus corresponding to moving it off. This means that, at least infinitesimally, there are holomorphic annuli in $Y$ with one boundary on $M$, along the knot $K$, and the other on $L_K$.
Adding D-branes on $L_K$ the open topological string computes
\eqn\topop{
Z^{top}(T^*M, L_K, V) = \sum_R Z^{top}_R(T^*M, L_K, Y)\, {\rm Tr}_R V.
}
Since $L_K$ is non-compact, in defining the theory it is natural to consider the gauge fields on $L_K$ as non-dynamical. In string theory terms, this corresponds to {\it not} summing over the degenerate maps to $L_K$, but only considering the finite ones. Above, $V = P \exp(\oint_{K} A')$ is the holonomy of the gauge field on $L_K$ at infinity,  which gets complexified by the mass of the bifundamental. It is natural to take a very large number of branes on $L_K$, so the sum in \topop\ is effectively over all Young diagrams $R$.

We can evaluate \topop\ as follows. If it were not for the branes on $L_K$, the theory would have been just pure Chern-Simons theory on the $S^3$. When we add branes on $L_K$, there is a new sector to the theory, corresponding to strings stretching between $L_K$ and $M$. In the topological string theory this gives a single
bifundamental on $K$, which is massless when $L_K$ and $M$ intersect, and gets a mass upon deformation. Since this is the heaviest mode around, it is natural to integrate it out first which gives\foot{Whether we get a $det$ upstairs or downstairs, i.e. whether the bifundamental is a fermion or a boson,  depends on the geometry and the relative orientation of $L_K$ and $M$.  Reversing the relative orientation of we flip from one choice or the other \tv . Here we simply made a convenient one. In addition, note that to be fateful to the bi-fundamental representation of the stretched string, we should have better written this as ${\det}^{-1}\bigl(1\otimes V -  U \otimes 1\bigr).$ Such factors are related to various normalizations. They will not matter until section 5. }
\eqn\ovd{
{\cal O}_K(U,V)={\det}^{-1}\bigl(1\otimes 1 -  U \otimes V^{-1}\bigr),
}
where $U = P \exp(\oint_{K} A)$ is the holonomy of the  gauge field on $M$. There are a few more ways of thinking about this amplitude, depending on the perspective. From the worldsheet point of view, it is natural to rewrite this as
\eqn\ovt{
{\cal O}_K(U,V)=\exp\bigl(\sum_{n=1}^{\infty} {1\over n}\,  {\rm Tr} U^n\, {\rm Tr} V^{-n}\bigr).
}
The right hand side of \ovt\ describes the contribution of one holomorphic annulus, a primitive curve embedded in $Y$, and the sum
over $n$ is the sum over its multi-covers. Thus, in the presence of branes on $L_K$, the open topological string is computing the Chern-Simons path integral on $M$ with insertion of ${\cal O}_K(U,V).$
To make contact with knot observables, one uses Schur-Weyl duality which implies that \ovt\ can also be written as
\eqn\ovm{
{\cal O}_K(U,V) = \sum_{R}  {\rm Tr}_R U \;{\rm Tr}_{R} V^{-1}
}
where the sum is over all the representations of $SU(N)$, the smaller of the two groups. Thus, the open topological string amplitude \topop\ is computed by Chern-Simons path integral on $M$ with Wilson loop along the knot $K$, in different representations,
$$
Z^{top}_{R}(Y,M, L_K) =  \langle {\rm Tr}_R U \rangle_M  =Z_{CS}(M; K, R)
$$
The full topological string amplitude is the expectation value of the operator \ovt\ on $M$.

One can replace $Y=T^*M$ with a more general Calabi-Yau $Y$ containing $M$ -- the neighborhood of the Lagrangian is still modeled on $T^*M$, but the global geometry may be different. In this more general setting there can be additional sectors in the open topological string theory coming from instantons, holomorphic curves in $Y$
with boundaries on $M$ \wst . This is analogous to what we had above, in the presence of extra D-branes.  Let us denote by
$Z^{top}_{inst}(M, Y; U, \lambda)$ the contribution of finite holomorphic maps to the open topological string partition function (evaluating these is the subject of open Gromov-Witten theory). $Z^{top}_{inst}$ depends on the holonomy $U = Pe^{i\oint A}$ of the gauge field coupling to the boundary of the instanton string worldsheet, and $\lambda = e^{-Area}$ measures its degree $Q$ in $H_2(Y, \IZ)$. As we will review in the next section, the instanton contributions have the form,
\eqn\unrefif{
Z^{top}_{inst}(Y, M; U) = \exp(\sum_{s_\ell,Q,R} \sum_{n=0}^\infty D^{s_\ell, Q}_{R}{
\;\;q^{2ns_\ell}  \over n(1- q^{n}) }\; \lambda^{nQ} \; {\rm Tr}_{R} U^n),
}
where $D^{s_\ell, Q}_{R}$ are integers, measuring the contribution of primitive holomorphic curves in class $Q$ with the boundary winding numbers related to $R$ \refs{\ovknot,\lmv}. This highly non-trivial statement follows from the relation of open topological string to M-theory, which we will review in the next section. For now, one should just observe that, in the presence of instantons, the full open topological string on $Y$, including both the finite and the degenerate holomorphic maps with boundaries, computes the expectation value
$$
Z^{top}(Y, M) = \langle \exp(\sum_{s_\ell,Q,R} \sum_{n=0}^\infty D^{s_\ell, Q}_{R}{
\;\;q^{2ns_\ell}  \over n(1- q^{n}) }\; \lambda^{nQ} \; {\rm Tr}_{R} U^n)\rangle_M
$$
of \unrefif\ in the Chern-Simons gauge theory on $M$.

\subsec{From Topological String to Chern-Simons theory}
Since $SU(N)$ Chern-Simons theory is the open topological string, one can use topological string methods to obtain Chern-Simons amplitudes. We will start this section by recalling aspects of canonical quantization of Chern-Simons theory from \ems\ (see also  \ref\douglas{
  M.~R.~Douglas,
  ``Chern-Simons-Witten theory as a topological Fermi liquid,''
[hep-th/9403119].
}), and then rederive them from the perspective of the open topological string.
This construction, taken from \refs{\amv,\akmv}, will be the starting point for the rest of this paper, where we will use the refined topological string and M-theory, to define and compute the amplitudes of the refined Chern-Simons theory.

In a topological field theory, one can use cutting and sewing to obtain amplitudes on more complicated manifolds from simpler ones.
Consider an $S^3$, viewing it as a $T^2$ fibration over the interval. Pick a basis of $1$ cycles of the two torus, so that
the $(1,0)$ cycle of the two torus degenerates over one the left end-point, and the  $(1,1)$ cycle degenerates over the right. Cutting the $S^3$ in the middle of the interval, we get two solid tori, $M_L$  and $M_R$, so that
inside $M_L$ the $(1,0)$ cycle of the $T^2$ fiber is filled in, and in $M_R$ the $(1,1)$ cycle. The Chern-Simons path integral on the $S^3$ is obtained from the path integral on $M_L$ and $M_R$ by gluing.

Since $M_L$ and $M_R$ are two three-manifolds with a $T^2$ boundary, the Chern-Simons path integral on them defines two states in the Hilbert space ${\cal H}_{T^2}$ of the theory on $T^2$. The Chern-Simons partition function on $M_L$ is well known from canonical quantization of the theory on ${\bf T}^2 \times {\IR}$ \ems.   A rough sketch of the derivation is as follows. If we denote by $t$ the ${\IR}$ time direction, integrating over $A_t$, the path integral localizes on the flat connections on $T^2$. As the fundamental group of $T^2$ is commutative, by a gauge transformation, we can set
$A = x d \theta_{0,1} + p d\theta_{1,0}$ where $x$ and $p$ are holonomies of the gauge field along the $(0,1)$ and $(1,0),$ cycles of the $T^2$.
Here $x$ and $p$ are periodic, with period $2\pi$. We can take them to depend only on $t$ and, moreover, classically commute. The fact that they commute implies we can diagonalize them simultaneously, so let $x_I$ and $p_J$ denote the corresponding eigenvalues.  Quantum mechanically, the Chern-Simons action \scs\ implies that $x$ and $p$ are canonically conjugate,
%
$$
[p_I, x_J ] = g_s \delta_{IJ}, \qquad I,J = 1, \ldots N
$$
where $g_s = {2\pi i \over {k+N}}.$ The shift of $k$ to $k+N$ comes about from carefully integrating out the off diagonal and the non-constant degrees of freedom \ems . The periodicity of holonomies implies that the phase space is compact, and the Hilbert space is finite dimensional.  The Chern-Simons partition function on solid torus $M_L$, with no insertions is a wave function obtained in \ems\ as
\eqn\lleft{
Z({M_L})(x_1, \ldots x_N) = \prod_{1\leq I<J \leq N} ( e^{(x_J-x_I)/2} -e^{(x_I-x_J)/2} ) = \Delta(x).
}
Here we are treating $x_I$ as a complex variables, which is natural as the holonomy gets complexified in string theory.
Written in the momentum variable $p_I$, this becomes the periodic delta function written in e.q. (4.12) of \ems . For a single D-brane, the partition function would have been simply $1$. This is the Fourier transform of the (periodic) delta function that sets $p=0$, and reflects the vanishing of the $(1,0)$ cycle of the $T^2$ in the interior of $M_L$. The wave function of $M_R,$ where $(1,1)$ cycle of the torus degenerates
\eqn\rright{
Z({M_R})(x_1, \ldots x_N) =   e^{- {\rm Tr}\, x^2/2g_s} \Delta(x),
}
differs from \lleft\ by the Gaussian factor. The gaussian factor corresponds to the fact that, in the interior of $M_R$ it is $(1,1)$ rather than the $(1,0)$ cycle that vanishes. The shift of the vanishing cycle, from $(1,0)$ to $(1,1)$ is implemented by the operator $e^{- {\rm Tr} x^2/2g_s}$, \ems\ which takes $p$ to $p+x$ leaving $x$ unchanged.
Consider now the partition function on the $S^3$, obtained by gluing the $M_L$ and $M_R$ together.
On the one hand, gluing $M_L$ to $M_R$ is accomplished by setting the corresponding holonomies equal, and integrating
\eqn\mm{
Z_{CS}(S^3) = \int d^N x \;Z^*_{CS}(M_L, x)\;Z_{CS}(M_R, x),
}
since this is the reverse of cutting the path integral into two by freezing the holonomies on the boundary.
Above $^*$ denotes complex conjugation, corresponding to reversing orientation of $M_L$ before gluing.
This gives
\eqn\mmt{
Z_{CS}(S^3) = {c_{TST}\over N!} \int d^N x\; \Delta^2(x)\; e^{-{1\over 2 g_s} {\rm Tr} x^2} .
}

\centerline{\includegraphics[width=6cm]{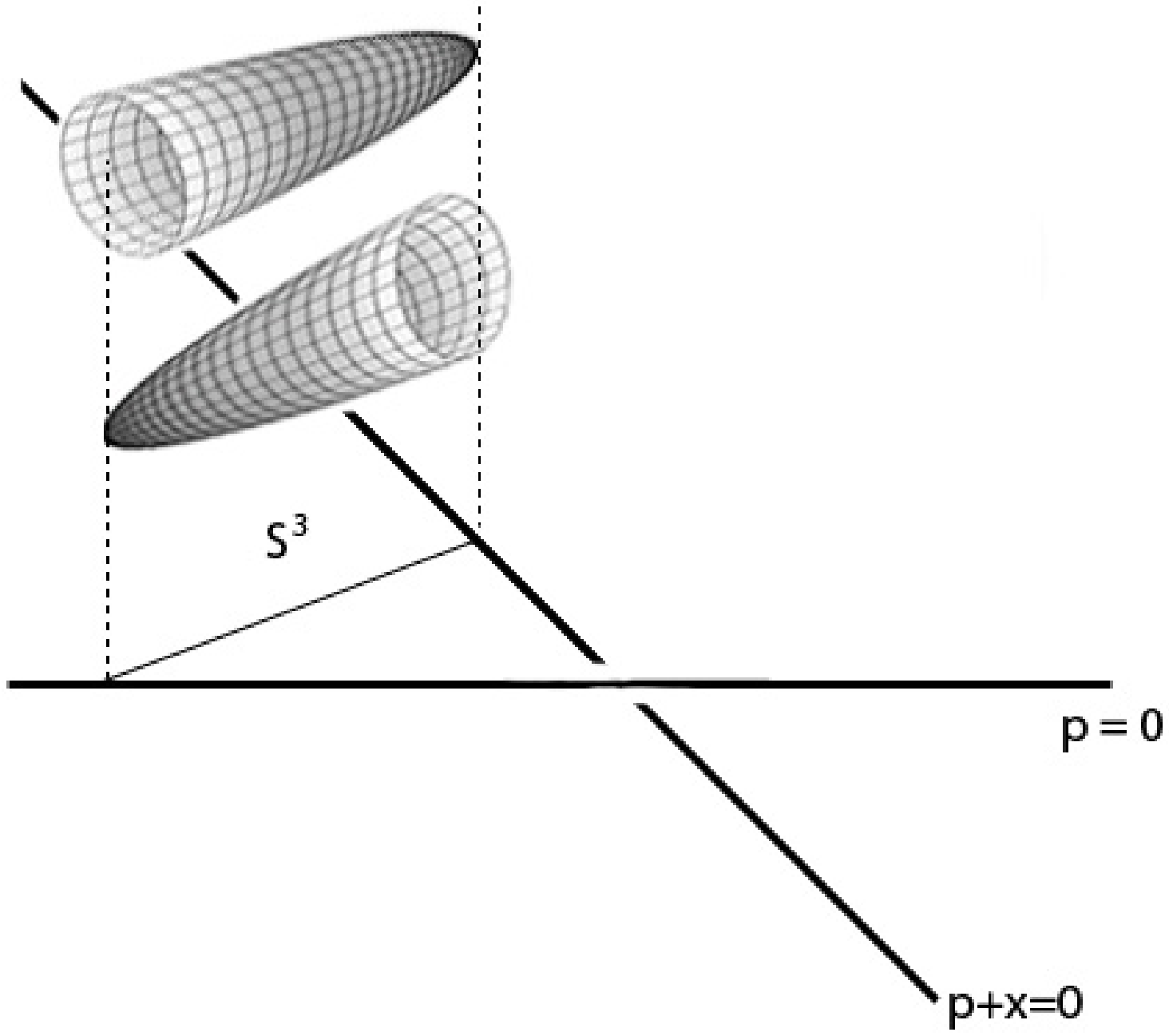}}
\noindent{\ninepoint \baselineskip=2pt {\bf Fig. 1.} {$Y=T^* S^3$ is a $T^2\times {\IR}$ fibration over ${\IR}^3$. The figure depicts the two lines in the base where the $(1,0)$ and $(1,1)$ cycles of the $T^2$ degenerate. Taking any path between them, together with the $T^2$ fiber over it, one obtains an $S^3$. Cutting the geometry in half, along the equator of the $S^3$ one obtains $Y_{L,R} = T^*(S^1 \times \IR^2)$.}}
\bigskip

On the other hand, geometrically, we are gluing together two solid tori, up to the action of mapping class group. The action of the boundary $T^2$ translates to action on the corresponding Hilbert space ${\cal H}_{T^2}$, so the path integral is computing a matrix element. If we denote by $S$ and $T$ the two generators of $SL(2,\IZ)$, then
$$
S: \qquad p\rightarrow x, \;\; x\rightarrow -p;  \qquad T: \qquad p \rightarrow p, \; \;x \rightarrow x+p
$$
so that the operation that sends $x$ and $p$ to $x$ and $x+p$ is $TST$,
\eqn\gl{
TST: \qquad x \rightarrow x, \qquad p \rightarrow p+x,
}
and therefore $Z_{CS}(S^3)$ equals the matrix element of this in the vacuum,
\eqn\gn{
Z_{CS}(S^3) = (TST)_{00}.
}


The $SU(N)$ Chern-Simons theory on the $S^3$ is the same as the topological string on $Y=T^*S^3$ with $N$ A-branes on $S^3$. $Y$ is a $T^2\times \IR$ fibration over the base $\IR^3$, parameterized by ${\rm Re}(x),{\rm Re}(p)$, and another variable ${\rm Re}(z)$. The $T^2$ fiber degenerates over two lines in the base: the ${\rm Re}(p)=0={\rm Re}(z)$ line, corresponding to the $(1,0)$ cycle degenerating, and ${\rm Re}(p)+{\rm Re}(x)=0$, ${\rm Re}(z)=1$ line corresponding to the $(1,1)$ cycle degenerating. Dividing the $S^3$ into $M_L$ and $M_R$ corresponds to dividing $Y$ into two halves, which we will call $Y_L$ and $Y_R$, containing $M_{L}$ and $M_R$, respectively. Both $Y_L$ and $Y_R$ are very simple, each is a copy of
$$
Y_{L,R} = T^*(S^1 \times {\IR}^2)=
{\,\IC}^* \times {\IC} \times {\IC}
$$
where $\;{\IC}^*$ is a cylinder, composed of the real line and the $S^1$ fiber that remains finite. This is also the $S^1$ that remains finite inside $M_{L,R}$. We have used here topological invariance of the theory, to view the solid torus, which is a disk $D$ times $S^1$ as ${\IR^2} \times S^1$.

\centerline{\includegraphics[width=5cm]{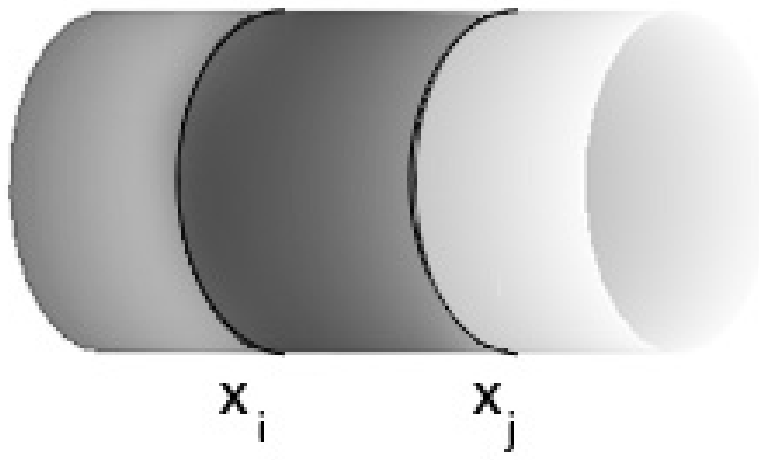}}
\noindent{\ninepoint \baselineskip=2pt {\bf Fig. 2.} {The D-branes wrap two $S^1$'s on the cylinder ${\IC}^*$ in $Y_L$. The  open topological string is counting the maps to the annulus between them.  }}
\bigskip

From the perspective of the open topological string,  \lleft\ and \rright\  arise as follows. Consider A-model on $Y_L$, with $N$ A-branes on $M_L$. Since $M_L$ is non-compact, the only contributions to the open topological string partition function come from instantons. Since the A-model topological string amplitudes depend holomorphicaly on Kahler moduli, we can evaluate the partition function for widely separated branes. For a pair of branes, one at $x=x_I$ and the other one at $x=x_J$, with $x_I<x_J$ one gets contributions from the maps to the portion of the $\;{\IC{^*}}$ cylinder in between the two branes. There are no additional contributions of any kind, since there are no other holomorphic curves of any kind in this geometry. The wavefunction \lleft\ than follows from considerations analogous to those leading to \ovd, together with the fact that now the $N$ A-branes are identical.

 Apart from the instantons, we should ask whether there may be any classical contributions to the partition function.  The classical contributions to the free energy come from the disk. The disk amplitude is per definition, the on-shell value of the Chern-Simons action on $M_L$. In terms of flat connections, the classical action is $\sum_I \int p_I d x_I$. We can evaluate this for each brane separately, since the interaction between the branes only comes from the one loop contribution. Classically, $p=0$ inside $M_L$, as the corresponding $S^1$ degenerates there. Thus, in $Y_L$ the disk contribution vanishes.

The same argument regarding the instantons holds in $Y_R$. However, now the disk amplitude contributes.  There, the disk contribution is again given by integrating $\sum_I \int p_I dx_I$, however, in this case, it does not vanish. Instead, since $p=-x$ there, it gives
$$\sum_I \int p_I dx_I = -\sum_I { x_I^2/2},
$$
in agreement with the above.

It is also easy to incorporate simple knots on the $S^3$ in the same formalism.
From the perspective of $Y$, consider introducing two sets of A-branes on non-compact Lagrangians
$L_1$ and $L_2$. Each set of branes corresponds to insertion of the operator \ovt\ in the matrix integral, with $x_I$ the eigenvalues of the unitary matrix $U$. From Chern-Simons perspective this corresponds to inserting,  along the $(0,1)$ cycle of the solid torus, a Wilson loop in representation $R_i$ in $M_L,$ and a Wilson loop in representation $R_j$ in $M_R$. This simply changes \lleft\  to
\eqn\leftW{
Z({M_L}; R_i)(x) =\Delta(x) \;{\rm Tr}_{R_i} e^{x};
}
and \rright\ to
\eqn\rightW{
Z({M_R}; R_j)(x) = \Delta(x)
\;e^{-{1\over 2 g_s}{\rm Tr} x^2} {\rm Tr}_{R_j} e^{x}.
}
Gluing these together, we get an $S^3$, with two linked unknots in representations $R_i$ and $R_j$, and linking number $1$. This is the Hopf link, in a specific framing.  The corresponding partition function is the $(TST)_{{\bar i}j}$ element of the $TST$ matrix,
\eqn\tst{
(TST)_{{\bar i} j } = {c_{TST}\over N!} \int d^N x \; \Delta^2(x)
\;e^{-{1\over 2 g_s}{\rm Tr}\, x^2} \;{\rm Tr}_{R_i} e^{-x}\; {\rm Tr}_{R_j} e^{x}.
}

\centerline{\includegraphics[width=7cm]{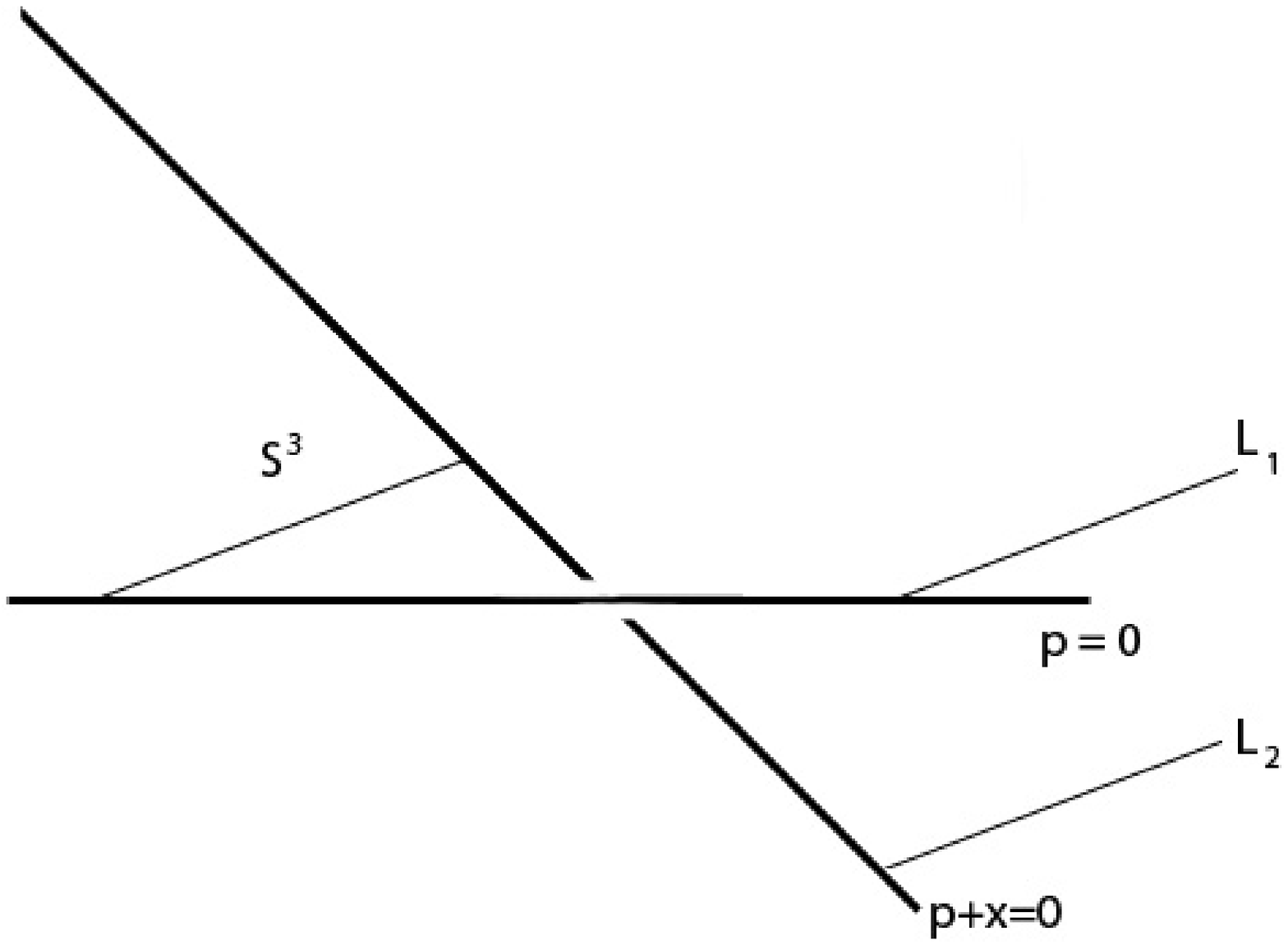}}
\noindent{\ninepoint \baselineskip=2pt {\bf Fig. 3.} {The $Y_L=T^*S^3$ geometry with two additional set of D-branes, in addition to the ones wrapping the $S^3$. }}
%
%
%
%

It easily follows \akmv\ that the right hand side of \tst\ gives the usual expression for the corresponding  $(TST)_{\bar ij}$ matrix element.
%
%
For future reference, it is useful to recall the matrix elements of $T$ and $S$ themselves
$$T_{\bar i, j } = T_i \delta _{{\bar i}, j} = c_T q^{{1\over 2}(\lambda_{R_i}+\rho, \lambda_{R_i}+\rho)}\delta_{{\bar i},j}, $$
and
$$
S_{{\bar i} j }/S_{00}  =s_{R_i}(q^{\rho})\; s_{R_j}(q^{\rho + \lambda_{R_j}}),
$$
where ${\rm Tr}_{R} e^{x} = s_{R}(e^x)$ is the Schur function $s_R$ in representation $R$ \macdonald ,
and
$$
S_{00} =   c_S \Delta(q^{\rho}) = c_S \prod_{i=1}^N(q^{i/2}-q^{-i/2} )^{N-i}
$$
is the partition function on the $S^3$, in vacuum.
In particular,
$$
Z(S^3) = (TST)_{00} = S_{00}.
$$
We will make the normalizations $c_S, c_T$ of the $S$ and $T$ matrices precise in section 5, where it becomes important.

Note that there is another way to view \mmt , namely as a matrix model, written in terms of eigenvalues $x_i$ of an $N$ by $N$ hermitian matrix describing the positions of the $N$ B-branes in wrapping the ${\IP}^1$ in $Y$. This matrix model generalizes the \dvI\ and \dvII\ matrix models to mirrors $Y$ of toric geometries \amv .
To summarize, the matrix integral \tst\ can be derived either from the Chern-Simons path integral on $S^3$, by viewing the latter as a $T^2$ fibration over an interval and reducing to zero modes, or by open topological string theory.


\newsec{M-theory and Refined Chern-Simons Theory}

In this section we will recall the relation of open topological string theory, and hence Chern-Simons theory, to M-theory with M5 branes. In the unrefined case, all sides of the story are well known. When we consider the refined case, the M-theory will provide the sole definition of the theory. Consider M-theory on
\eqn\background{
(Y \times TN \times S^1)_q\,,
}
where $Y$ is a Calabi-Yau manifold (for now $Y$ is arbitrary) and $TN$ is the Taub-NUT space.
The Taub-NUT space is twisted along the $S^1$, in the sense that going around the circle, the complex coordinates $z_1,z_2$ of the $TN$ rotate by
\eqn\rot{
z_1 \to q z_1,\qquad z_2 \to q^{-1} z_2,
}
so the space is not a direct product. We denoted this twist by a subscript $q$ in \background .
The M-theory partition function on this background \nikitaa\ is the same as the partition function of the closed topological A-model on $X$  \DVV, where one identifies $q = e^{g_s}$ with string coupling $g_s$.
To extend this to the open string \lref\cnv{
  S.~Cecotti, A.~Neitzke, C.~Vafa,
  ``R-Twisting and 4d/2d Correspondences,''
[arXiv:1006.3435 [hep-th]].
}\lref\cdv{
  M.~C.~N.~Cheng, R.~Dijkgraaf, C.~Vafa,
  ``Non-Perturbative Topological Strings And Conformal Blocks,''
[arXiv:1010.4573 [hep-th]].
} \refs{\AY, \cnv,\cdv}, we add $N$ M5 branes wrapping
$$
(M \times {\IC} \times S^1)_q
$$
where $M$ is a special Lagrangian 3-cycle in $Y.$
The branes wrap a $\;{\IC}$ subspace of $TN$ space fixed by the rotations \rot. We can take $\, {\IC}$ to correspond to the $z_1$ plane.
The partition function of the M5 branes on this background is
\eqn\mfive{
Z_{M}(Y, M)=  {\rm Tr}\;(-1)^{F} q^{S_1-S_2}.
}
Here  $S_1$ and $S_2$ are the generators of two $U(1)_{1,2}$ rotations in \rot , and $F=2S_1$ measures the fermion number. The M5 brane partition function \mfive\ is the same as the open topological string partition function on $Y$ with $N$ D-branes wrapping $M$,
$$
Z_{M}(Y, M,q) = Z^{top}_{open}(Y, M,g_s),
$$
where $q=e^{g_s}$ in terms of $g_s$, the topological string coupling.
There are two sets of contributions to the M5 brane partition function: the light modes on the M5 branes, and contributions coming from heavy BPS states of M2 branes ending on the M5 branes \cnv. This parallels the open string discussion of the previous section. The light modes correspond to perturbative Chern-Simons gauge fields, while the heavy modes correspond to the open topological string instantons \cnv .

It is natural to first integrate out the heavy M2 brane particles. The M2 branes are charged particles in three dimensions.  From the two-form $B$ on M5 brane world-volume one obtains, for each non-contractible 1-cycle $\alpha$ on $M$ and each M5 brane, a gauge field $A_I^{\alpha}$ in three dimensions, on ${\,\IC}\times S^1$.
\lref\avg{
  M.~Aganagic, C.~Vafa,
  ``G(2) manifolds, mirror symmetry and geometric engineering,''
[hep-th/0110171].
}  Each gauge field is in addition paired, by ${\cal N}=2$ supersymmetry of the 3d theory,  with a real scalar $\phi_I^{\alpha}$ corresponding to a massless modulus of the M5 brane in the Calabi-Yau directions. Let us denote the contribution to the index \mfive\ of the
massive charged BPS particles running around the $S^1$,
\eqn\index{Z^{BPS}_{M}(Y,M; q, U) =   {\rm Tr}_{{\cal H}_{BPS}} (-1)^{F} q^{S_1-S_2}\, \lambda^H\, U^{R} \,.
}
where $U_{I}^{\alpha} = e^{-\phi_I^{\alpha} - i \oint A_{I}^{\alpha}}$ is held fixed, and $\lambda$ is keeping track of the bulk M2 brane charges in $H_2(Y,\IZ).$
Here the trace is the trace in the Fock space of BPS M2 brane particles. Because of the twists in \rot\ in computing the trace, one should expand every 3d BPS particle $\Phi$ into modes on $\,{\IC}$
$$
\Phi^{s_1,s_2}(z_1) =  \sum_{n=0} \alpha^{s_1,s_2}_n z_1^{n+s_1}
$$
which contribute differently to the trace. Doing the trace, the index \index\ is easily found to have the structure \refs{\ovknot, \lmv, \AY, \cnv}
\eqn\unrefi{
Z^{BPS}_{M}(Y, M; U,q) = \exp(\sum_{s_\ell,Q,R} \sum_{n=0}^\infty D^{s_\ell, Q}_R{
\;\;q^{2ns_\ell}  \over n(1- q^{n}) }\; \lambda^{nQ} \; {\rm Tr}_R U^n),
}
where
$$D^{s_{\ell}, Q}_R = \sum_{s_1-s_2 = 2s_{\ell}}   D^{s_1,s_2, Q}_R$$
and
$
D^{s_1,s_2, Q}_R
$
is the number of BPS states of M2 branes of charge $Q \in H_2(Y, {\IZ})$, with
$U(1)_1 \times U(1)_2$ spin quantum numbers $s_1, s_2$ and transforming in representation $R$ of the symmetric group $S_N$ of $N$ elements. Here, $D^{s_1,s_2, Q}_R$ is the number of particles, taken with the plus sign for a bosonic particle, and the minus sign for a fermionic one.

The M-theory computation is related to the topological string by reducing on the $S^1$ to go down to IIA string theory on $Y$, with D4-branes wrapping $M$. The index gets related to computation of superpotential terms in the IIA theory on $Y$, which are captured by the topological string \refs{\cnv,\cdv}.  The M2 branes running around the $S^1$ map to string worldsheet instantons, and the gauge fields on the M5 branes coupling to them, map to perturbative topological string gauge fields.  Thus \index\ captures the contributions of worldsheet instantons $Z^{top}_{ inst}$ to the topological string partition function \unrefif . The relation between M-theory and the topological string explains the integrality of topological string amplitudes, observed first in the closed string context in \ref\CandelasRM{
  P.~Candelas, X.~C.~De La Ossa, P.~S.~Green, L.~Parkes,
  ``A Pair of Calabi-Yau manifolds as an exactly soluble superconformal theory,''
Nucl.\ Phys.\  {\bf B359}, 21-74 (1991).
}. In the open string case, the integrality is the statement that, written in the form \unrefi\  the coefficients $D^{s_{\ell}, Q}_R $ of the open topological string amplitudes \unrefif\ are integers. The integrality of the open topological string, via the relation to M-theory is also the key to the physical interpretation of the integrality of the Jones polynomial \ovknot . Since this involves another duality (the large $N$ duality) we will postpone discussing this until section 6.

We in addition need to integrate over the light modes on the M5 branes, with the action corresponding to the classical superpotential of the theory. The classical superpotential is the Chern-Simons functional on $M$,
\eqn\supp{W(A) = \int_M {\rm Tr}( A\wedge dA +A\wedge A\wedge A)
}
with the effective value of the coupling proportional to $g_s$, so one ends up computing the Chern-Simons path integral on $M$ with sources coming from holomorphic curves in $Y$, with boundaries on $M$ \refs{\cnv,\cdv}. One way to see this is to go down on the $S^1$ to IIA string theory, as we did before. The superpotential of the D4 brane theory is computed from the disk amplitude of the open topological string, and this is the Chern-Simons functional \supp .
An alternative, careful derivation appeared in \refs{\wr,\wrr,\wrrr} via the explicit analysis of the topologically twisted theory on the D4 branes on $M\times \IR^+$ and ending on a D6 brane wrapping $Y$. This is related to our setting
by going down to IIA on the Taub-Nut circle instead. At any rate, in the full theory, we end up computing the expectation value of the \unrefi\ in the Chern-Simons path integral on $M.$ Here $A$ should really be understood as a complex-valued connection, since the superpotential is holomorphic, and then one has to specify the contour of integration in the path integral. 
\lref\wpp{
  E.~Witten,
  ``Analytic Continuation Of Chern-Simons Theory,''
[arXiv:1001.2933 [hep-th]].
}The dependence of the path integral on the contour of integration is related to the choice of the vacuum, as discussed in \refs{\wpp,\wr,\cdv}.  Choosing the naive integration slice, the index \mfive\ becomes the bosonic path integral of $SU(N)$ Chern-Simons theory on $M$ with insertions coming from \index\ and integrating out massive M2 brane particles.

The equality of the open topological string and the M5 brane partition functions whether or not there are light modes present is also required by topological invariance of the theory. By cutting,  we can obtain from a three-manifold on which the M5 brane theory has light modes, and where the computation of the index cannot be reduced to counting BPS states, the manifolds where there are no light modes, and where the index simply counts BPS states of heavy M2 branes.  In the topological string the operation of obtaining the partition function on $Y$ from sewing partition functions on $Y_L$ an $Y_R$ is familiar, see for example \refs{\dvI,\tv, \adkmv}. On the M-theory side, these operations also have to make sense, by topological invariance that supersymmetry of the index implies. Moreover, if the partition functions of the M5 brane and the open topological string are the same on $Y_{L,R},$ they have to be also the same on $Y$, obtained from this by gluing. We will see an many example of this in the next sections, one of the simplest being $Y=T^*S^3$, and $Y_{L,R} = T^*(D\times S^1)$.

\subsec{Refined Chern-Simons Theory and the Refined Open Topological String}

In certain cases, M-theory on $Y$ can be used to define a refinement of the topological string \refs{\nikitaa, \no}\ref\hiv{
  T.~J.~Hollowood, A.~Iqbal, C.~Vafa,
  ``Matrix models, geometric engineering and elliptic genera,''
JHEP {\bf 0803}, 069 (2008).
[hep-th/0310272].
}. Consider, as before,
M-theory on $Y\times TN \times S^1$. We fiber $TN$ over the $S^1$, so that going around the circle, the coordinates $z_1$,$z_2$ of the TN space are twisted by
\eqn\rotref{
z_1 \to q z_1,\qquad z_2 \to t^{-1} z_2.
}
We will denote the resulting space by
$$
(Y\times TN \times S^1)_{q,t}.
$$
If $t \neq q$ this alone breaks supersymmetry. However, if the Calabi-Yau $Y$ is non-compact, M-theory on $Y$ gives rise to a five dimensional gauge theory. This has an additional $U(1)_R \subset SU(2)_R$ symmetry, and supersymmetry of the theory can be preserved provided, as one goes around the $S^1,$ one performs an additional $R$-symmetry twist \nikitaa . In this case, the BPS partition function depends on the additional parameter $t$, simply because the background \rotref\ does.
In particular, any state carrying a charge under the rotation of the $z_2$ plane will give a $t$ dependent contribution to the partition function.

Now consider adding $N$ M5 branes on
$$(M\times \,{\IC}\times S^1)_{q,t},$$
where $\,{\IC}$ corresponds to the $z_1$ plane.  The M5 brane configuration automatically preserves a $U(1)_1 \times U(1)_2$ symmetry rotating the $z_1$ and $z_2$ planes. The $U(1)_2$ symmetry is an R-symmetry of the theory that is always present. To define the index we need, we need a {\it sec ond} R-symmetry, which we will denote by $U(1)_R$. Let us focus for now on the main case of interest for us, when
$$
Y = T^*M.
$$
The theory has the $U(1)_R$ symmetry provided $M$ admits a free $U(1)$ action. More precisely, the action only needs to be semi-free: this corresponds to allowing a discrete subgroup of $U(1)$ to act with fixed points.\foot{We are indebted to Cumrun Vafa and Edward Witten for clarifying discussions regarding the construction of the $U(1)_R$ symmetry needed, given the free circle action on $M$.}

A $U(1)$ symmetry of $M$ is itself not an $R$ symmetry. An R-symmetry has to act in a non-trivial way on the directions transverse to the brane,  but a $U(1)$ symmetry of $M$ would not do that.  Instead, the $U(1)_R$ symmetry of the theory on the brane is a $U(1)$ rotation of $T^*M$, the normal bundle to $M$. Given a vector field $V$ on $M$ that generates the $U(1)$ action on $M$, at every point on $M$ one obtains a two-plane in the fiber $T^*M$, consisting of cotangent vectors orthogonal to $V$, in the pairing between the tangent and cotangent vectors provided by the symplectic form on $Y.$ The $U(1)_R$ symmetry is the rotation of this two-plane. More precisely, to get a nowhere vanishing vector field, it suffices to have a $U(1)$ action on $M$ which is semi-free -- some points may be fixed by finite $\IZ_p$ subgroups.
This implies that $M$ is a Seifert three-manifold. Seifert space is an $S^1$ fibration over a genus $g$ Riemann surface $\Sigma_g$,
$$
S^1 \rightarrow M \rightarrow \Sigma_g
$$
where the $U(1)$ action comes from the rotation of the fiber. The ordinary Chern-Simons theory on Seifert spaces was studied recently in \ref\bw{
  C.~Beasley, E.~Witten,
  ``Non-Abelian localization for Chern-Simons theory,''
J.\ Diff.\ Geom.\  {\bf 70}, 183-323 (2005).
[hep-th/0503126].
}\ref\BeasleyMB{
  C.~Beasley,
  ``Localization for Wilson Loops in Chern-Simons Theory,''
[arXiv:0911.2687 [hep-th]].
}\ref\BeasleyHM{
  C.~Beasley,
  ``Remarks on Wilson Loops and Seifert Loops in Chern-Simons Theory,''
[arXiv:1012.5064 [hep-th]].
}.

With the additional $U(1)_R$ symmetry preserved, the M5 brane partition function on $M \times {\IC} \times S^1$ defines an index,
\eqn\rindex{Z_M(T^*M, q,t)=   {\rm Tr}\, (-1)^{F}\, q^{S'_1}\, t^{-{S'}_2}.
}
Here $S_1$ and $S_2$ are combinations of
$U(1)_{1},$ $U(1)_2$ and $U(1)_R$ quantum numbers commuting with two of the four supercharges preserved by the M5 brane on $Y$. Let's denote these by $Q_{r}$, ${\bar Q}_r$. We can take $Q_r$ to have the quantum numbers $({1\over 2},{1\over 2},{1\over 2})$ under the three $U(1)'s$, respectively, and than ${\bar Q}_r$ transforms as $(-{1\over 2},-{1\over 2},-{1\over 2})$.\foot{Supersymmetries that get preserved by the brane have their $U(1)_{2}\times U(1)_R$ charges correlated. We can take the supercharges that survive to have $S_2=S_R$. Since both $S_2$ and $S_R$ act as R-symmetries, $S_2-S_R$ is a global symmetry. It is really the existence of this global symmetry that we need to define the refined index \rindex\ of the M5 brane theory on $M\times {\bf R}^2 \times S^1.$}
The first two quantum numbers are there in any case, see \lmv , the latter comes from the extra $U(1)_R$ symmetry. Then, defining
$$S_1' = S_1-S_R, \qquad S_2'= S_2-S_R.
$$
both $S_1'$ and $S_2'$ commute with $Q_r$, so \rindex\ defines an index. The index localizes on configurations that are annihilated by $Q_r, {\bar Q}_r$. Moreover, for $q=t$ it reduces to the unrefined index \mfive.
We will take the index \rindex\ as the definition of the refined $SU(N)$ Chern-Simons partition function on $M$:
$$
Z_{M}(T^*M,q,t) = Z_{CS}(M, q,t).
$$
This is analogous to the unrefined case, where the $N$ M5 brane partition function on $M$ in $T^*M$ equals the ordinary $SU(N)$ Chern-Simons partition function on $M$. Unlike in the unrefined case, we now do not have an alternative definition of the theory. Note that classical action of the Chern-Simons theory on $M$ is still given by the Chern-Simons functional \supp\ on $M$. Instead of $g_s$, it is now suppressed by $\ep$, where $q = e^{\ep}$ is the rotation parameter of the two-plane in \rotref\ wrapped by the M5 branes (see \refs{\GGV, \dimofte, \acdkv}).
While the classical action is unchanged, the measure of the path integral cannot be.\foot{We will shed some light onto how the measure changes once one abelianizes the theory, and writes it in terms of the flat connection on the $S^1$ fiber, somewhat analogously to \ref\aosv{
  M.~Aganagic, H.~Ooguri, N.~Saulina, C.~Vafa,
  ``Black holes, q-deformed 2d Yang-Mills, and non-perturbative topological strings,''
Nucl.\ Phys.\  {\bf B715}, 304-348 (2005).
[hep-th/0411280].
}.}

There are two directions in which one can generalize this. While we assumed so far that $Y = T^* M,$ we can allow more general geometries\foot{See \avg\ for such examples where $M=S^3$, or $S^3/\Gamma$. A large class of these is based on the dual description, where M-theory on the Calabi-Yau is replaced by IIB string theory on $\IR^3 \times TN \times S^1 \times S^1$ with $(p,q)$ five brane webs wrapping the $TN$ space give the 5d ${\cal N}=1$ gauge theory. The D3 branes stretching between them map to M5 branes on lens spaces. The M2 branes map to $(p,q)$ string webs ending on the D3 branes.}  where only the local neighborhood of $M$ in $Y$ is modeled on $T^*M$, as long as the $U(1)_R$ symmetry is preserved in the geometry. Then, in addition to the light modes captured by the refined Chern-Simons theory on $M$, there can also be massive M2 brane BPS states contributing to \rindex . Summing first over the heavy M2 brane particles while freezing the holonomy $U$, we get a refinement of the BPS counting in \index ,
\eqn\indexref{Z^{BPS}_{M}(Y,M; U, q,t) =   {\rm Tr}_{{\cal H}_{BPS}} (-1)^{F} q^{S'_1} t^{-S'_2}\, \lambda^H U^{R} \,.
}
This defines the refinement of the instanton contributions to the A-model open topological string
$$Z^{BPS}_{M}(Y,M; U, q,t) = Z^{top}_{inst}(Y, M; U,q,t).$$
The corresponding refined amplitude is easily seen to have the structure
\eqn\refi{
Z^{top}_{inst}(Y, M; U,q,t) = \exp(\sum_{s_1,s_2,Q,R} \sum_{n=0}^\infty D^{s_1, s_2,Q}_R{
\;\;q^{ns_1} t^{-n s_2}  \over n(1- q^{n}) }\; \lambda^{nQ} \; {\rm Tr}_R U^n).
}
For $q=t$, this reduces to the ordinary topological string partition function \unrefif . For $q\neq t$ the M-theory defines what we mean by the refined theory (one may call this the "refined open Gromov-Witten" theory).

We can also generalize this to include knot observables in the refined Chern-Simons theory. As we explained in the previous section, in the ordinary topological string, including a Wilson loop $K$ on a three-manifold $M$ in Chern-Simons theory corresponds to adding D-branes on a special Lagrangian $L_K$ in $T^*M$. To extend this to the refined case, both the knot $K$ and the three-manifold $M$ have to respect the $U(1)_R$ symmetry that is needed to define the theory. As explained in \ovknot, and reviewed in section 2, the Lagrangian $L_K$ is obtained as a total space of a rank two bundle over $K$. The fiber over a point on $K$ is the two-plane co-normal to the tangent vector $V$ to the knot at that point.
In the present  case, $L_K$ will preserve the $U(1)_R$ symmetry of $T^*M$ provided $K$ itself is the orbit of the $U(1)$ action on $M$ which we used in the construction. Then, also by construction, the rest of $L_K$ is invariant under $U(1)_R$. This in turn implies that the allowed knots and links in $M$ are the Seifert knots \BeasleyHM, wrapping the $S^1$ fibers over $\Sigma_g$ in $M$\foot{ Note that Seifert knots over different points on $\Sigma_g$ can be linked in non-trivial ways. An example of this is a Hopf-link in the $S^3.$ The $S^3$ is an $S^1$ bundle over $S^2.$ The Hopf link is obtained by taking the two $S^1$ fibers over the north and the south pole of the $S^2$ in the base.}. Note that the contribution of the bi-fundamental open strings that corresponded to inserting operators \ovd\ovt\ in the ordinary Chern-Simons case, will also need to get refined according to \refi , as we will discuss shortly.

\newsec{Refined Chern-Simons Theory from M-theory}

Having used M-theory to define the refined Chern-Simons theory on Seifert three-manifolds $M$, we will explain how to solve it. The basic idea is to use topological invariance of the theory to solve it on simple pieces first, and recover the rest by gluing. The key amplitude we will obtain in this way corresponds to taking
$$
M_L = S^1 \times {\IR}^2
$$
inside
$$
Y_L=T^*M_L = T^*(S^1 \times{\IR}^2) = {\IC}^* \times \,{\IC}^2.
$$
The free $U(1)$ action on $M_L$ corresponds to the rotation of the $S^1$, and the $U(1)_R$ symmetry one gets from it acts by rotating the fiber in $T^* {\IR}^2$. The Calabi-Yau $Y_L={\IC}^* \times \,{\IC}^2$ is essentially flat space.

Since $M_L$ is a non-compact three-manifold, the dynamics of the light modes is frozen by default, and the partition function depends on the choice of flat connection at infinity. We can, alternatively, view $M_L$ as a solid torus $S^1 \times D$ where the path integral computes a wave function depending on the boundary conditions we impose -- by topological invariance the two viewpoints are equivalent.  Since $b_1(M_L)=1$,  there is one complex modulus $x_I =\phi_I+ i \oint_{S^1} A_I $ on the M5 branes, which is related to the position of the branes on the ${\;\IC}^*$ cylinder in $Y_L$. With the branes widely separated, computing the partition function \indexref\ of the M5 brane theory on
\eqn\bmg{
(M_L \times {\IC} \times S^1)_{q,t}
}
inside
\eqn\cmg{(Y_L \times TN \times S^1)_{q,t}
}
amounts to counting BPS states of M2 branes stretching between them. As we will see, the spectrum is very simple, and the corresponding index \indexref\ is easily computable. We will then show, in this and the following section, that using topological invariance of the theory to put more complicated pieces together starting from $Y_L$ and $M_L$, we can recover everything else and indeed solve the theory on any Seifert three-manifold.

\subsec{Refined Chern-Simons Theory on $M_L$}
Consider M-theory on \cmg\
%
%
with $N$ M5 branes wrapping \bmg . We need to explain how the the partition function \lleft\  of the open topological string, at $q=t$,  arises from M theory, and then generalize this to arbitrary $q$ and $t$. What are the degeneracies
$$
D_R^{s_1,s_2},
$$
of the 3d particles coming from M2 branes wrapping holomorphic cycles in $Y_L$ and ending on the M5 branes? (We have suppressed the label $Q,$ since $H_2(Y_L,\IZ)$  is trivial.)

For a pair of M5 branes, one at $x=x_I$ and the other at $x=x_J$, the M2 branes wrap a portion of $\;{\IC}^*$ between the two M5 branes, $x_1\leq x \leq x_2$. What is the contribution to the index \indexref\ of these M2 branes?
To answer this question, it is easier to reformulate the problem in terms of IIA on $Y_L$, with D4 branes on $M_L$ instead \refs{\gvI,\gvII,\lmv}. There we need to find the BPS degeneracies of a D2 brane wrapping the annulus and ending on D4 branes. The holomorphic curve the D-branes wrap has no non-trivial moduli, so the only moduli space in the problem comes from the flat $U(1)$ bundle on the D2 brane. The moduli of the bundle on the annulus is an $S^1$, and quantizing this one gets two BPS particles, a boson and a fermion in the bifundamental representation of the gauge groups on the branes.
To determine the $U(1)$ charges of these states, one recalls \gvII\lmv\ that $Q_{\ell}$ acts non-trivially on the bundle moduli and trivially on the moduli of the curve, and $Q_r$ acts trivially on the moduli of the bundle, and non-trivially on the moduli of the curve. This implies that the $U(1)$ charges of the bosonic and the fermionic particle we get differ by the $U(1)$ charges of $Q_{\ell}$.
Note that $Q_{\ell}$ necessarily corresponds to super symmetries broken by the branes. Assigning to $Q_{\ell}$ the charges $({1\over 2}, -{1\over 2}, {1\over 2})$ under $U(1)_1\times U(1)_2\times U(1)_R$,
we deduce that
\eqn\bpsone{
D_{\tableau{1}, \bar{\tableau{1}}}^{0,0} =- 1, \qquad
D_{\tableau{1}, \bar{\tableau{1}}}^{0,{1}} = 1,
}
up to an ambiguous overall shift of the charge of the ground state.
Inserting this into \refi\ gives the partition function
$$
Z(M_L)(q,t) = \exp\Bigl(\sum_{n=1}^{\infty} {\;1-t^n\over n(1-q^n)} e^{n x_I-nx_J}\Bigr).
$$
Note that this is consistent with the known result for the unrefined case. Considering $N$ branes and symmetrizing, for $q=t$ we recover \lleft,
$$
Z({M_L})(q,q) = \prod_{1\leq I<J \leq N}\; (e^{(x_J - x_I)/2} - e^{(x_I-x_J)/2}).
$$
Now consider arbitrary $q$ and $t$. To avoid dealing with infinite products, we will specialize to
\eqn\betin{
 t = q^{\beta}, \qquad \beta \in \IN,
}
where $\beta$ is a positive integer. This specialization \betin\ is inessential -- while it will simplify the intermediate formulas, it is easy to reinstate arbitrary $\beta$ by inspection. In particular, we will see that the end result is independent of \betin. %
Thus the partition function of the refined $SU(N)$ Chern-Simons theory on a solid torus $M_L$ is
\eqn\llq{
Z({M_L})(x, q,t) = \prod_{m=0}^{\beta -1} \prod_{1\leq I<J \leq N}\; (q^{-m/2} e^{(x_J-x_I)/2}- q^{m/2} e^{(x_I-x_J)/2}) = \Delta_{q,t}(x).
}
This simple result is the key to the rest of this paper.

\subsec{Partition function on $S^3$}

The counting of BPS states of M2 branes computes directly only the partition functions of M5 branes on non-compact three-manifolds such as $M_L.$ However, from this by cutting and sewing, we can recover the partition functions on any other Seifert manifold.

Consider the refined Chern-Simons partition function on $S^3$.  $S^3$ is a Seifert three-manifold, viewed as an $S^1$ bundle over the $S^2$: we can describe the $S^3$ as a locus in ${\;\IC}^2$, with coordinates $z_1,z_2$, where
\eqn\threo{
|z_1|^2+|z_2|^2 = 1.
}
There is a free circle action $e^{i\theta} \in U(1)$ on this by, %
\eqn\actis{
(z_1, z_2) \rightarrow (e^{i\theta} z_1, e^{i \theta} z_2).
}
The orbit space of this is the $S^2$.

The partition function \indexref\ of $N$ M5 branes on
$$
(S^3 \times {\,\IC}\times S^1)_{q,t}
$$
inside
$$(T^*S^3 \times TN\times S^1)_{q,t}.$$
per definition, computes the $(TST)_{00}$ matrix element in the refined theory
$$
Z(S^3)(q,t) =(TST)_{00}.
$$

We can get the $S^3$ from $M_L$ and $M_R$, with $U(1)$ actions in each chosen to agree the free action of the $S^3$. Thus, the partition function on the $S^3$ in $Y$ can be obtained by sewing together the wave-functions of the M5 branes on $M_L$ in $Y_L=T^*M_L$, and the M5 branes on $M_R$ in $Y_R=T^*M_R$ by setting the holonomies at the boundary equal, and integrating.  We already obtained the wave-function on $Y_L$. For $Y_R$, we get the same set of BPS states contributing, as the geometries are the same, locally. The one difference, even in the unrefined $\beta=1$ case, is the classical contribution to the amplitude, the $\exp(-{\rm Tr} \, x^2/2 g_s)$ term.
The classical amplitudes of the topological string, coming from the sphere and the disk, are not affected by $\beta \neq 1$ deformation. The only thing the deformation does affect is the coefficient, which becomes $\ep$, where $q=e^{\ep}$. This comes about since the equivariant action is in essence compactification, on a circle of radius $1/\ep$. This gives the $z_1$ plane the brane wraps the volume $1/\ep_1$, and deforms the classical action to  $\exp(-{\rm Tr} \,x^2/2 \ep_1).$ In summary, we get from $M_R$,
\eqn\rrq{
Z({M_R})(x, q,t) =\Delta_{q,t}(x)\;e^{-{\rm Tr} x^2/2 \ep}.
}
The partition function of $N$ branes on $S^3$ in $T^*S^3$ is simply the overlap of wave functions from $Y_L$ and $Y_R$
\eqn\mmn{
Z(S^3) = \int d^N x \;Z^*(M_L, x)Z(M_R, x),
}
which equals
\eqn\refcs{
Z(S^3) ={c_{TST}\over N!} \int d^N x \;  \Delta^2_{q,t}(x)\;
\exp(-{\rm Tr} x^2/2 \ep).
}
This simple derivation gives the refined partition function of $SU(N)$  Chern-Simons theory on the $S^3$.
The matrix integral can be computed analytically -- we will give the derivation in the appendix B -- to find
\eqn\vac{
Z(S^3)(q,t) = (TST)_{00} =c_{TST} \prod_{m=0}^{\beta -1} \prod_{i=1}^{N-1}(q^{-m/2} t^{-i/2}-q^{m/2} t^{i/2} )^{N-i},
}
up to a normalization factor $c_{TST}$ that is unimportant for now. We will show, in section 6, that at large $N$, this agrees with the refined topological string partition function of $X = {\cal O}(-1) \oplus{\cal O}(-1) \rightarrow {\IP}^1$, the large $N$ dual geometry, as expected \refs{\gsv,\giv}. The matrix integral in \refcs\  gives the sought after generalization of the $\beta$-deformed matrix models
for refined B-model topological string \toda\ to A-model geometries.

\subsec{Incorporating simple knots on $S^3$}

Consider adding non-compact M5 branes on $T^*S^3$, in the M-theory realization of the construction in section 2.2.
We can make the branes on $L_1$, to intersect the $S^3$ along the unknot $K_1$, corresponding to setting $z_1=0$, and similarly, $L_2$ to intersect the $S^3$ along $K_2$, corresponding to setting $z_2 =0$:
$$
K_i = (L_i \cap S^3) : \qquad z_i =0, \qquad |z_1|^2+|z_2|^2 = 1.
$$
It is clear that both $K_{1,2}$ are invariant under the free $U(1)$ action \actis . The two knots at hand are linked in the $S^3$ and give a Hopf link.  Correspondingly, we can add the M5 branes on
$$(L_i \times {\;\IC}\times S^1)_{q,t}$$
while preserving the $U(1)_R$ symmetry. (For more examples of knots one can add to the theory, while preserving a $U(1)_R$ symmetry, see section 5.) We will take for now all three sets of branes to wrap the $z_1$ plane in the $TN$ space. The generalization to adding branes wrapping $z_2$ plane instead, will be discussed in section $6$.
We expect that the M5 brane partition function on the $S^3$, in presence of the M5 branes on $L_1$ and $L_2$ computes the matrix elements $(TST)_{{\bar i} j}$, now depending on $q$ and $t$.

The theory gets a new sector corresponding to BPS states of M2 branes stretching between the $S^3$ and $L_{1,2}$.  We need to start by finding the operator induced by integrating these particles out. This will generalize \ovd :
$$
{\cal O}_K(U,V, q, q)={\det}^{-1}\bigl(1 \otimes 1 -  U \otimes V^{-1}\bigr),
$$
to arbitrary $\beta$.
There is again only a single holomorphic curve with boundaries on $S^3$ and $L_1$ or $L_2$, respectively -- this is the annulus in Fig. 2. The spectrum of BPS states is essentially the same as \bpsone . More precisely, for both $L_1$ and $L_2$, and taking all spins into account, we get only two particles in the spectrum
\eqn\bpstwo{
D_{\tableau{1}, \bar{\tableau{1}}}^{0,0} = 1, \qquad
D_{\tableau{1}, \bar{\tableau{1}}}^{0,{1}} = -1.
}
This is the same as \bpsone, but with the spins exchanged, since the relative orientation of the branes has changed, as the unrefined answer also suggests. This implies that \ovd\ gets replaced by
\eqn\ovtq{
{\cal O}_K(U,V, q, t)=\prod_{m=0}^{\beta-1}{\det}^{-1}\bigl(1 \otimes 1 - q^m U \otimes V^{-1}\bigr).
}
In the refined Chern-Simons theory on the $S^3$, we thus end up computing the expectation value of a pair of these operator, where we keep the holonomies $V_{1,2}$ at infinity on $L_{1,2}$ frozen, and integrate over $U$.

In more detail, we would like identify which operator ${\cal O}_R$ corresponds to inserting a Wilson line in representation $R$, analogously to \ovm . We know that, at $\beta=1$ this has to reduce to ${\rm Tr}_R U= s_R(U)$, where $s_R$ is the Schur function \macdonald . It turns out the right choice is to identify ${\cal O}_R$ with insertions of the Macdonald polynomial $M_R(U)$ in representation $R$, evaluated at the holonomy $U$ \refs{\macda,\macdb, \macdonald}. Why this is the right Wilson loop observable will become clear in the next section.
Expanded in terms of Macdonald polynomials, one can show that \ovtq\ becomes simply
\eqn\ovmq{
{\cal O}_K(U,V, q, t)=\sum_{R} {1\over g_R}\; M_R(U) M_R(V^{-1})
}
where $g_R$ is the metric eigenvalue.

In summary,  the refinement of the wave functions \leftW\ that one gets in the presence of additional branes, is
\eqn\leftWr{
Z({M_L}; R_i)(x) =\Delta_{q,t}(x) \;M_{R_i} (e^{x});
}
and the refinement of \rright\ is
\eqn\rightWr{
Z({M_R}; R_j)(x) = \Delta_{q,t}(x)
\;e^{-{1\over 2 \ep}{\rm Tr} x^2} M_{R_j} (e^{x}).
}
The overlap of these computes the refinement of $(TST)_{{\bar i}j}$,
\eqn\mc{(TST)_{{\bar i} j } = {c_{TST}\over N!} \int d^N x \; \Delta_{q,t}^2(x)
\;e^{-{1\over 2 \ep}{\rm Tr} \, x^2} {M}_{R_i}(e^{-x}) M_{R_j}(e^{x}).
}
The integral has been evaluated in \ek , and the result can be written as follows
\eqn\tstre{
(TST)_{{\bar i} j } ={T}_i {S}_{{\bar i} j} {T}_j,
}
where $S_{{\bar i} j}$ is given by
\eqn\ssone{
S_{{\bar i}j}/S_{00} = M_{R_i}(t^{\rho}) M_{R_j}(q^{\lambda_{R_i}} t^{\rho}),
}
where $\lambda_R$ is the highest weight vector of the representation $R$ of $SU(N)$ and $\rho$ is the Weyl vector, defined as the sum over all the positive roots (see Appendix A for the list of standard group theory notations). This is, as we will see in section 5, the $S$-matrix of the refined Chern-Simons theory\foot{This expression agrees, for antisymmetric representations $R_i$ and $R_j$, with the expression obtained earlier \giv\ for the colored homological invariants of the Hopf link. For non-antisymmetric representations, as shown recently by \ikhopf, there is also a complete agreement with \giv\ after a change of basis of symmetric functions.}, as the notation we chose suggests. We have normalized it by the vacuum expectation value \vac\
\eqn\sstwo{
S_{00} =c_S \prod_{m=0}^{\beta -1} \prod_{i=1}^{N-1}(q^{-m/2} t^{-i/2}-q^{m/2}t^{i/2})^{N-i}.
}
Furthermore $T_i$ in \tstre\ is naturally thought off as the eigenvalues of the $T$-matrix of the refined Chern-Simons theory, given by
\eqn\ttone{
{T_{{\bar i} j}} = T_i \delta^i_ j =c_T  q^{{1\over 2}(\lambda_{R_i} + \beta \rho)^2} \delta^i_j.
}
We will give the precise overall normalizations $c_S$ and $c_T$ in the next section.

In the next section, we will show that the $S$ and the $T$ matrices given in \ssone,\sstwo,\ttone ,
$$
{S}_{{\bar i} j}, \qquad {T}_{\bar i j}
$$
and evaluated at $q= e^{2\pi i \over {k+\beta N}},$ $t= e^{2\pi \beta i \over {k+\beta N}},$ indeed are the $S$ and $T$ matrices acting on the Hilbert space ${\cal H}_{T^2}$ of a 3d topological field theory on a torus, consistent with topological invariance of the index they compute. For $\beta \neq 1$, they provide a refinement of the $S$ and $T$ matrices of the $SU(N)_k$ WZW model at $q = e^{2\pi i\over k+N}$, corresponding to replacing the ordinary Chern-Simons theory by the refined one. 

\subsec{Some Other Simple Amplitudes}

There are several more amplitudes that we obtain from this for free. Consider $N$ branes on $M=S^2\times S^1$,
inside
$$
T^*(S^2\times S^1).
$$
Cutting $M$ into two halves, along the equator of the $S^2$, we obtain two
copies of $Y_L$, glued together with trivial identifications. The state corresponding to each half is \lleft . Setting the holonomies equal and integrating, we obtain $Z(S^2\times S^1)$. We can add extra non-compact branes to this geometry, corresponding to inserting knots along the $S^1$, and at points on the $S^2$. This now manifestly preserves the free circle action, that simply rotates the $S^1$.  Adding a stack in each copy of $Y_L$, amounts to computing overlap of two copies of \leftWr, corresponding to representations $R_i$ and $R_j$. This gives
\eqn\spher{
Z(S^2, {\bar R}_i, R_j) =  {1\over N!} \int d^N x\, \Delta_{q,t}(x)^2 \;M_{R_i}( e^{-x}) \,M_{R_j} (e^{x}).
}
we get
\eqn\metric{
Z(S^2, {\bar R}_i, R_j) =g_{{\bar i}j} = g_i {\delta^{i}}_j,
}
where $g_i=g_i(q,t)$ is the Macdonald norm, given in more detail in the next section, and in the appendix A.
%

Consider now adding more D-branes on $T^*(S^2\times S^1)$, corresponding to inserting
Wilson lines in representations $R_i$, $R_j$ and $R_k$ at three points on the $S^2$, and winding around the $S^1$,
\eqn\vcd{
Z(S^2\times S^1, R_i,R_j, R_k)=
 {1\over N!} \int d^N x\, \Delta_{q,t}(x)^2 \;M_{R_i}( e^{x}) \,M_{R_j} (e^{x})\,M_{R_k} (e^{x}).
}
We get from this
\eqn\vc{
Z(S^2\times S^1, R_i,R_j, R_k)= N_{i j k}
}
where $N_{ijk}$ can be computed either by an explicit integration in \vcd\  or using \metric\
from tensor-product coefficients ${N^{k}}_{ij}$ for Macdonald polynomials. Defining the latter by
$$
M_{R_i}(e^x) M_{R_j}(e^x) = \sum_{k} {N^k}_{ij} M_{R_k}(e^x),
$$
we get immediately,
$$
N_{ijk} = \sum_{\ell} g_{i \ell} {N^{\ell}}_{jk}.
$$
Naturally, ${N^i}_{jk}$ are the refinement of the Verlinde coefficients of the $SU(N)_k$ WZW model.

\newsec{Refined Chern-Simons Theory}

The theory of $N$ M5 branes on $(M\times \,{\IC} \times S^1)_{q,t}$ defines an index \rindex\ when $M$ is a Seifert three-manifold. The existence of the index guarantees that the theory on $M$ has a supercharge $Q$ (or more precisely a pair of those) which satisfies $Q^2=0$. Existence of such a symmetry often comes hand in hand with topological invariance, although in our case, we expect the theory to be independent only of certain metric deformations, those that preserve the circle symmetry. As we will now show, refined Chern-Simons theory indeed is a topological theory in this sense:
from its $S$ and $T$ matrices, we automatically obtain
topological invariants of Seifert three-manifolds with Seifert knots in them. 

In any topological field theory, the path integral on a manifold $M$ with a boundary $B$, $\del M = B$,  gives a state $|M\rangle \in {\cal H}_B$ in the Hilbert space ${\cal H}_B$ obtained by quantizing theory on $B\times \IR$.  Reversing the orientation of the boundary, one gets elements of the dual Hilbert space ${\cal H}^*_B.$  The overlap of states $|M_R \rangle$ in ${\cal H}_B$ and $\langle M_L| \in {\cal H}^*_B$, computes the partition function on the closed three-manifold $M$, obtained by gluing $M_R$ to $M_L$ with the orientation of the latter reversed. Moreover, diffeomorphisms of the boundary $B$ have a unitary representation on ${\cal H}_B$.

For a three dimensional topological theory, and especially one based on Seifert manifolds, a particularly important instance of this construction deals with the Hilbert space on the torus, $B=T^2$. A theory such as ours is in essence two dimensional, as all of its topological data can be phrased in terms of $\Sigma$ \refs{\aosv,\bw}. The importance of the Hilbert space on $T^2$ thus  originates in the fact that the Hilbert space of a topological theory on $\Sigma$ is based on an $S^1$.
The basis of the Hilbert space ${\cal H}_{T^2}$ can be obtained by taking a solid torus and placing Wilson lines in various representations in its interior in a particular way. More precisely, choosing a basis of $H_1(T^2)$ of the boundary torus, and taking the $(1,0)$ cycle of the $T^2$ to be contractible in the interior, one defines a state
\eqn\hilb{
|R_i \rangle \;\; \in \;\; {\cal H}_{T^2}
}
by the path integral on the solid torus with a  Wilson line in representation $R_i$ running along the $(0,1)$ cycle of the torus. Taking two solid tori, with Wilson lines in representations ${R_i}$, $R_j$ inside, and gluing together, one obtains the path integral on $S^2\times S^1$, with two Wilson lines,
\eqn\met{
\langle R_i | R_j \rangle = g_{{\bar i} j} =Z(S^2 \times S^1, {\bar R_i} , R_j).
}
This can equivalently be viewed as defining a hermitian metric
$g_{{\bar i} j} = \langle R_i | R_j \rangle $. The metric is hermitian, since exchanging the roles of $R_i$ and $R_j$ corresponds to orientation reversal of the manifold,
$$
g_{{\bar j} i}= \langle R_j | R_i \rangle = \langle R_i | R_j \rangle^* = g_{{\bar i} j}^*,
$$
where $^*$ denotes complex conjugation.
Moreover, on the boundary of the $T^2$, one gets the action of $SL(2,\IZ)$ corresponding to the mapping class group of the torus. An element
$K$  of $SL(2, \IZ)$ acts on the basis states by
$$
K |R_i\rangle  = \sum_j {K^j}_i |R_j\rangle
$$
simply corresponding to the fact that the Hilbert space is finite dimensional, and $K$ acts on it.
The representation of $SL(2,\IZ)$ acting on ${\cal H}_{T^2}$, is generated by $S$ and $T$ matrices,
$$
{S^{i}}_j, \qquad {T^{i}}_{j},
$$
satisfying the defining relations of $SL(2,\IZ)$,
\eqn\sltz{
S^4= 1,\qquad (ST)^{3}= S^2.
}
The indices are raised and lowered by the metric $g$; for example, defining
$$
\langle R_j|K |R_i\rangle  =K_{{\bar j} i}
$$
we have, using the definitions,
$$
K_{{\bar j} i} = \langle R_j|K |R_i\rangle =\sum_k {{K}^k}_i\langle R_j|R_k\rangle=  \sum_k {K^k}_i \,g_{{\bar j} k}.
$$
The representation has to be unitary,
since otherwise
3d general covariance would have been lost.
Namely $\langle K R_j | R_i \rangle = \langle R_j | K^{-1}R_i \rangle$ implies
\eqn\unit{
K_{{\bar i} j}^* = {K^{-1}}_{{\bar j} i},
}
since
$\langle K R_j | R_i \rangle=\langle R_i| K R_j  \rangle^*$.
Topological invariance further constrains the representation. For example, the $S$ matrix has to be symmetric and satisfy,
\eqn\ssym{
S^{-1}_{{\bar i}j} = S^*_{{\bar i} j}.
}
This follows since $S_{{\bar i},j}$ is the amplitude of a Hopf link in $S^3$, obtained by gluing two solid tori, with
Wilson lines corresponding to states $|R_i\rangle$ and $|R_j\rangle $ we defined before. Gluing these with
an $S$ transformation of the boundary we get the amplitude corresponding to two linked knots in the $S^3,$ the Hopf link:
$$S_{{\bar i} j} = \langle R_i | S | R_j\rangle.
$$
%
%
The fact that we can smoothly re-arrange the link so that the roles of $R_i$ and $R_j$ get exchanged implies that $S$ has to be symmetric, $S_{{\bar i} j} = S_{{\bar j} i}$. Unitarity then implies
\ssym.

For any topological field theory in three dimensions, one can define Verlinde coefficients $N_{ijk}$ by the partition function of the theory on $S^2\times S^1$ with Wilson lines in representations $R_i$, $R_j$ and $R_k$ inserted at three points on the $S^2$, and winding around the $S^1$.
\eqn\vcd{
N_{ijk} = Z(S^2\times S^1, R_i,R_j, R_k)=\langle 0 | R_i R_j R_k \rangle,
}
If one defines an operator ${\cal O}_{R_i}$ inserting the Wilson line in representation $R_i$ long the $(0,1)$ cycle of the solid torus, i.e.
$$
{\cal O}_{R_i}|0\rangle = |R_i \rangle,
$$
then the Verlinde coefficients come from fusing
$$
{\cal O}_{R_i} {\cal O}_{R_j} = \sum_{k} {N^{k}}_{ij} {\cal O}_{R_k}.
$$

The coefficients are no longer integers, for $q\neq t$ but rational functions of $q,t.$ Moreover, $N_{ijk}$ satisfy the Verlinde formula \VerlindeSN\
\eqn\vf{
S_{{\bar k }i}\, S_{{\bar k} j}/S_{{\bar k} 0 } = \sum_{\ell}{ N^{\ell}}_{i j} S_{{\bar k }\ell}
}
or equivalently,
\eqn\vftwo{
N_{{i} { j} {\bar k}} = \sum_{\ell} {S_{{\bar\ell}i} \, S_{{\bar \ell}j}\,{{( S^*)_{\bar k}}^{{\bar \ell}}\,}/S_{{\bar \ell}0} }.
}
by computing the same amplitudes in two different ways. A derivation of this given in \wcs\ uses an ingredient  we will not have in general, namely braiding. A different derivation using only the ingredients at hand is presented in \ref\AganagicJS{
  M.~Aganagic, H.~Ooguri, N.~Saulina, C.~Vafa,
  ``Black holes, q-deformed 2d Yang-Mills, and non-perturbative topological strings,''
Nucl.\ Phys.\  {\bf B715}, 304-348 (2005).
[hep-th/0411280].
}, and we review it in appendix C.

\subsec{Chern-Simons Theory, a Review}

In the case of $SU(N)$ Chern-Simons theory, the basis of Hilbert space ${\cal H}_{T^2}$ is provided by the conformal blocks on $T^2$ of $SU(N)_k$ $WZW$ model \wcs .  Only a subset of $SU(N)$ representations enter, those that correspond to integrable highest weight representations of the affine lie algebra. This can be phrased in terms of a constraint on the corresponding Young diagram that $0\leq R_1 \leq k$, where $R_1$ is the length of the first row.
%
%
The basis of the Hilbert space \hilb\ provided by the path integral on the solid torus with Wilson lines, is automatically orthonormal:
\eqn\ort{
g_{{\bar i}j } = \langle R_i|R_j\rangle = {\delta^i}_j.
}
This is because $g_{{\bar i} j} = Z(S^2\times S^1, {\bar R_i}, R_j)$ is the partition function of the theory on $S^2\times S^1$ with Wilson lines in representations ${\bar R}_i$ and $R_j$ inserted at points on the $S^2,$ and wrapping the $S^1$. This is either 1 or zero depending on whether or not the $R_i$ and $R_j$ are the same representation.
The $S$ and $T$ matrices are given by
$$
S_{{\bar i}j} =d_{R_i}(q^{\rho})s_{R_j}(q^{\rho +\lambda_{R_i}}),
$$
and
$$T_{{\bar i} j} = {\delta^i}_j q^{{1\over 2}(\lambda_{R_i} + 2 \rho, \lambda_{R_i})-{k\over 2 N}(\rho,\rho)},
$$
with
$$
d_{R_i}(q^{\rho}) = {{i}^{N(N-1)/2}\over N^{1\over 2}(k+N)^{N-1\over 2}}
\; \prod_{\alpha>0}(q^{-(\alpha, \rho)/2} - q^{(\alpha, \rho)/2}) \;s_{R_i}(q^{\rho})
$$
where the product is over all positive roots $\alpha$, and $\lambda_{R_i}$ is the highest weight of representation $R_i$.
In all of this, $q$ is a root of unity
$$
q = e^{2\pi i \over k+N}.
$$
Here $S$ and $T$ are finite dimensional matrices, since the space of allowed representations is finite. Moreover they explicitly satisfy the properties \sltz\unit\ssym\ listed above.

The Verlinde coefficients ${N^{i}}_{jk}$ are simply tensor product coefficients, restricted to the allowed set of representations corresponding to states in ${\cal H}_{T^2}$. This corresponds to the fact that the operator
${\cal O}_{R_i}$ that inserts Wilson loop in representation $R_i$ is simply the trace of the holonomy around the $(0,1)$ cycle ${\cal O}_{R_i} = s_{R_i}(e^x)$. The fact that the $S$ matrix satisfies the Verlinde formula formula follows, since it is written in terms of Schur functions, evaluated at roots of unity, as explained in \VerlindeSN .

\subsec{Refined Chern-Simons Theory}

As we have seen in section 4, M-theory provides a way to derive the Hilbert space of the theory on the torus, and any amplitudes obtained from this by cutting and gluing. In particular, the partition function on $S^2\times S^1$,
obtained by gluing two copies of \leftW\ with the corresponding Wilson lines computes the metric which is diagonal, but no longer trivial\foot{We could have normalized the Macdonald polynomials in a way to set $g_i$ to one. We will refrain from doing so, since normalized Macdonald polynomials will contain square roots, and we prefer working in terms of rational functions of $q$ and $t$.}
$$\langle R_i |R_j\rangle =g_{{\bar i} j} = g_i {\delta^{i}}_j.
$$
where
\eqn\metd{
g_i = \prod_{m=0}^{\beta -1}
\prod_{\alpha>0}{q^{-{1\over 2}( \alpha, \lambda_R)}t^{ -{1\over 2} (\alpha,\rho)} q^{-{m\over 2}}-
q^{{1\over 2} (\alpha, \lambda_R)} t^{ {1\over 2} (\alpha,\rho)} q^{m\over 2}\over
q^{-{1\over 2}( \alpha, \lambda_R)}t^{ -{1\over 2} (\alpha,\rho)} q^{m\over 2}-
q^{{1\over 2} (\alpha, \lambda_R)} t^{ {1\over 2} (\alpha,\rho)} q^{-{m\over 2}}}
}
The metric is hermitian, $g_{{\bar i}{ j}} = g^*_{{\bar j}{ i}}$, and $g_i$ is real
provided
$$q^* = q^{-1}, \qquad t^* = t^{-1},
$$
where $t=q^{\beta}$, as before. While the hermiticity is manifest here, the fact that one can continue these expressions to arbitrary real $\beta$ is not. The equivalent form of $g_i$ presented in \metd\ makes that fact manifest.

The Hilbert space ${\cal H}_{T^2}$ is unchanged from its unrefined values, and in particular, the allowed representations $R_i$ still correspond to the integrable highest weight representations of $SU(N)_k$. Indeed, setting
\eqn\rtu{
q = e^{2\pi i \over k+\beta N}, \qquad t = e^{2 \pi i \beta \over k+\beta N}.
}
the metric vanishes for representations other than the those whose Young tableau fits in box of width $k$,
This is to be expected, since $\beta$ is arbitrary and we can change it away from the unrefined value $\beta=1$ adiabatically -- in a finite dimensional Hilbert space, the states have nowhere to go to.

The $S$ and the $T$ matrices are given by
\eqn\nsm{
S_{{\bar i}j} =d_{R_i}(t^{\rho})M_{R_j}(t^{\rho} q^{ \lambda_{R_i}}),
}
and
\eqn\ntm{T_{{\bar i} j} = {\delta^i}_j \;q^{{1\over 2}(\lambda_{R_i}, \lambda_{R_i})}\,t^{(\lambda_{R_i}, \rho)}t^{{\beta-1\over 2}(\rho, \rho)}q^{-{k\over 2 N}(\rho,\rho)},
}
where
$$
d_{R_i}(t^{\rho}) = {{i}^{N(N-1)/2}\over N^{1\over 2}(k+\beta N)^{N-1\over 2}}
\; \prod_{m=0}^{\beta-1}\prod_{\alpha>0}(q^{-m/2} t^{-(\alpha,  \rho)/2} - q^{m/2} t^{(\alpha,  \rho)/2}) \;M_{R_i}(t^{\rho})
$$
Setting $\beta$ to $1$, $q$ and $t$ coincide, Macdonald polynomials become Schur functions, and the refined Chern-Simons amplitudes reduce to ordinary ones, as they should.

The fact that $S_{{\bar i} j}$ is symmetric, which is crucial for the three dimensional interpretation of it as expectation value of the colored Hopf link,
\eqn\ssi{
S_{{\bar i} j} = S_{{\bar j} i}
}
is not obvious from the formula \nsm , just like in the unrefined case. However, it is obvious from the M-theory derivation of it, see \mc .
%
%
In fact, symmetry of the formula \ssym\ was one of the original Macdonald conjectures \refs{\macda,\macdb} , and the formula \mc\ was derived in \ek\ as a means of proving it (for earlier proofs, see \refs{\cheredniko, \cherednikt, \kirillov})! Here we have discovered the same statements from M-theory.
The forms of the metric the $S$ matrix and $T$ are slight rewriting of what we had in section 4, motivated by \kirillov , now with normalizations carefully fixed.

We can furthermore see that $S$ and $T$ provide a unitary representation of $SL(2,\IZ)$, acting on the Hilbert space ${\cal H}_{T^2}$. Since the metric $g_{{\bar i} { j}}$ is not identity, $S_{{\bar i} j}$ and ${S^{i}}_j$ are not the same, instead,
$$
S_{{\bar i} {j}} = \sum_{k} g_{{\bar i}k} {S^{k}}_j = g_{i} {S^{i}}_j,
$$
and similarly for $T$.
In particular, while $S_{{\bar i}j}$ is symmetric, ${S^{i}}_{j}$, is not. From our discussion above, for $S$ and $T$ to provide a unitary representation on $SL(2,\IZ)$,
$$
{S^{i}}_j, \qquad {T^{i}}_{j},
$$
should satisfy
$$
S^4= 1,\qquad (ST)^{3}= S^2.
$$
They indeed do, one can checked explicitly in examples. The general statement was proven in \refs{\cheredniko, \kirillov, \cherednikt }, for integer $\beta$. It can also be shown that $S$ and $T$ are unitary, so that
\eqn\unit{
{(S^{-1})^i}_j = {{(S^*)}^i}_j, \qquad {(T^{-1})^i}_j = {{(T^*)}^i}_j.
}

The quantum dimension of the representation $R_i$ generalizes to
$$
S_{{\bar i}0}/S_{00} =
\prod_{m=0}^{\beta -1} \prod_{1\leq i<j\leq N}
{q^{{R_j - R_i\over 2}} t^{{i-j\over 2}} q^{-{m\over 2}}-q^{R_i - R_j\over 2} t^{j-i\over 2} q^{m\over 2}\over
 t^{{i-j\over 2}} q^{-{m\over 2}}-t^{j-i\over 2} q^{m\over 2}}
=   M_{R_i}(t^{\rho}).
$$

The Verlinde coefficients ${N^i}_{jk}$ similarly follow, as in section 4, from explicitly computing the algebra of operators
${\cal O}_{R_i}$ which are in this case given by Macdonald polynomials, as functions of holonomies, $M_{R_i}(e^x)$.
Unlike in the ordinary Chern-Simons case, they are no longer integral. They are rational functions of $q$ and $t$ instead. Finally, since the $S$ matrix itself is obtained by evaluating the Macdonald polynomials at special points, the Verlinde formula \vf\
$$
S_{{\bar k }i}\, S_{{\bar k} j}/S_{{\bar k} 0 } = \sum_{\ell}{ N^{\ell}}_{i j} S_{{\bar k }\ell}
$$
also follows, see \co .

\subsec{Three-manifold Invariants from Refined Chern-Simons Theory}

A topological field theory in three dimensions assigns to a three-manifold $M$ a topological invariant,
the value of the partition function on $M$. In any topological field theory in three dimensions, the three-manifold invariant of $M$ can be obtained from that of any other three-manifold, say $S^3$, by surgery. This is an operation where one cuts out a tubular neighborhood of a knot $K$ in $S^3$ -- this is a solid torus-- and glues it back in up to an $SL(2,{\IZ})$ transformation of the boundary. To get an arbitrary three-manifold invariant in this way one needs both $S$ and $T$ matrices, and the braiding matrix \refs{\wcs,\ms}.
In the present case, we only have the $S$ and the $T$ matrices -- the braiding matrix is not obtainable from M-theory. This is just as well, since a three-manifold invariant corresponding to a Seifert space can be written in terms of $S$ and $T$ only.

A compact Seifert three-manifold, fibered over the Riemann surface of genus $g$, carries labels (the description of the geometry is borrowed from \bw )
$$
( g, n; (\alpha_1, \beta_1), \ldots,(\alpha_r,\beta_r)),
$$
where $n$ is the degree of the circle bundle, and $(\alpha_i,\beta_i)$ are integers, parameterizing the type of special fibers that occur. Let $z$ be the local coordinate on the base, centered at the point where the fiber is special, and view the circle bundle over $\Sigma$ as a line bundle, with complex coordinate $s$ parameterizing the fiber. Then, near the point with fiber of type $(\alpha_j, \beta_j)$, the total space is modeled on ${\;\IC}\times {\IC}/\IZ_{\alpha_j}$, where $\IZ_{\alpha_j}$ acts by
$$
z\rightarrow \zeta z , \qquad s \rightarrow \zeta^{\beta_j} s, \qquad \zeta = e^{2\pi i\over \alpha_j}.
$$
The $S^1$ action that rotates the fiber is no longer free -- the ${\IZ_{\alpha_j}}$ subgroup of it fixes the fiber above $z=0$. Similarly, the base of the, obtained by forgetting the $S^1$ fiber, is an orbifold. But, the total space is smooth, and in particular the $U(1)$ action defines a nowhere vanishing vector field $V$ on $M$, as we needed in section 3.

The three-manifold invariant of this can be computed\foot{See for example, \ref\hn{S. K. Hansen, "Reshetikhin-Turaev Invariants of Seifert 3-Manifolds and a Rational
Surgery Formula," Algebr. Geom. Topol. 1 (2001) 627�686, math.GT/0111057.} for details. Earlier work on three-manifold invariants of Seifert spaces includes \lref\seifa{R. Lawrence and L. Rozansky, "Witten-Reshetikhin-Turaev Invariants of Seifert Manifolds," Commun. Math. Phys. 205 (1999) 287�314}
\lref\seifb{L. Rozansky, "Residue Formulas for the Large k Asymptotics of Witten's Invariants of Seifert Manifolds: The Case of SU(2)," Commun. Math. Phys. 178 (1996) 27�60, hep-th/9412075}
\lref\seifc{D. Freed and R. Gompf, "Computer Calculation of Witten�s 3-Manifold Invariant," Commun. Math. Phys. 141 (1991) 79�117}
\lref\seifd{L. Jeffrey, "On Some Aspects of Chern-Simons Gauge Theory," D.Phil. thesis, University of Oxford, 1991}
\lref\seife{L. Jeffrey, "Chern-Simons-Witten Invariants of Lens Spaces and Torus Bundles," and the Semiclassical Approximation,  Commun. Math. Phys. 147 (1992) 563�604.}
\lref\seiff{ S. Garoufalidis, "Relations Among 3-Manifold Invariants," Ph.D. thesis, University of
Chicago, 1992.}
\lref\seifg{ J.R. Neil, "Combinatorial Calculation of the Various Normalizations of the Witten
Invariants for 3-Manifolds," J. Knot Theory Ramifications 1 (1992) 407�449.}
\lref\seifh{ L. Rozansky, "A Large k Asymptotics of Witten's Invariant of Seifert Manifolds,"
Comm. Math. Phys. 171 (1995) 279�322, hep-th/9303099.} \refs{\seifa,\seifb,\seifc,\seifd,\seife,\seiff,\seifg,\seifh}.} by surgery on a torus link in the $S^3$. The end result is
$$
Z(g, n; (\alpha_1, \beta_1), \ldots,(\alpha_r,\beta_r)) = \sum_{j}T_j^{n}  (g_j)^{g-1}  (S_{0,j})^{2-r-2g}\Bigl( \prod_{i=1}^r (SK^{(\alpha_i, \beta_i)})_{0, j}\Bigr)
$$
where $K^{(\alpha_i, \beta_i)}$ is an $SL(2,{\IZ})$ matrix whose first column is ${(\alpha_i, \beta_i)}$. Each such matrix, as an element of $SL(2, \IZ)$ can be written as a product of $S$ and $T$ matrices. The $U(1)$ bundle over $\Sigma_g$ has first Chern class $n - \sum_{i=1}^r \beta_i/\alpha_i$.
In the simple case without special fibers $(g,n)$, the three-manifold is a circle bundle over a smooth Riemann surface $\Sigma_g$ of degree $n$, the corresponding partition function is
$$ Z(g,n) = \sum_i { (g_i)^{g-1}  T_i^n (S_{0i})^{2-2g}}.
$$
We can obtain $(g, n; (\alpha_1, \beta_1), \ldots,(\alpha_r,\beta_r))$ from this by cutting out the neighborhoods of $r$ knots wrapping the $S^1$ fibers over points on $\Sigma$, and gluing back corresponding solid tori by $SL(2, {\IZ})$ transformation of their boundaries, corresponding to $K^{(\alpha_i, \beta_i)}$.

Written in terms of $S$ and $T$, the three-manifold invariant no longer depends on the underlying topological field theory, but only on $M$. The dependence on the theory enters only through $S$ and $T$. To get a three-manifold invariant corresponding to $SU(N)$ Chern-Simons theory on $M$, one would use the $S$ and the $T$ matrices of the $SU(N)_k$ WZW model. To get the path integral of the refined Chern-Simons theory on $M$ instead, one uses the refined $S$ and $T$ matrices of the previous subsection.

\subsec{Knot Invariants in Operator Formalism and the Refined Chern-Simons Theory}

In addition to three-manifold invariants, a topological field theory in three dimensions also gives rise to invariants of knots in three-manifolds. A beautiful result of \wcs\ is the realization that knot invariant that one gets from $SU(2)_k$ Chern-Simons theory on $S^3$, with Wilson loops in fundamental representation of $SU(2)$, is the Jones polynomial \jones . The knot invariants arising from our theory on $S^3$, with as we will see in section 7, compute the dimensions of knot homologies instead.

 As we explained in section 3, in the refined Chern-Simons theory one can consider any knot or a link that is fixed by a semi-free $S^1$ action on the Seifert manifold $M.$ Viewing $M$ as an $S^1$ fibration over a Riemann surface (there may be more than one way to think about a given three-manifold in this way), the knots preserved by the $S^1$ action are those wrapping the fiber.  The operator formulation of these knot invariants involves the $S$ and $T$ matrices alone, and the fusion coefficients
$N_{ijk}$. The later can also be written in terms of $S$, by the Verlinde formula. Written in terms of $S$ and $T$, this no longer depends on the details of the theory, although the answers that one gets of course do.\foot{The way we normalized the Macdonald polynomials, the metric $g_{{\bar i} j} = g_i {\delta^{i}}_j$  also enters. However, this is just a matter of convention, as we could have set the metric to identity, by rescaling the MacDonald polynomials.}
In the previous sections we found the $S$ and $T$ of the refined Chern-Simons theory, form M-theory. With them in hand, we can compute the knot invariants.

We will explain how to compute invariants of torus knots in the $S^3$\foot{In the unrefined case, these were considered from perspective of string theory recently in \ref\bem{
  A.~Brini, B.~Eynard, M.~Marino,
  ``Torus knots and mirror symmetry,''
[arXiv:1105.2012 [hep-th]].
}.}. Knots and topological invariants associated to other three-manifolds can be computed equally easily, but in aiming to connect with the work of Khovanov, we will focus on the $S^3$. A torus knot in $S^3$ is a knot that can be obtained from an unknot by an $SL(2, {\IZ})$ transformation.
We view the $S^3$ as a locus in ${\;\IC}^2$, with coordinates $z_1,z_2$, where
\eqn\thre{
|z_1|^2+|z_2|^2 = 1
}
In this way, the $S^3$ is $T^2$ fibration over the interval, where the $T^2$ fibers correspond to phases of $z_1$ and $z_2$. We denote by $(0,1)$ cycle of the $T^2$ the $S^1$ generated by the phase of $z_1$, and by $(1,0)$ the $S^1$ generated by the phase of $z_2$.
An $(n,m)$ torus knot is described by \thre\ together with the equation
$$
z_1^n=z_2^m
$$
This is invariant under the $U(1)$ action that takes $(z_1, z_2)\rightarrow (\zeta^m z_1, \zeta^n z_2)$, with $\zeta=e^{i \theta}.$ This $U(1)$ action acts freely on the $S^3$, except for a ${\IZ}_m$ subgroup, generated by $\zeta= e^{2 \pi i /m}$, that has fixed points at $z_2=0$, and a  ${\IZ}_n$ subgroup that similarly has fixed points at $z_1=0$.

We view the $S^3$ as obtained by gluing two solid tori.
Take a solid torus, corresponding to the "right half" of the $S^3$ in \thre . This is a $T^2$ fibration over the interval, in such a way that $(1,0)$ cycle shrinks in the interior, and $(0,1)$ cycle stays finite, and denote by
$$|0\rangle,$$
the state in ${\cal H}_{T^2}$ obtained by computing the path integral on it. To get the $S^3$, we glue together two copies of this, by an $S$
transformation of the boundary, exchanging the $(1,0)$ and the $(0,1)$ cycle,
$$
\langle 0 | S|0\rangle.
$$
Let us denote by ${\cal O}^{(0,1)}_{R_i}$ an operator that inserts, along the $(0,1)$ cycle of torus, an unknot in representation $R_i$. Then, per definition, the operator we just defined creates from the vacuum $|0\rangle$, the state $|R_i\rangle $,
$$
{\cal O}^{(0,1)}_{R_i}|0\rangle = |R_i\rangle.
$$
Inserting a $(0,1)$ knot in the $S^3$, corresponds in the operator language to computing
$$
\langle 0 | {\cal O}^{(0,1)}_{R_i} S|0\rangle = \langle R_i|S|0\rangle = S_{i0}.
$$
The operator ${\cal O}^{(n,m)}_{R_i}$ that inserts the $(n,m)$ knot instead is related to the operator ${\cal O}^{(0,1)}_{R_i}$ by an operation that takes the torus boundary, and maps the $(0,1)$ cycle into $(n,m)$ cycle. In other words
$$
{\cal O}^{(n,m)}_{R_i} =K{\cal O}^{(0,1)}_{R_i} K^{-1}
$$
where $K$ is an element of $SL(2,\IZ)$ that takes the $(1,0)$ cycle to the cycle $(a,b)$ cycle dual to $(n,m)$, i.e.
$$
K = \pmatrix{ a & n \cr
              b &  m \cr} \;\;  \in \;\; SL(2,{\bf Z}).
$$
with $am-nb =1$ since then the action of ${\cal O}_{R_i}^{(0,1)}$ on $|0\rangle$, and ${\cal O}^{(n,m)}_{R_i}$ on $ K|0\rangle$ agree. Any such $K$ can be written explicitly in terms of strings of $S$ and $T$ matrices
$$
S = \pmatrix{ 0 & -1 \cr
              1 &  0 \cr} , \;\; T = \pmatrix{1& 1\cr 0&1} \;\; \in \;\; SL(2,{\bf Z}).
$$
%
We can write this out explicitly, in terms of the Verlinde coefficients
$${\cal O}^{(0,1)}_{R_i}|R_j\rangle =\sum_{k} {N^{k}}_{ij} |R_k\rangle.
$$
and the matrices expressing the action of $K$ on the Hilbert space, as follows:
\eqn\tn{
K{\cal O}^{(0,1)}_{R_i}K^{-1} |0\rangle = \sum_{j,k,\ell} {{(K)^{\ell}}_k}{N^{k}}_{ij}{(K^{-1})^{j}}_0|R_{\ell}\rangle.
}

We want to use this to compute knots in the $S^3$, so we glue together the two solid tori, one of which contains our knot, by an $S$ transformation of their boundary. For, example, for computing the knot invariant of the $(n,m)$ torus knot in $S^3$, we compute
$$
\langle 0 |{\cal O}^{(n,m)}_{R_i} S |0\rangle =
\langle 0 | K {\cal O}^{(0,1)}_{R_i}{K}^{-1} S|0\rangle =
\sum_{j,k,\ell} K_{0k}{N^{k}}_{ij}{(K^{-1})^{j}}_{\ell} {S^{\ell}}_0,
$$
Similarly, considering prime links in the $S^3$ whose components are torus knots we consider two copies of \tn\ with the suitable choices of the $K$ matrices, and glue them together by an $S$ transformation. For example, linking the $(1,0)$ knot with an $(n,m)$ link in the $S^3$, one would compute
$$\langle 0 | {\cal O}^{(n,m)}_{R_i} S{\cal O}^{(0,1)}_{R_p}|0\rangle =
\langle 0|K{\cal O}^{(0,1)}_{R_i}K^{-1} S |R_p\rangle = \sum_{j,k,\ell} K_{0k}{N^{k}}_{ij}{(K^{-1})^{j}}_{\ell} {S^{\ell}}_p,
$$
taking into account that the $S$ transformation exchanges the $(0,1)$ and $(1,0)$ knots. We will show in the next sections that the knot invariants arising from our theory are related to computing dimensions of knot homologies.

\newsec{Large $N$ Duality and Refined Chern-Simons Theory}

In this section will first review large $N$ duality in the ordinary topological string, and then explain how it extends to the refined theory.

Recall, from section 2, that $SU(N)$ Chern-Simons theory on $S^3$ is the same as the open topological string on
$$Y=T^*S^3,$$
with $N$ D-branes on the $S^3$.
Gopakumar and Vafa showed this has a large $N$ dual, the ordinary topological string theory on
$$
{X} = {\cal O}(-1) \oplus {\cal O}(-1) \rightarrow {\IP}^1.
$$
The duality is a large $N$ duality in the sense of 't Hooft \ref\thooft{G. �t Hooft, �A Planar Diagram Theory for Strong Interactions,� Nucl. Phys. 72 (1974) 461.}. The topological string coupling $g_s$ is the same on both sides -- it is related to the level $k$ in Chern-Simons theory by
$g_s = {2\pi i \over k+N}$. The later is the effective coupling constant of Chern-Simons theory, due to the famous shift of
$k\rightarrow k+N$, generated by quantum effects. The rank $N$ of the gauge theory is related to the area of the ${\IP^1}$ in $X$ by
$\lambda = e^{-Area({\IP^1})}$ by $\lambda = q^N$ where $q = e^{g_s}.$
This is a large $N$ duality, since when the geometry of $X$ is classical, $\lambda$ is continuous, and this is only true in the limit of large $N$. The duality in this case also has a beautiful geometric interpretation: it is a geometric transition that shrinks the $S^3$ and  grows the ${\IP^1}$ at the apex of the conifold, thereby taking $Y$ to $X$ \gvI. The duality has been checked, at the level of partition functions,  to all orders in the $1/N$ expansion \gvI .

As we explained in section $2$, adding knots on the $S^3$ corresponds to introducing non-compact branes on a Lagrangian $L_K$ in $Y$, intersecting the $S^3$ along a knot $K$,
$$
K = L_K \cap S^3.
$$
The geometric transition affects the interior of $X$ and $Y$, but not their asymptotics, which are the same.
A non-compact Lagrangian $L_K$ on $Y$, goes through the transition to give a Lagrangian $L_K$ on $X$.

\subsec{Large $N$ Duality and the Refined Chern-Simons Theory}

The large $N$ duality relating open topological string on $Y=T^*S^3$, i.e, Chern-Simons theory on the $S^3$ to topological string on $X= {\cal O}(-1) \oplus {\cal O}(-1) \rightarrow {\IP}^1$is expected to extend to the refined topological string \refs{\civ,\gsv, \toda}.
With the refined Chern-Simons theory in hand, we will give strong evidence that the refined theory indeed inherits the large $N$ duality.
Analogous results for the B-model topological string were provided in \toda .

Consider first the pure refined Chern-Simons theory, without knot observables. In this case, on $Y$, we are computing the partition function of the theory on $S^3$, $Z(S^3)$. On $X$ we are computing the partition function of the refined closed topological string. The latter is well known.  Let $\lambda$ parameterize the size of the ${\IP}^1$ in $X$ as  $\lambda =e^{-Area({\IP}^1)}$, were by the area we really mean the mass of a BPS state wrapping the $\IP^1$. The partition function of the closed topological string in this background is  \civ\
\eqn\refc{
Z^{top}(X, \lambda, q,t) = \exp\Bigl(-\sum_{n=0}^{\infty} {\lambda^n \over n (q^{n/2} - q^{-n/2})(t^{n/2} - t^{-n/2})}\Bigr),
}
up to simple factors that correspond to classical intersection numbers on the Calabi-Yau, and are ambiguous.

On the other hand, we obtained the refined Chern-Simons partition function on $S^3$ in \vac . With $N$ branes on the $S^3$ this is
$$
Z(S^3, q,t)=S_{00} = \prod_{m=0}^{\beta -1} \prod_{i=1}^{N-1}(1-t^{N-i } q^{m})^{i},
$$
if we set
$$
t=q^{\beta}, \qquad \beta \in \IN
$$
(as emphasized before, we can easily analytically continue everything away from integral $\beta$, but the explicit formulas are simpler and easier to deal with if we do not).
Taking the large $N$ limit of this expression, we find
\eqn\lncs{
Z(S^3,q,t)\sim \prod_{m=0}^{\beta -1}{ \prod_{i=1}^{\infty}(1-t^{N-i} q^m)^{i}}.
}
up to terms with trivial $N$ dependence.\foot{All of the terms we drop are ambiguous due to the non-compactness of the Calabi-Yau. There are perturbative terms, that would have been related to triple intersection numbers if the Calabi-Yau was compact.
There are also terms that are due to D0 branes -- BPS states with no D2 brane charge on $X =  {\cal O}(-1) \oplus {\cal O}(-1) \rightarrow {\IP}^1$ \civ . These are also ambiguous, since D0 brane can be anywhere on $X$, so feels the non-compactness of the manifold. This is unlike the D2 brane (or M2 brane) which is fixed to wrap the minimal ${\IP}^1$ at  the apex of the conifold, if it is to be BPS.}
%
%
Rewriting \lncs\ as follows
\eqn\lncsa{\eqalign{
Z(S^3,q,t)&=\exp\Big(- \sum_{m=0}^{\beta -1} \sum_{i,k =1}^{\infty} {i\over k} {t^{(N-i) k }q^{mk} }\Big), \cr
& =\exp\Big( -\sum_{m=0}^{\beta -1} \sum_{k =1}^{\infty} {1\over k} {t^{N k}\,t^{k/2}\,q^{-k/2} \over (t^{k/2} -t^{- k/2})(q^{k/2} - q^{-k/2})}\Big) ,
}}
we find that \lncs\ and \refc\ agree,
$$
Z(S^3,q,t, N) = Z^{top}(X, q,t, \lambda),
$$
provided we identify
\eqn\map{\lambda =  t^{N} t^{1\over 2} q^{-{1\over 2}}.
}
The fact that $N$ $\ep_1$-branes on the $S^3$ back-react on the geometry in such a way to open up a ${\IP}^1$ of size $N\ep_2$, is as expected on general grounds, as discussed recently in detail in \acdkv . By $\ep_1$ branes we mean the branes wrapping the $z_1$ plane rotated by the parameter $q=e^{\ep_1}$, and by $\ep_2$ branes, the branes wrapping the $z_2$ plane, rotated by $t = e^{\ep_2}$.  The quantum shifts of the moduli by $(\ep_2-\ep_1)/2,$ which vanish in the unrefined theory at $q=t$, are typical \refs{\toda,\acdkv}.

Now that we have identified the large $N$ dual of the $SU(N)$ refined Chern-Simons theory on the $S^3$ as the refined topological string on $X$, we can introduce knots on the $S^3$. As we explained in sections $2$ and $3$, adding knots on the $S^3$ corresponds to introducing non-compact branes on the Lagrangian $L_K$, intersecting the $S^3$ along a knot $K$,
$$
K = L_K \cap S^3,
$$
where we require that $K$ be fixed by a (semi-)free $U(1)$ action on the $S^3$.
The refined topological string amplitude on $X$, with the branes on $L_K$ is given
by first integrating out the contributions to the refined index coming from massive BPS states of M2 branes stretching between the M5 branes $S^3$ and $L_K$. Then, integrating over the light modes on the M5 branes, one ends up computing in the refined Chern-Simons theory on the $S^3$, the expectation value of the corresponding operator \indexref, or in this case \ovtq .

More precisely, in the refined topological string, M-theory, there are two different ways to introduce the additional M5-branes, and hence two different ways define the expectation value of the knot (in addition to the usual framing ambiguity). Namely, while the Lagrangian $L_K$ is fixed in $Y$, in M-theory we have to choose which of the $z_1$, $z_2$ planes in $TN$ space the brane wraps. This results in two different kinds of branes of the refined topological string \refs{\GGV,\acdkv}. If it were not for the branes on $S^3$, the two choices we would be related by a symmetry of the theory that exchanges $q$ and $t$. With the branes on $S^3$, this is no longer the case. If we take, as in sections 2 and 3, the branes on the $S^3$ to wrap $z_1$ plane, then by having the branes on $L_K$ wrap $z_1$ plane as well, we end up computing the expectation value of the operator ${\cal O}_K(U,V, q, t)$ \ovtq , as in section 3, since there we manifestly dealt with only one kind of brane
\eqn\ovtqa{\eqalign{
{\cal O}^{\ep_1}_K(U,V)&=\prod_{m=0}^{\beta-1}{\det}^{-1}\bigl(1 \otimes 1 - q^m U \otimes V^{-1}\bigr).\cr
&=\sum_{R} {1\over g_R}\; M_R(U) M_R(V^{-1}).
}}
Here, $U$ is the holonomy on the $S^3$, that gets integrated over.
On the other hand, choosing the branes on $L_K$ to wrap the $z_2$ plane instead, we get $\ep_2$ branes on $L_K$, and the operator \ovtqa\  is replaced by
\eqn\ovtqb{\eqalign{
{\cal O}^{\ep_2}_{K}(U,V)&={\det}\bigl(1 \otimes 1 - U \otimes V^{-1}\bigr),\cr
=&\sum_{R}\; (-1)^R M_R(U) M_{R^T}(V^{-1}).
}}
The difference between the two operators \ovtqa\ and \ovtqb\ corresponding to inserting $\ep_1$ or $\ep_2$ branes is analogous to the B-model case discussed in \acdkv .
Thus, depending on the brane one inserts, the topological string is computing
\eqn\fullt{
Z_{\ep_{1,2}}(S^3, K, V) = \langle 0| {\cal O}^{\ep_{1,2}}_{K}(U,V) S|0\rangle
}
Before we make mathematical predictions, we need to explain what question in M-theory corresponds to computing the Wilson loop expectation value in the refined Chern-Simons theory in some representation $R$. The natural object, analogous to what one did in the unrefined case, is simply to extract the coefficient in front of $M_R(V)$ from \fullt . This will depend on which brane we insert, since in one case we would be computing
$\langle M_R(U)\rangle_{S^3} m_R$, corresponding to $\ep_1$ brane while in the other  $\langle M_{R^T}(U)\rangle_{S^3}$, corresponding to $\ep_2$ brane. For example, in the latter $\ep_2$ brane case
the expectation value of the unknot on the $S^3$, in the fundamental representation is given by
$$
Z_{\ep_2}(S^3, \bigcirc, R)/Z(S^3) = S_{0\tableau{1}}/S_{00}.
$$
From our explicit formulas for the $S$ matrix in terms of Macdonald polynomials, we find
$$
Z_{\ep_2}(S^3, \bigcirc)/Z(S^3)= {t^{N/2} - t^{-N/2}\over t^{1/2} - t^{-1/2}}.
$$
While this agrees with the Khovanov-Rozansky polynomial for the unknot \refs{\gsv,\giv}  the answer is not canonical. Had we chosen $\ep_1$ brane instead, as was implicit in \giv ,  we would have obtained

$$
Z_{\ep_1}(S^3, \bigcirc)/Z(S^3) = {1\over g_{\tableau{1}}} S_{0\tableau{1}}/S_{00} = {t^{N/2} - t^{-N/2}\over q^{1/2} - q^{-1/2}}.
$$
\vskip 0.3cm

To avoid dealing with these subtle issues, the natural objects for us to compute are the reduced knot invariants, where one normalizes the expectation value of the unknot to identity.
The canonical objects to compute are the reduced knot invariants, where one normalizes the expectation value of the unknot in representation $R$ to identity,
$$
Z(S^3, K, R)/Z(S^3, \bigcirc, R)
$$
This no longer depends on which brane we choose, so we can drop the $\ep_{1,2}$ index.
\newsec{Knot Homology, M-theory and Refined Chern-Simons Theory}


To explain the integrality of the Jones polynomial,  Khovanov \kh\ introduced the idea of associating a bi-graded homology theory to a knot $K$ in $S^3$, in such a way that the Euler characteristic of it, taken with respect to one of the gradings, reproduces the Jones polynomial $J(K)$ of the knot $K,$
$$J(K)({\bf q}) = \sum_{i,j} (-1)^i {\bf q}^j \dim {\cal H}^{i,j}(K).
$$
The Jones polynomial corresponds to expectation value of a Wilson line in fundamental representation of $SU(2)$ Chern-Simons theory on $S^3$, at level $k$, and  ${\bf q}=e^{\pi i /k+2}$.
We will work with the reduced homology, so that $J(K)({\bf q})$ is normalized in such a way that the expectation value of the unknot is $J(\bigcirc)=1$.
The homology theory clearly has more information than its Euler characteristic; in terms of its ability to distinguish knots, the Poincar\'e polynomial
$$
Kn(K)({\bf q}, {\bf t}) =\sum_{i,j} {\bf t}^i {\bf q}^j \dim {\cal H}^{i,j}(K).
$$
which gives the Euler characteristic at ${\bf t}=-1$, is a stronger invariant.
\lref\khh{M. Khovanov, Triply-graded link homology and Hochschild homology
of Soergel bimodules, math.GT/0510265.}
This has been generalized in \kro\ to knot homology theory categorifying $SU(N)$ knot invariants,
$$
Kn_N(K)({\bf q}, {\bf t}) =\sum_{i,j} {\bf t}^i {\bf q}^j \dim {\cal H}_N^{i,j}(K).
$$
In \krt\ a generalization to a triply-graded knot homology theory categorifying the HOMFLY polynomial
$$
H(K)({\bf q}, {\bf a}) = \sum_{i,j} (-1)^i {\bf q}^j {\bf a}^k \dim {\cal H}^{i,j, k}(K).
$$
was found. The Poincar\'e polynomial of the triply graded homology theory
\eqn\super{
{\cal P}(K)({\bf q},{\bf a}, {\bf t}) = \sum_{i,j} {\bf t}^i {\bf q}^j {\bf a}^k \dim {\cal H}^{i,j, k}(K),
}
was called "superpolynomial" in \gr .
The HOMFLY polynomial of the knot $H(K)({\bf q}, {\bf a})$ arizes by specializing the super polynomial to ${\bf t}=-1$. As explained in \gr\  the Poincar\'e polynomials of $SU(N)$ knot homology theory, and the triply graded theory, are related by simply specializing the latter ${\bf a}={\bf q}^N$
$$Kh_N(K)({\bf q}, {\bf t}) ={\cal P}(K)({\bf q},{\bf q}^n, {\bf t})  =  \sum_{i,j, k} {\bf t}^i {\bf q}^j  \dim {\cal H}_n^{i,j}(K).
$$
{\it only} for sufficiently large $N$\foot{The authors are grateful to E. Gorsky and S. Gukov for explanations and clarifications of this point.}.
For small $N$ (depending on the knot), an important role is played by the so called $d_N$ differential: the $SU(N)$ knot homology emerges from the HOMFLY knot homology only upon taking the cohomology with respect to $d_N$.


\subsec{Spaces of BPS states and Knot Homologies}

Nearly simultaneously with Khovanov's work another explanation for the integrality of the Jones polynomial was put forward, from physics \ovknot . Recall the two duality relations, the large $N$ duality relating $SU(N)$ Chern-Simons theory on the $S^3$ to the topological string on $X = {\cal O}(-1)\oplus {\cal O}(-1)\rightarrow \IP^1$, and the duality of the topological string on $X$ with to M-theory on $(X \times TN\times S^1)_q$.
In \ovknot\ the authors showed that, taken together, the two dualities imply that computing the invariants of a knot $K$ on $S^3$, such as the Jones or the HOMFLY polynomial is related to the index \mfive\ counting of BPS states of M2 branes on ending on M5 branes  wrapping $L_K$ in $X$,
$$
Z_{CS}(K, S^3, V, q) = Z_M(L_K, X, V, q).
$$
The right hand side is the partition function of M5 branes wrapping the Lagrangial $L_K$ associated to the knot $K$:
$$(L_K \times \IC \times S^1)_q \qquad \in \qquad (X\times TN \times S^1)_q.
$$
Since
$L_K$ is non-compact, there are no light modes on the M5 brane, and the partition function is computed by the index \index\
\eqn\parti{
Z_M(L_K,X; q, V) =   {\rm Tr}_{{\cal H}_{BPS}} (-1)^{F} q^{S_1-S_2} }
counting the BPS states of M2 branes ending on the M5 branes on $L_K$. Here $V$ is the holonomy at infinity on $L_K$.\foot{Recall that $b_1(L_K)=1$ for any knot $K$, by construction in section 2.}
%
%
%
%
The fact that the knot invariants get related to a problem of counting BPS particles explains their integrality.

In \gsv\ this was extended to a conjecture that the spaces of BPS states of M2 branes ending on the M5 branes wrapping $L_K$ in $X$ are the vector spaces that arise in computing homologies of the knot $K$. In particular, the knot homology ${\cal H}^{i, j,k}(K)$ categorifying the HOMFLY polynomial is conjectured to be the same, up to re-grading, as the space of M2 brane BPS states in the fundamental representation of the gauge group on M5 branes,  ${\cal H}_{BPS}^{s_1,s_2,Q}(L_K)$.
The latter
%
%
is graded by the M2 brane charge $Q$ in $H_2(X, \IZ)$ and
$U(1)_1\times U(1)_2$ spins $s_1,$  and $s_2$. The two $U(1)$'s are the two rotations of the Taub-Nut space in \rotref .
%

For a general knot $K$ in the $S^3$, the best one can hope to do is construct explicitly the spaces of BPS states. This is perfectly good in terms of making contact with Khovanov homology, since the spaces of BPS states is all that is needed. While one can define a Poincar\'e polynomial counting the dimensions of the spaces of BPS states
%
by refining the counting in \parti\ to
$$
 {\rm Tr}_{{\cal H}_{BPS}} q^{S_1} t^{-S_2}
$$
by hand,
%
%
{\it the later is not an index} for a general knot $K$ in the $S^3$. This implies that one cannot relate it to the partition function of M-theory, as in any attempt to do so, non-BPS states would contribute\foot{We are grateful to Edward Witten for discussions and explanations tied to this point.}.

\subsec{Refined Chern-Simons theory and an index on knot homologies}

The Poincar\'e polynomial on the space of BPS states is not an index.
However, we argued above that, provided the theory has an additional $U(1)_R$ symmetry, one can define the refined index \indexref, computed by refined Chern-Simons theory, which is the M-theory partition function
\eqn\trui{
   {\rm Tr}_{{\cal H}_{BPS}} (-1)^{F} q^{S_1-S_R} t^{S_R-S_2}\,.\,
}
This index arises either from M-theory on $Y=T^*S^3$, before the geometric transition (or more generally $Y=T^*M$, with $M$ a Seifert manifold), or after the transition, on $X={\cal O}(-1)\oplus {\cal O}(-1)\rightarrow {\bf P}^1$. Taken together with conjecture of \gsv\ equating the spaces of BPS states with knot homologies, it leads us to conjecture existence of a refined index on knot homology groups: on $SL_N$ homology, before the transition, and HOMFLY homology after the transition.
The existence of the extra $U(1)_R$ symmetry implies that the knot homology groups (corresponding to knots colored by arbitrary representation) should admit an additional grading, beyond the usual ${\bf q}-$grading and the homological grade
$$
{\cal H}_{ij} = \oplus_{k}{\cal H}_{ijk}.
$$
This allows one to define a refined index written in terms of knot theory variables abstractly as,
\eqn\rcci{
{\cal P}_{R_i}(K)=\sum_{i,j,k} (-1)^k {\bf q}^i {\bf t}^{j+k} {\rm {dim}} {\cal H}_{ijk}.
}
Finally, we conjecture that the refined index is computed by refined Chern-Simons theory: for $SL_N$ homologies, by the finite $N$ refined Chern-Simons theory, and for HOMFLY homologies, by the refined Chern-Simons theory at large $N$ (in particular, the HOMFLY homology groups would require four gradings). The refined index has more information about knot homology than the Euler characteristic computed by the ordinary $SU(N)$ Chern-Simons theory, but in general less than the Poincar\'e polynomial of the knot homology theory
$$
\sum_{i,j, k} {\bf q}^i {\bf t}^{j} {\rm {dim}} {\cal H}_{ijk} =
\sum_{i,j} {\bf q}^i {\bf t}^{j} {\rm {dim}}{ \cal H}_{ij}  .
$$
However, while computing dimensions of knot homology groups is hard, the index can, by contrast, be obtained simply, by cutting and gluing, from refined Chern-Simons theory.

\subsec{Relation to the work of \wr\foot{Added in the revised version. }}

In \wr\  a physical approach to knot homology was proposed, based on studying gauge theory on D4-branes wrapping a four-manifolds with a boundary on the three-manifold $M$, where the knots live, times a thermal $S^1$ (there were other duality frames studied in \wr\ as well, but we will focus on this one, as it is closest to us).  The advantage of the approach initialized in \wr\ is that it provides one a way to get at knot homology groups themselves, not relying on indices that exist when $M$ is special. But, nevertheless it is important to note that the physical setting of \wr\ and the one we use here are related by a simple duality.

To define the refined Chern-Simons theory on a three-manifold $M$, we needed to study M-theory on $Y\times TN \times S^1$, where $Y = T^*M$ with $N$ M5 branes on $M\times {\IC}\times S^1$. Consider a dual description of this, by dimensionally reducing on the $S^1$ of the Taub-Nut space. Without M5 branes, we would obtain IIA string theory on the geometry,
$$
Y\times {\IR}^{3} \times S^{1}
$$
with a D6 brane wrapping  $Y \times S^{1}$ and sitting at the origin of ${\IR}^{3}.$   Adding the $N$ M5 branes on $M\times {\,\IC}\times S^1$, we get IIA string theory with the addition of $N$ D4 branes, wrapping $M\times {S^1}$ times a half-line ${\IR}_+$ in ${\IR}^3$, ending on the D6 brane. This is a D4 brane on a four manifold ${\IR}_+\times M\times S^1$, with the specific boundary condition imposed by the D6 brane. This setup is the same as that in \wr\ (see the discussion on the bottom of p. 13 of \wr\ and else where in the paper). Now consider how the symmetry generators map between the two pictures. From this we will deduce the index in IIA corresponding to the refined Chern-Simons partition function computed in M-theory, and recover in the unrefined limit, the index computation in \wr\ that gave rise to the Jones polynomial.

Before we add branes, the Taub-NUT geometry has an $SU(2)_{\ell} \times SU(2)_{r}$ isometry. We used the $U(1)_{\ell}\times U(1)_r$ subgroup of it in the definition of the index. The $U(1)_{\ell} \times U(1)_{r}$ act one the complex coordinates $(z_1, z_2)$ of the TN space by $(e^{i (\theta_{\ell} + \theta_r)/2}z_1, e^{i (-\theta_{\ell} + \theta_r)/2}z_2).$  Asymptotically, Taub-NUT looks like $S^{1} \times {\IR}^{3}$ and the $U(1)_{\ell}$ isometry rotates the $S^{1}$, while the $SU(2)_{r}$ rotates the base geometry.  So upon dimensional reduction, the charge under $U(1)_{{\ell}}$ becomes the D0-brane charge, while the charge under $SU(2)_{r}$ becomes the spin in the base ${\IR}^{3}$.
In addition to the this, IIA and M-theory have a common $SU(2)_R$ R-symmetry of a five-dimensional gauge theory.

The branes we add preserve the $U(1)_r$ subgroup of the $SU(2)_r$ rotation group, for any $M$. For any $M$, setting $q=t=q_0$, the partition function of the M5 brane theory \mfive\ equals the partition function of the D4 brane theory in this background
$$
Z_{D4}(T^*M, q_0)=   {\rm Tr}\, (-1)^{F}\, q_0^{Q_0}
$$
and both equal to the partition function of the ordinary $SU(N)$ Chern-Simons theory on $M$. In \wr, the Chern-Simons level arises due to non-zero value of the Wilson line of the RR 1-form potential $C$ in IIA string theory, $\int_{S^1} C$. This couples to D0 brane charge. It is the same as $\log q_0$, the chemical potential for the D0 branes turned on in our setting.
When $M$ is a Seifert three-manifold both the M5 brane, and the D4 brane theories should also preserve a $U(1)$ subgroup of the $SU(2)_R$ R-symmetry group of the five dimensional background, by the duality. Then, we can define the refined index \rindex , giving rise to the refined Chern-Simons theory, and depending on one more parameter.
The refined partition function \rindex\ becomes the partition function of the theory on $N$ D4 branes in this background
$$
Z_{D4}(T^*M, q_0,y)=   {\rm Tr}\, (-1)^{F}\, q_0^{Q_0}\, y^{2 J_3 -2{S}_R}.
$$
Here $q_{0}=\sqrt{qt}$, $y=\sqrt{q/t}$, $Q_{0}$ is the D0 brane charge, and $J_{3}$ is the generator of the rotation group in
${\IR}^{3}$, and $S_R$ is the generator of the $U(1)$.

\subsec{Superpolynomials From Refined Chern-Simons Theory}

We will see in this section that one can recover, in the large number of examples, the HOMFLY superpolynomial of \gr\ from our refined Chern-Simons theory. Since the super polynomial is a Poincar\'e polynomial, in this case of the knot homology theory categorifying the HOMFLY polynomial, whilst refined Chern-Simons theory computes an index, this is not a priori expected. However, it is not unprecedented either -- it simply requires that the states contributing to the index have vanishing $U(1)_R$ charge.\foot{Note added: A further explanation of this fact is provided in  \ref\AganagicNE{
  M.~Aganagic and S.~Shakirov,
  ``Refined Chern-Simons Theory and Knot Homology,''
[arXiv:1202.2489 [hep-th]].
}.} For now, based on a large number of examples,  we will simply conjecture that for Wilson lines in the
$R = \tableau{1}
$
fundamental representation, this is always the case. To relate our knot invariants to the superpolynomial one proceeds as follows.

\vskip 0.1cm

\item{1.}{ Compute the normalized expectation value of the knot observable of the refined $SU(N)$ Chern-Simons theory at level $k$, where we divide by the expectation value of the unknot,}
$$
 Z(S^3, K)/Z(S^3, \bigcirc).
$$
The result is a rational function of $q = e^{2\pi i/k+\beta N}$, $t = e^{2\pi i \beta /k+\beta N}$ and $t^N$. Here $\beta$ is the refinement parameter -- one recovers ordinary Chern-Simons theory at $\beta=1$.
\vskip 0.2 cm
\item{2.}{ Write the results by absorbing all the $N$ dependence into a parameter
\eqn\mapagain{\lambda = t^{N} t^{1/2}q^{-1/2}.}
\vskip 0.2 cm
\item{3.}{Rewrite the knot invariants in terms of new variables  $\bf{q}$, ${\bf t}$ and ${\bf a}$, related to the original ones by
\eqn\changeofv{{\bf q} = \sqrt{t}, \qquad {\bf t} = -\sqrt{q/t}, \qquad {\bf a} = \sqrt{\lambda },}
and denote the resulting knot invariant by
$$
Z(S^3, K)/Z(S^3, \bigcirc)={\cal P}(K)({\bf q}, {\bf a}, {\bf t}).
$$%
For ${\bf t} = -1$, the ${\cal P}(K)$ manifestly reduces to Jones polynomial by setting ${\bf a} = {\bf q}^2$, or the $SU(N)$  knot invariant, by setting ${\bf a} = {\bf q}^N$.
\vskip 0.2 cm
\item{4. }  The polynomial ${\cal P}(K)({\bf a}, {\bf q}, {\bf t})$, we conjecture, is the
superpolynomial, the Poincar\'e polynomial of knot homology theory categorifying the HOMFLY polynomial\foot{In mathematical literature, the knot polynomials where the unknot expectation value is set to one are termed "reduced". In this sense, our Poincar\'e polynomial is categorifying the reduced HOMFLY polynomial.}
$$
{\cal P}({\bf q},{\bf a}, {\bf t}) = \sum_{i,j} {\bf t}^i {\bf q}^j {\bf a}^k \dim {\cal H}^{i,j, k}(K).
$$

 \vskip 0.5 cm

There is a natural physical interpretation to the step 2. To relate to HOMFLY, one has to go to the large $N$ dual, the refined topological string on $X$. The latter is per definition counting BPS states of M2 branes ending on $L_K$ in $X$. If one recalls that
$
\lambda =  t^{N} t^{1/2}q^{-1/2}
$
in \mapagain\ is the degree counting parameter in $H_2(X, \IZ)$, c.f. \map , than it is clear that the step 2 is just the statement of large $N$ duality:  we reinterpret the refined Chern-Simons computations in the terms of the theory on $X$. In fact, were it not for the large $N$ duality, it would not have been clear that the change of variables is sensible.

Below, we will work out in detail the examples of $(2, 2k+1)$ torus knots for low values of $k$, then consider the example of a $(3, 4)$ torus knot, and conclude with a example of $(2,3)$ knot (trefoil), colored with the representation $\tableau{2}$ of $SU(N)$. The results of all these computations support our conjecture. We have also checked the conjecture for the $(3,5),(3,7)$ and $(3,8)$ knots.

\subsec{Trefoil knot}

\centerline{\includegraphics[width=3cm]{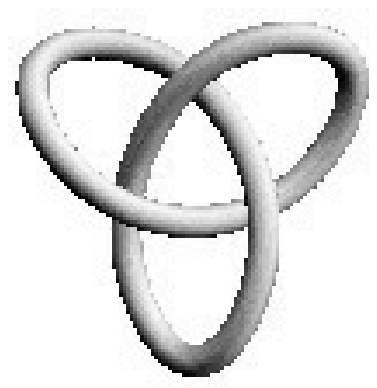}}
\bigskip

One of the simplest examples is the trefoil, which is a torus knot with winding numbers $(2,3)$. In this case, the matrix $K$ can be taken as $K = S^{-1} T^{-2} S^{-1} T^{-2}$ since it is easy to check that in the defining representation, where $S,T$ matrices have a form
\eqn\sldef{
S = \pmatrix{ 0 & -1 \cr
              1 &  0 \cr} , \;\; T = \pmatrix{1& 1\cr 0&1}
}
the $K$ matrix takes the $(1,0)$-cycle to the $(2,3)$-cycle
$$                    K =              S^{-1} T^{-2} S^{-1} T^{-2}   = \pmatrix{ -1 & 2 \cr
              -2 & 3 \cr}
$$
as it should. The amplitude then takes a form
$$
\langle 0 |{\cal O}^{(2,3)}_{\tableau{1}} S |0\rangle =   \sum_{j,k} {{(S^{-1} T^{-2} S^{-1} T^{-2} )^{0}}_k} \ {N^{k}}_{\tableau{1}, j} \ {(T^2 S^{-1} T^2)^{j}}_0.
$$
and the normalized amplitude takes a form
$$
Z(K_{2,3}) = \frac{\langle 0 |{\cal O}^{(2,3)}_{\tableau{1}} S |0\rangle}{\langle 0 |S|0\rangle}
$$
This amplitude is a finite sum of Macdonald polynomials, which can be efficiently evaluated with any computer algebra system. Using the refined $S$ and $T$ matrices given in s.5.2., it is easy to compute that
$$
\frac{Z(K_{2,3})}{Z(\bigcirc)} \Big|_{SU(N)} = t^{2-2N} q^{-3 + 3/N} + t^{3-2N} q^{-2+3/N} - t^{2-N} q^{-2+3/N}
$$
Making here a change of variables from $N,t,q$ to

$$
{\bf a}^2 = t^N \sqrt{t/q}, \ \ \ {\bf q} = \sqrt{t}, \ \ {\bf t} = -\sqrt{q/t}
$$
we find

$$
\frac{Z(K_{2,3})}{Z(\bigcirc)} \Big|_{SU(N)} = \left(\frac{t}{q}\right)^{5/2} \frac{q^{3/N}}{t^{3N}} \; {\cal P}(K_{2,3})
$$
where

$$
{\cal P}(K_{2,3})({\bf q}, {\bf a}, {\bf t}) = {\bf a}^2 {\bf q}^{-2} + {\bf a}^2 {\bf q}^2 {\bf t}^2 + {\bf a}^4 {\bf t}^3
$$
The trefoil knot invariant ${\cal P}(K_{2,3})({\bf q}, {\bf a}, {\bf t})$ of our refined Chern-Simons theory, agrees with the Poincar\'e polynomial of the knot homology theory categorifying the colored HOMFLY polynomial computed in \refs{\gr,\rasa,\rasb} . This quantity is called the superpolynomial of trefoil in \gr . See table 5.7 of that paper for comparison. We conclude that this superpolynomial is given by the corresponding amplitude of refined Chern-Simons theory. The prefactor here is easily related to framing: recall that
$$
T_{\tableau{1}} = \sqrt{\frac{t^{N+1}}{q^{1/N+1}}}
$$
so the prefactor corresponds to -6 units of framing
$$
\left(\frac{t}{q}\right)^{5/2} \frac{q^{3/N}}{t^{3N}} = \sqrt{\frac{q}{t}} \ T_{\tableau{1}}^{-6}
$$
modulo a $\sqrt{q/t}$ factor, which vanishes in the unrefined case. This factor should be related to a shift in the open flat coordinate $V$ on the brane wrapping $L_K$. One should be able to verify this by considering the higher representations.

\subsec{Example: the $(2,5)$ torus knot}

\centerline{\includegraphics[width=2.5cm]{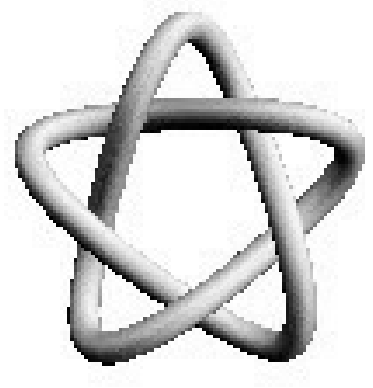}}
\bigskip

Next example along these lines is the torus knot with winding numbers $(2,5)$. The matrix $K$ can be taken as $K = S^{-1} T^{-3} S^{-1} T^{-2}$, since in the defining representation \sldef\
$$
 K = S^{-1} T^{-3} S^{-1} T^{-2} = \pmatrix{ -1 & 2 \cr
              -3 & 5 \cr}.
$$
The amplitude then takes a form
$$
Z(K_{2,5}) = \langle 0 | {\cal O}^{(2,5)}_{\tableau{1}} S |0\rangle =   \sum_{j,k} {(S^{-1} T^{-3} S^{-1} T^{-2} )}^{0}_{k} \ {N^{k}}_{\tableau{1}, j} \ {(T^2 S T^3S^2)^{j}}_0.
$$
Again, for any particular integer $N$ and $k$ this is a finite sum, which one can compute for many low values of $N$ and $k$ and then deduce the functional dependence of these parameters. Say, for $N=2$ we find, after a series of checks for $k = 1 \ldots 7$, the formula
$$
Z(K_{2,5})\Big|_{SU(2)} = i \big( q^{(7k-10)/4} - q^{(5k-10)/4} - q^{(5k-6)/4} + q^{(3k-2)/4} + q^{(k-6)/4} - q^{(-k-2)/4} \big),
$$
which, normalized by the unknot and rephrased in terms of $q,t$, states that
$$
\frac{Z(K_{2,5})}{Z(\bigcirc)} \Big|_{SU(2)} = q^{-1} t^{3} \big( 1 + tq + t^2 q^2 - t^2 q - t^3 q^2 \big)
$$
Repeating the same calculation for higher ranks gives
$$
\frac{Z(K_{2,5})}{Z(\bigcirc)} \Big|_{SU(N)} = \frac{q}{t} \ T^{-10}_{\tableau{1}} \ \big( t^{2N-1} q^{-1} + t^{2N} + t^{2N+1} q - t^{3N-1} - t^{3N} q \big)
$$
or, in terms of knot-theoretic variables $\bf a, \bf q, \bf t$
$$
\frac{Z(K_{2,5})}{Z(\bigcirc)} \Big|_{SU(N)} = \frac{q}{t} \ T^{-10}_{\tableau{1}} \ {\cal P}(K_{2,5})
$$
where

$$
{\cal P}(K_{2,5})\left({\bf q}, {\bf a}, {\bf t} \right) = {\bf a}^4 {\bf q}^{-4} + {\bf a}^4 {\bf t}^2 + {\bf a}^4 {\bf q}^4 {\bf t}^4 + {\bf a}^6 {\bf q}^{-2} {\bf t}^3 + {\bf a}^6 {\bf q}^2 {\bf t}^5
$$
is known (\gr , table 5.7) as the superpolynomial of (2,5) knot. This provides another check that refined CS theory amplitudes, associated to knots, give their superpolynomials.

\subsec{Example: the $(2,2m-1)$ torus knot}

\centerline{\includegraphics[width=2.5cm]{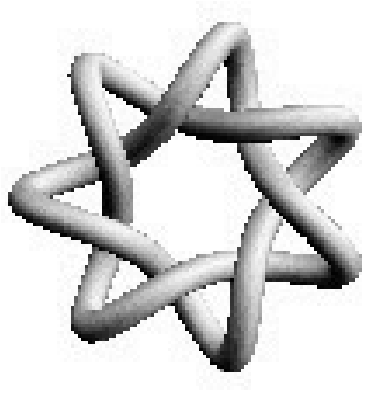}}
\bigskip

Generalizing the above example, it is just as easy to construct the amplitude for the $(2,2m-1)$ torus knot. For this, one can take $K = S^{-1} T^{-m} S^{-1} T^{-2}$, so that
$$
 K = T^{2} S T^{m} = \pmatrix{ -1 & 2 \cr
              -m & 2m-1 \cr}
$$
The amplitude then takes form
$$
Z(K_{2,2m-1}) = \langle 0 |S {\cal O}^{(2,2m-1)}_{\tableau{1}} |0\rangle = \sum_{j,k} {{(S^{-1} T^{-m} S^{-1} T^{-2} )^{0}}_k} \ {N^{k}}_{\tableau{1}, j} \ {(T^2 S T^m)^{j}}_0.
$$
Direct computation gives the following expression for this amplitude:
$$
\frac{Z(K_{2,2m-1})}{Z(\bigcirc)} \Big|_{SU(N)} = T^{-2(2m-1)}_{\tableau{1}} \ t^{Nm-m} \ \left[ t^{N+1} \sum\limits_{i=0}^{m-1} (tq)^i - \sum\limits_{i=1}^{m-1} (tq)^i \right].
$$
Expressed in terms of knot-theoretic variables $\bf a, \bf q, \bf t$, it takes form
$$
\frac{Z(K_{2,2m-1})}{Z(\bigcirc)} \Big|_{SU(N)} = \left(\sqrt{\frac{q}{t}}\right)^{m-1} \ T^{-2(2m-1)}_{\tableau{1}} \ {\cal P}(K_{2,2m-1}),
$$
where
$$
{\cal P}(K_{2,2m-1})\left({\bf q}, {\bf a}, {\bf t} \right) = {a}^{2m-2} \sum\limits_{i=0}^{m-1} {\bf q}^{4m-2i+2} {\bf t}^{2i} + {\bf a}^{2m} \sum\limits_{i=1}^{m-1} {\bf q}^{4i-2m} {\bf t}^{2i+1}.
$$
This function ${\cal P}(K_{2,2m-1})\left({\bf q}, {\bf a}, {\bf t} \right)$, obtained from refined Chern-Simons calculation, exactly agrees with existing formula for superpolynomials of (2,2m-1) knots (\gr , eq. (83)). This is already quite a non-trivial check of the general correspondence between triply graded superpolynomials of knots, and the refined Chern-Simons amplitudes.

\subsec{Example: the $(3,4)$ torus knot}

\centerline{\includegraphics[width=2.5cm]{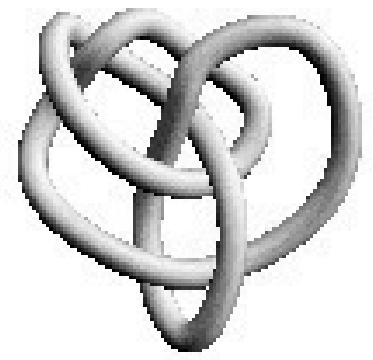}}

\bigskip
This is the first example where the first winding number is not 2. In this case, the matrix $K$ can be realized as $K = S^{-1} T^{-1} S T^{3}$ since in defining representation
$$
  K = S^{-1} T^{-1} S T^{3} = \pmatrix{ 1 & 3 \cr
              1 & 4 \cr}
$$
The amplitude then takes a form
$$
Z(K_{3,4}) = \langle 0 |S {\cal O}^{(3,4)}_{\tableau{1}} |0\rangle = \sum_{j,k} {{(S^{-1} T^{-1} S T^{3} )^{0}}_k} \ {N^{k}}_{\tableau{1}, j} \ {(T^{-3} S^{-1} T)^{j}}_0.
$$
Direct computation gives the following expression for this amplitude:
$$
\frac{Z(K_{3,4})}{Z(\bigcirc)} \Big|_{SU(N)} = \left(\frac{q}{t}\right)^{3/2} T^{-12}_{ \tableau{1}} \Big(
t^{3N-1/2} q^{1/2} + t^{3N+3/2} q^{1/2} + t^{3N+1/2} q^{1/2} +
$$
$$
+ t^{3N-1/2} q^{-1/2} + t^{3N-3/2} q^{5/2} - t^{4N+1/2} q^{3/2} - t^{4N-1/2} q^{3/2} - t^{4N-1/2} q^{1/2}
$$
$$
- t^{4N-3/2} q^{1/2} - t^{4N-3/2} q^{-1/2} + t^{5N-3/2} q^{3/2}
 \Big)$$
Expressed in terms of knot-theoretic variables $\bf a, \bf q, \bf t$, this takes form
$$
\frac{Z(K_{3,4})}{Z(\bigcirc)} \Big|_{SU(N)} = \left(\frac{q}{t}\right)^{3/2} \ T^{-12}_{ \tableau{1}} \ {\cal P}(K_{3,4})
$$
where
$$
{\cal P}(K_{3,4})\left({\bf q}, {\bf a}, {\bf t} \right) = {\bf a}^6 {\bf t}^4+{\bf a}^6 {\bf q}^6 {\bf t}^6+{\bf a}^6 {\bf q}^2 {\bf t}^4+{\bf a}^6 {\bf q}^{-2} {\bf t}^2+{\bf a}^6 {\bf q}^{-6}+
$$
$$
+ {\bf a}^8 {\bf q}^4 {\bf t}^7+{\bf a}^8 {\bf q}^2 {\bf t}^7+{\bf a}^8 {\bf t}^5+{\bf a}^8 {\bf q}^{-2} {\bf t}^5+{\bf a}^8 {\bf q}^{-4} {\bf t}^3+{\bf a}^{10} {\bf t}^8
$$
So obtained function ${\cal P}(K_{3,4})\left({\bf q}, {\bf a}, {\bf t} \right)$ exactly agrees with existing formula for the superpolynomial of the (3,4) knot (\gr , eq. (92)).

\subsec{Prediction: colored superpolynomial of the trefoil}

The amplitudes, calculated above, all correspond to the fundamental representation of the gauge group. This is in part because the fundamental representation is the most studied one, and there are a lot of knot-theory results to compare with. The aim of this section is to show that the method of present paper allows easy generalization to other representations: one just has to insert operators ${\cal O}_R$ corresponding to representations $R$. To illustrate this, we compute the trefoil amplitude in representation $\tableau{2}$ of $SU(N)$
$$
Z_{\tableau{2}}(K_{2,3}) = \frac{\langle 0 |{\cal O}^{(2,3)}_{\tableau{2}} S |0\rangle}{\langle 0 |S|0\rangle}
$$
which we normalize to the unknot in the same representation:
$$
Z_{\tableau{2}}(\bigcirc) = \frac{(t^{N/2}-t^{-N/2})(q t^{N/2}-t^{-N/2})}{(t^{1/2}-t^{-1/2})(q t^{1/2}-t^{-1/2})}
$$
Direct computation gives the following expression for this amplitude:
$$
\frac{Z_{\tableau{2}}(K_{2,3})}{Z_{\tableau{2}}(\bigcirc)} \Big|_{SU(N)} = \frac{q}{t} \ T_{\tableau{2}}^{-6} {\cal P}_{\tableau{2}}(K_{2,3})
$$
where
$$
{\cal P}_{\tableau{2}}(K_{2,3}) = q t^{2 N-3}+q^3 t^{2 N-2}-q^3 t^{3 N-3}-q^4 t^{3 N-3}+
$$
$$
+ q^4 t^{2 N-2}-q^5 t^{3 N-2}+q^5 t^{2 N-1}-q^6 t^{3 N-2}+q^6 t^{4 N-3} = $$
$$
= {\bf q}^4 {\bf t}^6 {\bf a}^4 + {\bf q}^{-4} {\bf a}^4+{\bf q}^2 {\bf t}^4 {\bf a}^4+{\bf t}^{8} {\bf q}^8 {\bf a}^4+{\bf q}^8 {\bf t}^{11} {\bf a}^6+{\bf q}^6 {\bf t}^{9} {\bf a}^6+{\bf t}^5 {\bf a}^6+{\bf q}^2 {\bf t}^7 {\bf a}^6+{\bf q}^6 {\bf t}^{12} {\bf a}^8
$$
Function ${\cal P}_{\tableau{2}}(K_{2,3})$ is our prediction for the $\tableau{2}$-colored superpolynomial of the trefoil, that coincides with results obtained by different methods (see, e.g. \coloredtrefoil).

\newsec{Some Future Directions}

Let us finish this paper with some directions for future work.

One generalization of our work corresponds to replacing $G=SU(N)$ by an arbitrary gauge group $G$. It is natural to conjecture that to compute the homological invariants based on the gauge group $G$, all one has to do is reinterpret the expressions for the $S$ and $T$ matrices we gave in section 5, in terms of the corresponding root system. The only essential change is that the effective value of $q$ becomes $q = e^{2\pi i \over k + \beta h}$, and $t =  e^{2\pi i \beta \over k + \beta h}$ where $h$ is the corresponding dual Coxeter number. The fact that the corresponding $S$ and $T$ matrices still provide a unitary representation of $SL(2,\IZ)$, for any semi-simple Lie algebra, was proven in \refs{\cheredniko, \kirila}. This alone is strong evidence in support of the conjecture. For classical groups this should follow from a fairly a straight forward modification of our setup.  To get a uniform description of the theory for all cases, one would presumably need to work with the more exotic realizations of the 6d $(2,0)$ theory that one finds in type IIB string theory compactified on the A-D-E singularities.\foot{Note added: The fact that refined Chern-Simons theory indeed exists, for arbitrary ADE group, was shown in \ref\AganagicAU{
  M.~Aganagic and K.~Schaeffer,
  ``Orientifolds and the Refined Topological String,''
[arXiv:1202.4456 [hep-th]].
}.}

Recently, \ref\rast{
  A.~Gadde, L.~Rastelli, S.~S.~Razamat, W.~Yan,
  ``The 4d Superconformal Index from q-deformed 2d Yang-Mills,''
[arXiv:1104.3850 [hep-th]].
}\foot{See also \ref\GaddeTE{
  A.~Gadde, L.~Rastelli, S.~S.~Razamat, W.~Yan,
  ``The Superconformal Index of the $E_6$ SCFT,''
JHEP {\bf 1008}, 107 (2010).
[arXiv:1003.4244 [hep-th]].
},\ref\GaddeEN{
  A.~Gadde, L.~Rastelli, S.~S.~Razamat, W.~Yan,
  ``On the Superconformal Index of N=1 IR Fixed Points: A Holographic Check,''
JHEP {\bf 1103}, 041 (2011).
[arXiv:1011.5278 [hep-th]].
}, for previous work.
}
\lref\nn{
  N.~A.~Nekrasov,
  ``Seiberg-Witten prepotential from instanton counting,''
Adv.\ Theor.\ Math.\ Phys.\  {\bf 7}, 831-864 (2004).
[hep-th/0206161].
}
computed a supersymmetric partition function of $N$ M5 branes on a Riemann surface $\Sigma$ times $S^3\times S^1$, the ${\cal N}=2$ superconformal index \ref\sci{J. Kinney, J. M. Maldacena, S. Minwalla and S. Raju, Commun. Math. Phys. 275, 209 (2007) [arXiv:hep-th/0510251]; C. Romelsberger, Nucl. Phys. B 747, 329 (2006) [arXiv:hep-th/0510060].}. For a specific choice of chemical potentials, the partition function is the same as the partition function of $SU(N)$ Chern-Simons theory on $\Sigma\times S^1$ (the partition function of two dimensional qYM theory, which \rast\ use, is the same as the Chern-Simons partition function, analytically continued to arbitrary $q$).  It is natural to conjecture that, with a more general choice of chemical potentials, one will recover the partition function of the refined Chern-Simons theory, on $\Sigma \times S^1$. This is natural, since the latter is the partition function of $N$ M5 branes on $\Sigma\times S^1$, as we saw in this paper.  This should also provide a physical explanation for why the answer in \rast\ is  the 2d qYM partition function, although some steps need to be filled in.

This paper opens up an avenue for understanding the refined topological string as a theory of Lagrangian D-branes. In the original topological vertex of \refs{\tv, \civ} ordinary Chern-Simons theory played a crucial role in solving the theory. The refined Chern-Simons theory should play an analogous role in the refined topological string. This should complement the beautiful recent work of \ref\nob{N. Nekrasov, A Okounkov, "The Index of M-theory", in preparation.} where the partition function of the refined topological string on arbitrary toric Calabi-Yau manifold $X$ was computed as the refined partition function of a D6 brane wrapping $X$.

\newsec{Acknowledgments}

We thank Sergei Cherkis,  Ivan Cherednik, Robbert Dijkgraaf, Eugene Gorsky, Sergei Gukov, Marcos Marino, Andrei Okounkov, Nicolai Reshetikhin, Steve Shenker, Cumrun Vafa, Ben Webster and Edward Witten for valuable discussions. We are especially grateful to Cumrun Vafa and Edward Witten for their insightful comments on a draft of this paper. The pictures of knots in section 7 are courtesy of E.Weisstein's World of Mathematics website. The research of MA is supported in part by the Berkeley Center for Theoretical Physics, by the National Science Foundation (award number 0855653), by the Institute for the Physics and Mathematics of the Universe, and by the US Department of Energy under Contract DE-AC02-05CH11231. The research of SS is supported in part by Ministry of Education and Science of the Russian Federation under contract 14.740.11.5194, by RFBR grant 10-01-00536 and by joint grants 09-02-93105-CNRSL, 09-02-91005-ANF.

\appendix{A}{ Macdonald Polynomials }

Macdonald polynomials $M_R$, parametrized by Young diagrams $R = (R_1 \geq R_2 \geq \ldots)$ of various size $|R| = \sum_i R_i $ form a useful basis in the space of symmetric polynomials. The purpose of this Appendix is to describe these polynomials and, at the same time, give disambiguations for all the basic notations and collect most important formulas. We start by giving two definitions of Macdonald polynomials. These are the definitions most commonly used in physical applications.

\subsec{Definition as orthogonal polynomials}

Polynomials $M_R$ form a unique basis orthogonal with respect to the integral Macdonald scalar product, defined on the space of polynomials in $N$ variables:
$$
\Big< f, g \Big> = \int\limits_{0}^{2\pi} d x_1 \ldots d x_N \ \Delta_{q,t} \ f\big( e^{i x_1}, \ldots, e^{i x_N} \big) g\left(e^{- i x_1}, \ldots, e^{- i x_N} \right)
$$
where $\Delta_{q,t}$ is the Macdonald measure
$$
\Delta_{q,t} = \prod\limits_{m = 0}^{\beta-1} \prod\limits_{I \neq J} \left(  e^{i (x_I - x_J)/2} - q^{m} e^{i (x_J - x_I)/2} \right)
$$
The orthogonality condition states
$$
< M_R, M_{R^{\prime}} > = g_R \delta_{R, R^{\prime}}
$$
where the quantity $g_R$ is known as the (integral) quadratic norm of Macdonald polynomials. In the context of present paper, we sometimes also call $g$ the metric, for the reasons which should be clear from the main text.

Explicitly, $g_R$ can be expressed in two equivalent forms. The first one is combinatorial:
$$
g_R = g_{\emptyset} \prod\limits_{(i,j) \in R} \frac{1 - t^{R^T_j-i} q^{R_i-j+1} }{1 - t^{R^T_j-i+1} q^{R_i-j} } \ \frac{1 - (t/q) t^{N-i} q^{j} }{1 - t^{N-i} q^{j}}
$$
where where $g_{\emptyset} = <1,1>$ is a constant, the product goes over all boxes $(i,j)$ of the Young diagram $R$ (namely, $1 \leq j \leq R_i$, $1 \leq i \leq {\rm length}(R)$) and $R^T$ is the transposed diagram to $R$ (namely, $(R^T)_j = $ the number of entries $\leq j$ in $R$).
The second one is rather Lie-theoretical:
$$
g_R = N! \prod_{m=0}^{\beta-1}
\prod_{\alpha>0}{q^{-{1\over 2}( \alpha, \lambda_R)}t^{ -{1\over 2} (\alpha,\rho)} q^{-{m\over 2}}-
q^{{1\over 2} (\alpha, \lambda_R)} t^{ {1\over 2} (\alpha,\rho)} q^{m\over 2}\over
q^{-{1\over 2}( \alpha, \lambda_R)}t^{ -{1\over 2} (\alpha,\rho)} q^{m\over 2}-
q^{{1\over 2} (\alpha, \lambda_R)} t^{ {1\over 2} (\alpha,\rho)} q^{-{m\over 2}}}
$$
where the product goes over all positive roots $\alpha$ of $SU(N)$ (namely, over $N(N-1)/2$ vectors $\alpha = e_I - e_J, \ I < J$ where $e_I$ are the basis vectors $(e_I)_j = \delta_{I,j}$), Weyl vector $\rho$ is the sum of all positive roots (namely, $\rho_j = (N+1)/2 - j$) the bracket is just the simple Euclidean product (namely, $(\alpha, v) = (e_I - e_J, v) = v_I - v_J$) and $\lambda_R$ is the highest weight vector in representation $R$ of $SU(N)$ (namely, $(\lambda_R)_j = R_j - |R|/N$).

\subsec{Generalized Cauchy-Stanley expansion}

Various functions can be expanded in the basis of Macdonald polynomials. One of the most basic such expansions is the expansion of the bilinear exponential:
$$
\exp\left( \sum\limits_{k = 1}^{\infty} \frac{1}{k} \frac{1-t^k}{1-q^k} \ p_k {\widetilde p}_k \right) = \sum\limits_{R} m_R M_R(p) M_R({\widetilde p})
$$
which is well-known as (generalized) Cauchy-Stanley identity. Here
$$
m_R = \prod\limits_{(i,j) \in R} \frac{1 - t^{R^T_j-i+1} q^{R_i-j} }{1 - t^{R^T_j-i} q^{R_i-j+1} }
$$
corresponds to the large $N$ limit of $g^{-1}_R$.

\subsec{Generalized Littlewood-Richardson coefficients}

Since Macdonald polynomials form a basis in the space of all symmetric polynomials, a product of two Macdonald polynomials should be expandable in this basis. This gives rise to a set of structure constants $N^{Y}_{P Q}$, the generalized Littlewood-Richardson coefficients:
$$
M_P M_Q = \sum\limits_{Y} N^{Y}_{P Q} M_Y
$$
A few first of these coefficients are:
$$
N^{\tableau{2}}_{\tableau{1}, \tableau{1}} = 1, \ \ \ \ N^{\tableau{1 1}}_{\tableau{1}, \tableau{1}} = \frac{(1+t)(1-q)}{(1-tq)}
$$
$$
N^{\tableau{3}}_{\tableau{2}, \tableau{1}} = 1, \ \ \ \ N^{\tableau{2 1}}_{\tableau{2}, \tableau{1}} = \frac{(1-qt^2)(1-q^2)}{(1-tq^2)(1-tq)}, \ \ \ \ N^{\tableau{1 1 1}}_{\tableau{2}, \tableau{1}} = 0
$$
$$
N^{\tableau{3}}_{\tableau{1 1}, \tableau{1}} = 0, \ \ \ \ N^{\tableau{2 1}}_{\tableau{1 1}, \tableau{1}} = 1, \ \ \ \ N^{\tableau{1 1 1}}_{\tableau{1 1}, \tableau{1}} = \frac{(1+t+t^2)(1-q)}{1-qt^2}
$$
Natural restriction of these coefficients to the set of representations of $SL(N)$ (i.e. diagrams with less than $N$ rows) gives the Verlinde coefficients, which we discussed in the main text.

%
%
%
%

\subsec{Specializations in $t,q$}

Macdonald polynomials generalize several previously known simpler bases of orthogonal polynomials. If one puts $t = q^{\beta}$ and then takes the limit $q \rightarrow 1$, one recovers the basis of Jack symmetric polynomials $J_R$

$$
\lim\limits_{q \rightarrow 1} M_R \Big|_{ t = q } = J_R \ \ \ {\rm associated\ with\ the\ measure} \ \prod\limits_{i<j}(z_i-z_j)^{2\beta}
$$
If, further, one takes $\beta = 1$, one recovers the Schur polynomials $S_R$, also denoted as $\chi_R$

$$
\lim\limits_{q \rightarrow 1} M_R \Big|_{ t = q } = \chi_R \ \ \ {\rm associated\ with\ the\ measure} \ \prod\limits_{i<j}(z_i-z_j)^{2}
$$
Notably, for Schur polynomials there is no need to take the $q \rightarrow 1$ limit: in fact, $M_R|_{ t = q } = \chi_R$ and does not depend on $q$. Let us also mention other important classes of symmetric polynomials, which can be recovered as particular cases of the Macdonald ones: the monomial symmetric polynomials (corresponding to the case $t = 1$) and the Hall-Littlewood polynomials (corresponding to the case $q = 0$).

\subsec{Specializations in $z$-variables}

In certain points, Macdonald polynomials take simple values. For example,
$$
M_R( t^{\rho} ) = \prod_{m=0}^{\beta - 1} \prod_{1\leq i<j\leq N}
{q^{{R_j - R_i\over 2}} t^{{i-j\over 2}} q^{-{m\over 2}}-q^{R_i - R_j\over 2} t^{j-i\over 2} q^{m\over 2}\over
 t^{{i-j\over 2}} q^{-{m\over 2}}-t^{j-i\over 2} q^{m\over 2}}
$$
and this is a generalization (refinement) of the well-known quantum dimension formula. A different formula for the same value is
$$
M_R( t^{\rho} ) = t^{||R^T||/2 - N |R|/2} \prod\limits_{(i,j) \in R} \frac{ 1 - (t/q) t^{N-i} q^{j} }{1 - t^{R^T_j-i+1} q^{R_i-j} }
$$
the product goes over all boxes $(i,j)$ of the Young diagram $R$.

\subsec{Specializations in $N$: the case $N = 2$ }

The case $N=2$ is exceptionally simple and for this reason noteworthy. In this case, it is possible to give an explicit formula for generic Macdonald polynomial:
$$
M_{[R_1, R_2]}(z_1, z_2) = z_1^{R_1} z_2^{R_2} \sum\limits_{l = 0}^{R_1 + R_2} \left(\frac{z_2}{z_1}\right)^{l} \prod\limits_{i = 0}^{l-1} \frac{[R_1 + R_2 -i]_q}{[R_1 + R_2 -i+\beta-1]_q} \frac{[i+\beta]_q}{[i+1]_q}
$$
where
$$
[x]_q = \frac{q^{x/2} - q^{-x/2}}{q^{1/2} - q^{-1/2}}
$$
is a $q$-number. These polynomials $M_{[R_1,R_2]}(z_1,z_2)$ are also known as $q$-ultraspherical polynomials \ref\ultrasp{
R.Askey and M.E.H.Ismail,
  ``A generalization of ultraspherical polynomials,''
  Studies in Pure Mathematics, Birkhauser, 1982, pp. 55-78.
}. Note, that at $\beta = 1$ they reduce to geometric sums, what gives for Schur polynomials an even shorter answer:
$$
\chi_{[R_1, R_2]}(z_1, z_2) = z_1^{R_1} z_2^{R_2} \sum\limits_{l = 0}^{R_1 + R_2} \left(\frac{z_2}{z_1}\right)^{l} =
\left(\frac{z_2}{z_1}\right)^{R_2} \frac{z_1^{|R|+1} - z_2^{|R|+1}}{z_1 - z_2}
$$
The existence of such simple formulas for $N=2$ implies that all the objects, constructed from Macdonald polynomials, for $N = 2$ can be likewise expressed by explicit formulas. For example, the $S$-matrix that we considered will have a form
$$
S_{[n][m]}/S_{00} = M_{[n]}(t^{\rho}) M_{[m]}(q^{[n]} t^{\rho}) = M_{[n]}(t^{1/2}, t^{-1/2}) M_{[m]}(t^{1/2} q^n, t^{-1/2} )
$$
where $M_{[n]}$ and $M_{[m]}$ are given by the explicit $q$-ultraspherical formula above. Note, that in the unrefined case $\beta = 1$ they again reduce to geometric sums, what gives
$$
S_{[n][m]}/S_{00} = \frac{q^{(n+1)/2} - q^{-(n+1)/2}}{q^{1/2}-q^{-1/2}} \frac{q^{(n+1/2)(m+1)} - q^{-(m+1)/2}}{q^{n+1/2}-q^{-1/2}} = q^{mn/2} [(m+1)(n+1)]_q
$$
-- the conventional S-matrix of unrefined Chern-Simons theory.

\appendix{B}{ The refined Chern-Simons matrix model}

The original Chern-Simons matrix model, which captures correctly the partition function of ordinary Chern-Simons TQFT on $S^3$, is given by
$$
Z_{N} = \frac{ (2\pi g)^{-N/2}}{N!} \int\limits_{-\infty}^{\infty} du_1 \ldots \int\limits_{-\infty}^{\infty} du_N \ \prod\limits_{i \neq j} \Big( e^{(u_i - u_j)/2} - e^{(u_j - u_i)/2} \Big) \ \exp\left( - \sum\limits_{i = 1}^{N} \frac{u_i^2}{2g} \right)
$$
with trigonometric Vandermonde factor. For natural $N$, partition function $Z_N$ is a polynomial in $q = e^{g}$ of degree $N(N^2-1)/6$, which is quite simple and nicely factorizable:
$$
Z_{N} = \prod\limits_{k = 1}^{N} \big( 1 - q^k \big)^{N-k}
$$
A naive refinement of this is to take the Vandermonde to power $\beta$ following what was done in \toda\ for some related, but simpler theories :
$$
\frac{ (2\pi g)^{-N/2}}{N!} \int\limits_{-\infty}^{\infty} du_1 \ldots \int\limits_{-\infty}^{\infty} du_N \ \prod\limits_{i \neq j} \Big( e^{(u_i - u_j)/2} - e^{(u_j - u_i)/2} \Big)^{\beta} \ \exp\left( - \sum\limits_{i = 1}^{N} \frac{u_i^2}{2g} \right)
$$
For natural $N,\beta$, as pointed out in \refmm , this partition function is a polynomial in $q = e^{g}$ of degree $\beta^2 N(N^2-1)/6$, but it is not be factorizable and quite messy -- it does not have the properties one would expect from the refined Chern-Simons theory. The correct matrix model is derived in section 3, using M-theory.

This also has an explanation in purely matrix-model terms. The explanation is related to $q$-nature of the Chern-Simons matrix model. There are two important deformations of matrix models -- $\beta$-deformation and $q$-deformation, and the original Chern-Simons model behaves as if it's already $q$-deformed. This is far from evident from the point of view of the integral itself -- the integral looks just like an ordinary eigenvalue matrix model, there are no $q$-integrations (Jackson sums) present. The reason to think that CS matrix model is a $q$-matrix model is just the structure of answers: the partition function $Z_N$ is a product of $q$-numbers, and so are the correllators in the model.
Following this point of view, it is clear
$$
\prod\limits_{i \neq j} (\lambda_i - \lambda_j) \ \ \mapsto \prod\limits_{i \neq j} (\lambda_i - \lambda_j)^{\beta}
$$
is not what is needed. The correct $\beta$-deformation is
$$
\prod\limits_{i \neq j} (\lambda_i - \lambda_j) \ \ \mapsto \prod\limits_{m = 0}^{\beta-1} \prod\limits_{i \neq j} (\lambda_i - q^m \lambda_j)
$$
which is typical for $q$-matrix models. The only not quite typical detail is that in this particular model parameter $q$ is fixed relative to the genus expansion parameter $g$: $q = e^g$. So we find the following  $\beta$-deformed Chern-Simons model:
$$
Z_{N,\beta} = \frac{ (2\pi g)^{-N/2}}{N!} \int\limits_{-\infty}^{\infty} du_1 \ldots \int\limits_{-\infty}^{\infty} du_N \ \prod\limits_{m = 0}^{\beta-1} \prod\limits_{i \neq j} \Big( e^{(u_i - u_j)/2} - q^m e^{(u_j - u_i)/2} \Big) \ \exp\left( - \sum\limits_{i = 1}^{N} \frac{u_i^2}{2g} \right)
$$
where $q = e^g$. Calculations confirm this expectation: the partition function equals

$$
Z_{N,\beta} = \prod\limits_{k = 1}^{N} \prod\limits_{m=0}^{\beta-1} \big( 1 - q^{\beta k + m} \big)^{N-k}
$$
the 1-Macdonald correllator equals
$$
\big< M_A \big> = t^{- |A|/2} \ \frac{(\sqrt{q})^{||A||}}{(\sqrt{t})^{||A^T||}} \ M_A \big( t^{i} \big)
$$
and the 2-Macdonald correllator equals
$$
\big< M_A M_B \big> = t^{N|B|} t^{|B|/2 - |A|/2} \ \frac{(\sqrt{q})^{||A|| + ||B||}}{(\sqrt{t})^{||A^T|| + ||B^T||}} \ M_A \big( t^{i} \big) M_B \big( q^{A} t^{-i} \big)
$$
These three equations (the second is a corollary of the third) summarize the finite $N$ solution of the refined Chern-Simons matrix model. These are all nicely factorisable results, which are in accordance with the higher-dimensional considerations, as explained in the main part of the paper.

It should be emphasized that the above formulas for the partition function and for Macdonald averages are not new:
in mathematics literature they are known as Cherednik-Macdonald-Mehta constant term identities \ek, an important chapter of modern representation theory. At this point it may be unclear to the reader, what is the relation between our integrals and constant-term evaluations. To illustrate this relation, let us actually prove the simplest of these results -- formula for the partition function. This simultaneously explains the connection to constant term identities.

First of all, changing variables via $u_i = \log x_i$, we rewrite the partition function as
$$
Z_{N,\beta} = \frac{ (2\pi g)^{-N/2}}{N!} \int\limits_{0}^{\infty} dx_1 \ldots \int\limits_{0}^{\infty} dx_N \ \prod\limits_{m = 0}^{\beta-1} \prod\limits_{i \neq j} \left( 1 - q^m \frac{x_i}{x_j} \right) \ \exp\left( - \sum\limits_{i = 1}^{N} \frac{(\log x_i)^2}{2g} \right)
$$
More convenient is to get rid of logarithms, using a simple identity of integrals
$$
\frac{1}{\sqrt{2\pi g}} \int\limits_{0}^{\infty} dx \ x^d \ \exp\left( -\frac{(\log x)^2}{2g}\right) = q^{d^2/2} = \oint\limits_{|z|=1} \frac{dz}{z} \ z^d \ \theta_{q}(z), \ \ \ d \in {\IZ}
$$
where
$$
\theta_{q}(z) = \prod\limits_{k=1}^{\infty} \Big(1-q^k\Big) \Big( 1 + q^k \frac{z}{\sqrt{q}} \Big) \Big( 1 + q^k \frac{1}{z\sqrt{q}} \Big) = \sum\limits_{n=-\infty}^{\infty} q^{n^2/2} z^n
$$
is the Jacobi theta function (called $\gamma$ in \ek). Using this, we obtain
$$
Z_{N,\beta} = \frac{1}{N!} \oint\limits_{|z_1|=1} \frac{dz_1}{z_1} \ldots \oint\limits_{|z_N|=1} \frac{dz_N}{z_N} \prod\limits_{m = 0}^{\beta-1} \prod\limits_{i \neq j} \left( 1 - q^m \frac{z_i}{z_j} \right) \ \prod\limits_{i = 1}^{N} \theta_q(z_i)
$$
Note, that in this form the partition function is reminishent of another important class of matrix models, namely the Selberg integrals
\ref\selb1{
A.Selberg,
"Bemerkningar om et multipelt integral",
Norsk. Mat. Tisdskr. 24 (1944) 71.
},
\ref\selb2{
P. J. Forrester and S. O. Warnaar,
"On the importance of Selberg integral",
Bull. Amer. Math. Soc. (N.S.) 45 (2008), 489-534,
[arXiv:0710.3981]
},
\ref\selb2{
A. Mironov, Al. Morozov and And. Morozov,
"Matrix model version of AGT conjecture and generalized Selberg integrals",
Nucl.Phys.B843:534-557,2011,
[arXiv:1003.5752]
}.

At this point it should be already obvious that the contour integrals just pick the constant-term contributions. Let us denote them as ${\rm C.T.}$, i.e. "the constant term of":
$$
{\rm C.T.} \ f(z_1, \ldots, z_N) \equiv \oint\limits_{|z_1|=1} \frac{dz_1}{z_1} \ldots \oint\limits_{|z_N|=1} \frac{dz_N}{z_N} f(z_1, \ldots, z_N)
$$
The answer is slightly simplified if the q-Pochhammer symbol is introduced:
$$
(a; q)_k = \prod\limits_{j = 0}^{k-1} (1 - a q^j)
$$
The Vandermonde then takes form
$$
\prod\limits_{m = 0}^{\beta-1} \prod\limits_{i \neq j} \left( 1 - q^m \frac{z_i}{z_j} \right) = \prod\limits_{i < j} \left( \frac{z_i}{z_j}; q \right)_\beta \left( \frac{z_j}{z_i}; q \right)_\beta
$$
the Jacobi theta function takes form
$$
\theta_{q}(z) = \left( -\sqrt{q} z; q\right)_\infty \left( \frac{-\sqrt{q}}{z}; q\right)_\infty (q;q)_\infty
$$
and, ultimately, the whole partition function takes form
$$
Z_{N,\beta} = \frac{1}{N!} \ {\rm C.T.} \ \prod\limits_{i < j} \left( \frac{z_i}{z_j}; q \right)_\beta \left( \frac{z_j}{z_i}; q \right)_\beta \ \prod\limits_{i = 1}^{N} \left( -\sqrt{q}z_i; q\right)_\infty \left( \frac{-\sqrt{q}}{z_i}; q\right)_\infty (q;q)_\infty
$$
and a simple algebraic transformation can be used to bring it into a form
$$
Z_{N,\beta} = \frac{(q;q)_\infty^N (1-q^\beta)^N}{(q^\beta;q^\beta)_N} \ {\rm C.T.} \ \prod\limits_{i < j} \left( \frac{z_i}{z_j}; q \right)_\beta \left( \frac{qz_j}{z_i}; q \right)_\beta \ \prod\limits_{i = 1}^{N} \left( \frac{z_i}{\sqrt{q}}; q\right)_\infty \left( \frac{1}{z_i\sqrt{q}}; q\right)_\infty
$$
This constant term evaluation is a particular case of the so-called Morris-Kadell formula (see Theorem 3 of \ref\kadell{
K.W.J.Kadell, "A proof of Askey's conjectured q-analogue of Selberg integral and a conjecture of Morris", SIAM J.Math.Anal.19, 1988, 969-986.}) for this C.T. in a more general situation, when the infinite products are substituted by finite ones of order $a$ and $b$:
$$
{\rm C.T.} \ \prod\limits_{i < j} \left( \frac{z_i}{z_j}; q \right)_\beta \left( \frac{qz_j}{z_i}; q \right)_\beta \ \prod\limits_{i = 1}^{N} \left( \frac{z_i}{\sqrt{q}}; q\right)_a \left( \frac{1}{z_i\sqrt{q}}; q\right)_b =
$$
$$ =
\frac{(q^\beta; q^\beta)_N}{(1-q^\beta)^N} \prod\limits_{i=1}^{N} \frac{(q;q)_{a+b+(i-1)\beta}}{(q;q)_{a+(n-i)\beta} (q;q)_{b+(i-1)\beta}} \prod\limits_{1\leq i<j\leq N} (q^{\beta(j-i)}; q)_\beta
$$
Accordingly, the partition function is recovered in the limit of $a,b\rightarrow \infty$:
$$
Z_{N,\beta} = (q;q)_\infty^N \prod\limits_{1\leq i<j\leq N} (q^{\beta(j-i)}; q)_\beta  \lim\limits_{a,b\rightarrow \infty}\prod\limits_{i=1}^{N} \frac{(q;q)_{a+b+(i-1)\beta}}{(q;q)_{a+(n-i)\beta} (q;q)_{b+(i-1)\beta}}
$$
The rightmost factor trivializes in this limit
$$
\lim\limits_{a,b\rightarrow \infty} \prod\limits_{i=1}^{N} \frac{(q;q)_{a+b+(i-1)\beta}}{(q;q)_{a+(n-i)\beta} (q;q)_{b+(i-1)\beta}} = \prod\limits_{i=1}^{N} \frac{(q;q)_{\infty}}{(q;q)_{\infty} (q;q)_{\infty}} = \frac{1}{(q;q)_{\infty}^N}
$$
and we finally obtain
$$
Z_{N,\beta} = \prod\limits_{1\leq i<j\leq N} (q^{\beta(j-i)}; q)_\beta = \prod\limits_{k = 1}^{N} \prod\limits_{m=0}^{\beta-1} \big( 1 - q^{\beta k + m} \big)^{N-k}
$$
what completes the proof. A more generic proof of the integral with two Macdonald polynomial insertions can be found in \ek.

\appendix{C}{Verlinde Formula}

The derivation is instructive so we will explain it. Consider, for definiteness
\vftwo. Per definition, $N_{ij{\bar k}}$ is the amplitude on $S^2\times S^1$ with knots in representations $R_i$, $R_j$ and ${\bar R}_k$ inserted, as we explained above. On the other hand, we can compute the same amplitude by
starting with the partition function on $T\times S^1$, where $T$ is a trinion, with wilson lines inserted around the three holes. The trinion itself computes a state in
${\cal H}_{T^2} \times {\cal H}_{T^2} \times {\cal H}^*_{T^2}$. This is becomes the amplitude we want upon capping off the three holes of the trinion with three solid tori, viewed as $D\times S^1$, where $D$ is a disk, and with Wilson lines $R_i$, ${R_j}$ and ${\bar R}_k$ inserted, one per disk. Make use of this, we first need to evaluate the trinion and the disk amplitude. We can proceed as follows. Consider first the amplitude on the solid torus, $D\times S^1$, with Wilson line around the boundary of the disk inserted. This differs from the amplitude
$|R_i\rangle$ we defined above in that the Wilson line is around the contractable $(1,0)$ cycle instead of the $(0,1)$ cycle, hence, it is given by
$$S|R_i\rangle.
$$

\centerline{\includegraphics[width=3.5cm]{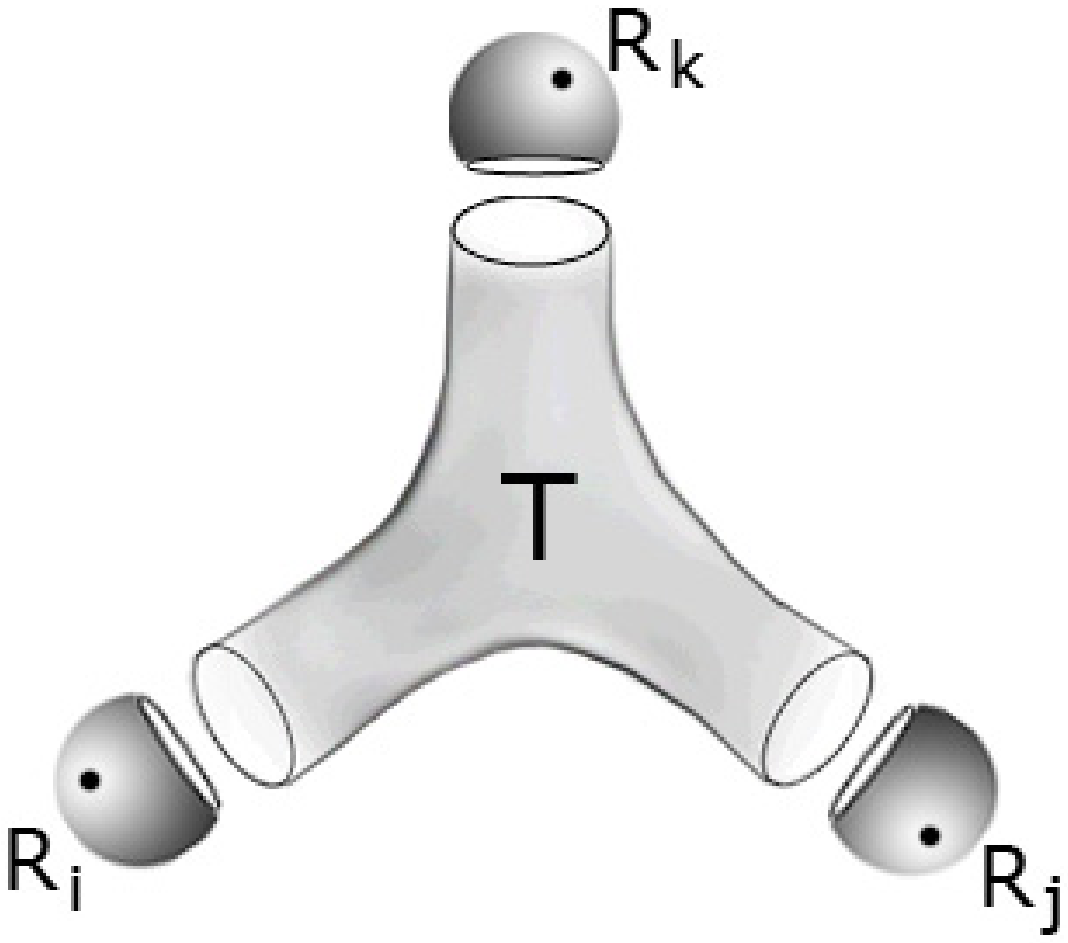} \ \ \ \ \ \ \ \ \ \ \ \ \ \ \ \ \ \includegraphics[width=6cm]{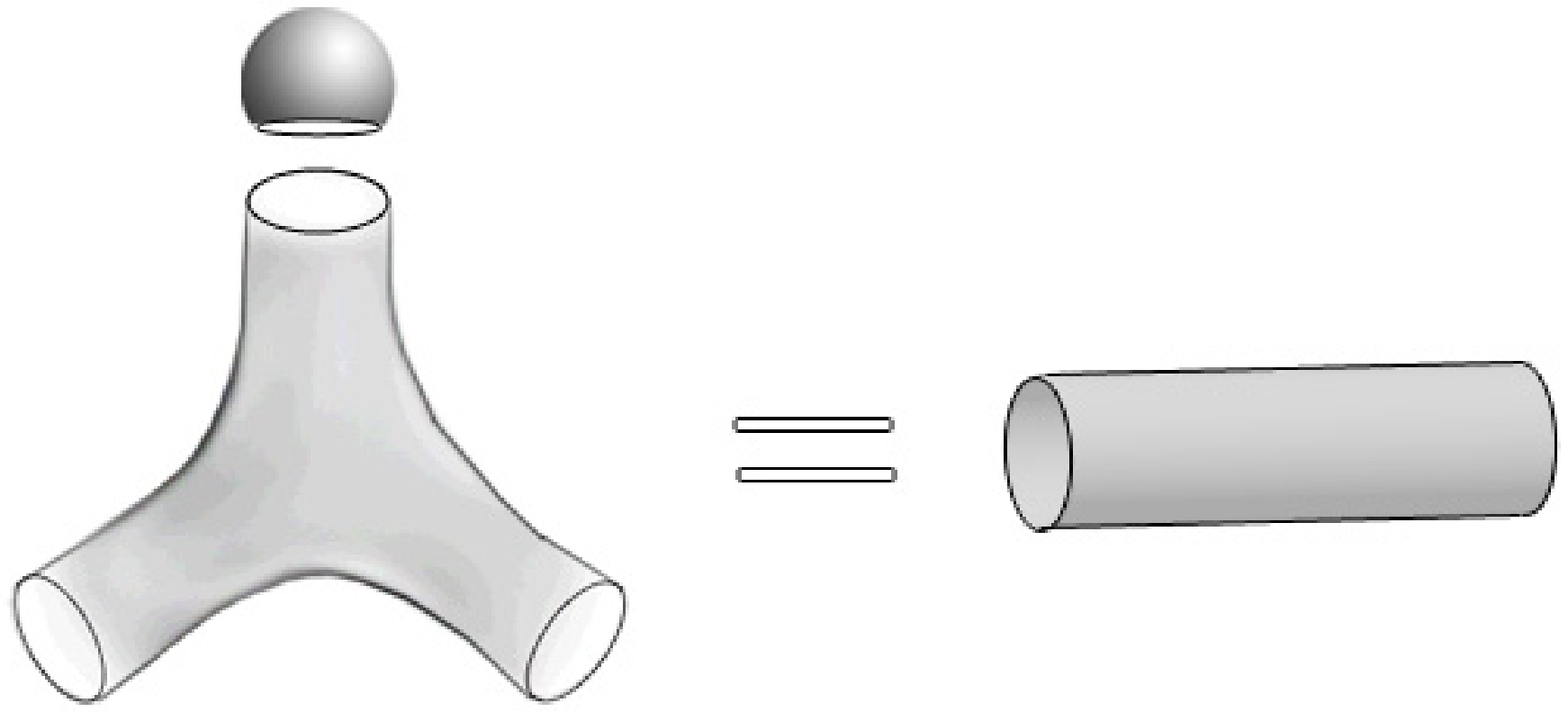}}
\noindent{\ninepoint \baselineskip=2pt {} }}
\bigskip

Next, consider the trinion. We would like to compute the state
\eqn\trin{
 \sum_{i,j,k} T^{ij {\bar k}} |R_i\rangle |R_j \rangle \langle R_k|
}
in ${\cal H}_{T^2} \times {\cal H}_{T^2} \times {\cal H}^*_{T^2}$ corresponding to it, where we define this in terms of the Wilson lines $R_i$ as inserted on $T$. To find the state we are after, consider capping off the hole corresponding to $R_i$ first. If we do cap it off, the result is per definition either a state in  ${\cal H}_{T^2}  \times {\cal H}^*_{T^2}$ corresponding to the propagator
$$
\sum_{j k} g^{j {\bar k} }|R_j \rangle \langle R_k|.
$$
where $g^{j {\bar k}}$ is the inverse metric, $g^{j {\bar k}} = (g_j)^{-1} \delta_{jk},$
or the state
$$
\sum_{j k} g^{j {\bar k} }|R_j \rangle  |R_k\rangle =\sum_{j} (g_{j})^{-1} |R_j \rangle  |R_j\rangle
$$
if we close off the outgoing hole.

Capping off, on the other hand corresponds to inserting $D \times S^1$, without any Wilson lines. Since the boundary in question is a torus, we can glue in either $|0\rangle$ or $S|0\rangle$, depending on which of the two circles we want capped off. It is not hard to see that the latter is the correct choice. Equating the two ways to look at the same amplitude,
%
%
one finds that the state  \trin\ can be written as
\eqn\trina{\eqalign{
 \sum_{i,j,k} T^{ij {\bar k}} |R_i\rangle |R_j \rangle \langle R_k|&=\sum_{i,k} {g^{i {\bar k}}\over S_{0 i} }\; |R_i\rangle\,|R_i \rangle \,\langle R_k|\cr &=\sum_{i} {( g_i S_{0 i})^{-1} } |R_i\rangle |R_i \rangle \langle R_i|
}}
Finally, to compute from this the $S^2\times S^1$ amplitude with Wilson lines in representations $R_i$, $R_j$ and $R_k$, we glue in three disks with these wilson lines, in such a way as to to cap off all three holes, as we described above. This gives, the right hand side of \vftwo .
This derivation used topological invariance of the three dimensional theory and nothing else.

\listrefs
\end